\documentclass[12pt]{article}
\pdfoutput=1
\usepackage{jheppub}
\usepackage{ytableau}
\usepackage{comment}
\usepackage[vcentermath]{youngtab}
\usepackage{tikz}

\newcommand\be{\begin{equation}}
\newcommand\ee{\end{equation}}

\newcommand\Tr{\mathrm{Tr}}

\preprint{
RUP-23-7
}

\title{Exact $\mathcal{N}=2^{*}$ Schur line defect correlators
}
\abstract{
We study the Schur line defect correlation functions in $\mathcal{N}=4$ and $\mathcal{N}=2^*$ $U(N)$ super Yang-Mills (SYM) theory. 
We find exact closed-form formulae of the correlation functions of the Wilson line operators in the fundamental, antisymmetric and symmetric representations 
via the Fermi-gas method in the canonical and grand canonical ensembles. 
All the Schur line defect correlators are shown to be expressible in terms of multiple series that generalizes the Kronecker theta function. 
From the large $N$ correlators we obtain generating functions for the spectra of the D5-brane giant and the D3-brane dual giant and find a correspondence between the fluctuation modes and the plane partition diamonds. 
}
\author[a]{Yasuyuki Hatsuda}
\author[b]{and Tadashi Okazaki}

\emailAdd{yhatsuda@rikkyo.ac.jp, tokazaki@seu.edu.cn}
\affiliation[a]{Department of Physics, Rikkyo University, Toshima, Tokyo 171-8501, Japan}
\affiliation[b]{
Shing-Tung Yau Center of Southeast University,\\
Yifu Architecture Building, No.2 Sipailou, Xuanwu district, Nanjing, Jiangsu, 210096, China}
\begin{document}
\maketitle

\section{Introduction and summary}
\label{sec_intro}

The superconformal indices \cite{Romelsberger:2005eg,Kinney:2005ej} of four-dimensional $\mathcal{N}= 2$ supersymmetric field theories 
allow for a specialization, known as the Schur indices \cite{Gadde:2011ik,Gadde:2011uv}. 
They can be viewed as supersymmetric partition functions on $S^1\times S^3$ that enumerate the BPS local operators 
annihilated by four supercharges. 
The Schur indices are identified with the vacuum characters of the associated chiral algebras \cite{Beem:2013sza}. 
For a class $\mathcal{S}$ theory they can be viewed as correlation functions of 2d TQFT on a Riemann surface \cite{Gadde:2011ik,Gadde:2009kb} and the closed-form expressions have been explored \cite{Pan:2021mrw,Beem:2021zvt}. 
The Schur indices can be decorated by adding the BPS line defects wrapping the $S^1$ and sitting at a point along a great circle in the $S^3$ \cite{Dimofte:2011py,Gang:2012yr,Drukker:2015spa,Cordova:2016uwk,Neitzke:2017cxz,Gaiotto:2020vqj}. 
\footnote{
The decoration can be also achieved by inserting BPS local operators 
\cite{Pan:2019bor,Dedushenko:2019yiw,Wang:2020oxs}. 
}
They can count the BPS local operators sitting at the endpoints of the supersymmetric line defects \cite{Cordova:2016uwk}, 
which we call the Schur line defect correlation functions. 

In this paper we study the Schur line defect correlation functions of $\mathcal{N}=2^*$ $U(N)$ super Yang-Mills (SYM) theory 
that is obtained by adding the mass term for the adjoint hypermultiplet in the $\mathcal{N}=4$ vector multiplet.%
\footnote{
See \cite{Buchel:2013id,Bobev:2013cja,Chen-Lin:2014dvz,Zarembo:2014ooa,Chen-Lin:2015dfa,Chen-Lin:2015xlh,Chen-Lin:2017pay,Liu:2017fiq} 
for the study of the correlation functions of Wilson loops in $\mathcal{N}=2^*$ SYM theory on $S^4$. 
}
They involve the fugacity associated to the adjoint mass parameter related to the R-symmetry 
so that they can be also understood as the flavored Schur line defect correlation functions of $\mathcal{N}=4$ $U(N)$ SYM theory. 
The Schur index and line defect correlators of $\mathcal{N}=4$ $U(N)$ SYM theory admit a systematic analysis based on the Fermi-gas formalism \cite{Bourdier:2015wda,Bourdier:2015sga,Drukker:2015spa,Gaiotto:2020vqj,Hatsuda:2022xdv}. 
They can be identified with canonical partition functions of quantum free Fermi-gas consisting of $N$ particles on a circle. 
We derive an exact closed-form formula of the line defect correlators via the Fermi-gas method. 
They can be expressed in terms of multiple series which generalizes the Kronecker theta function 
\cite{zbMATH02706826,MR1723749,MR1106744,MR2796409}. 
We refer to it as \textit{multiple Kronecker theta series}. 
It can be expanded with respect to 
the twisted Weierstrass functions \cite{Dong:1997ea,Mason:2008zzb}. 
From the Fermi-gas approach we further examine the grand canonical ensemble of the Schur line correlation functions. 
The grand canonical line defect correlation functions are shown to be expressed 
in terms of the generating functions of the multiple Kronecker theta series as well as the multiple Kronecker theta series themselves. 
They obey differential equations which lead to recursion relations of the canonical correlation functions. 
From our exact formulae, we also 
study the large $N$ limits of the Schur line defect correlation functions of $\mathcal{N}=4$ $U(N)$ SYM theory. 
They should encode the spectra of the excitations of the holographic dual $AdS_2$ geometry proposed in 
\cite{Maldacena:1998im,Rey:1998ik,Drukker:2005kx,Yamaguchi:2006tq,Gomis:2006sb,Rodriguez-Gomez:2006fmx,Hartnoll:2006hr,Gomis:2006im,Yamaguchi:2007ps}. 
We conjecture the exact forms for the large $N$ limit of the 2-point functions of the charged Wilson line operators 
and those in the rank-$m$ antisymmetric and symmetric representations. 
We find that the 2-point function of the Wilson line operators in the rank-$m$ symmetric and antisymmetric representation for $\mathcal{N}=4$ $U(N)$ SYM theory 
coincides with the generating function for the Schmidt type partitions, known as the \textit{plane partition diamonds} \cite{MR1868964,MR4370530} as $N\rightarrow \infty$ and $m\rightarrow \infty$. 
This leads to a correspondence between 
the fluctuation modes of the holographic dual D5-brane wrapping $AdS_2\times S^4$, \textit{D5-brane giant} or 
the D3-brane wrapping $AdS_2\times S^2$, \textit{D3-brane dual giant} and the plane partition diamonds. 

\subsection{Structure}
The organization of the paper is as follows. 
In section \ref{sec_setup} we start with the description of the Schur line defect correlation functions as matrix integrals including 
symmetric functions in the integrands. 
We summarize several useful formulae and properties of symmetric functions. 
We argue that in the half-BPS limit the flavored Schur index of $\mathcal{N}=4$ $U(N)$ SYM theory reduces to 
the measure of the Hall-Littlewood functions. 
In this limit, the closed-form formula of the Schur line defect correlators can be obtained in terms of Kostka-Foulkes polynomials. 
In section \ref{sec_Fermigas} we study the Fermi-gas formulation of the Schur line defect correlation functions of $\mathcal{N}=2^*$ $U(N)$ SYM theory. 
We show that the Schur line defect correlators can be expressed in terms of the multiple Kronecker theta series 
which generalizes the Kronecker theta function and that they can be also expressed in terms of the twisted Weierstrass functions. 
In section \ref{sec_GC} we analyze the grand canonical ensemble of the Schur line defect correlation functions. 
We find the exact closed-form expressions of the grand canonical Schur line defect correlation functions 
and the differential equations which lead to the recursion relations of the canonical correlators. 
In section \ref{sec_largeN} we investigate the large $N$ limit of the Schur line defect correlation functions. 
We discuss the holographic dual and combinatorial aspects of the large $N$ correlators. 
In Appendix \ref{app_def} we summarize the notations and definitions of the functions in this paper. 
In Appendix \ref{app_qxx} examples of the multiple Kronecker theta series and its relation to the twisted Weierstrass function are shown. 
In Appendix \ref{app_spectralZ} we present spectral zeta functions with higher orders. 

\subsection{Open questions}
There remain several interesting future works which we do not pursue in this paper. 
We expect that they can be addressed by using the closed formulae which we present in this work. 
\begin{itemize}

\item
The elliptic version of the Cauchy determinant formula plays a central role in the Fermi-gas analysis   
of the Schur indices and the Schur line defect correlators. 
In fact, there are some sort of generalizations of the Frobenius determinant formula 
\cite{MR2304342,MR3984000}. 
Also it would be intriguing to generalize our analysis to the cases with other gauge groups 
as well as other $\mathcal{N}=2$ supersymmetric gauge theories. 

\item 
Upon S-duality of $\mathcal{N}=4$ SYM theory the Wilson line operators map to the 't Hooft line and dyonic line operators. 
It would be nice to check S-duality of line operators by reproducing our analytic expressions from the dual descriptions 
by using difference operators \cite{MR1354144} or/and the monopole bubbling indices 
\cite{Ito:2011ea,Mekareeya:2013ija,Brennan:2018yuj,Brennan:2018rcn,Hayashi:2020ofu}. 

\item 
While $\mathcal{N}=2^*$ SYM theory is not conformal, its holographically dual supergravity background has been investigated \cite{Pilch:2000ue,Buchel:2000cn,Chen-Lin:2015xlh}. 
It would be interesting to examine the ratio of the index to the large $N$ index 
which can lead to a giant graviton expansion \cite{Gaiotto:2021xce}. 

\item 
$\mathcal{N}=2^*$ SYM theory possesses a rich phase structure \cite{Russo:2013qaa,Zarembo:2014ooa}. 
The correspondence between the Schur line defect correlators and the canonical partition functions of the quantum free Fermi-gas 
allows for various techniques of the Wigner method in quantum statistical mechanics. 
We hope to report the detailed analysis of the phase structure. 

\item 
The half-indices \cite{Dimofte:2011py,Gang:2012yr,Cordova:2016uwk,Gaiotto:2019jvo,Okazaki:2019ony} and quarter-indices \cite{Gaiotto:2019jvo,Okazaki:2019bok} of $\mathcal{N}=4$ SYM theory 
can be also decorated by the line defects. 
It would be interesting to explore their exact formulae and examine their analytic properties. 

\item
The twisted Weierstrass function \cite{Dong:1997ea,Mason:2008zzb} 
which appears in the expression of the Schur index and line defect correlation functions generates the quasi-Jacobi forms \cite{MR2796409}. 
It would be interesting to study the modular properties of the Schur line defect correlators and their physical implications of the BPS spectra. 

\item 
While the unflavored Schur indices are equivalent to the vacuum characters of the associated vertex operator algebras (VOAs) \cite{Beem:2013sza}, 
the unflavored Schur line correlation functions can be expressed as a linear combination of characters of certain modules for the VOAs \cite{Cordova:2016uwk}. 
We hope to investigate the connection to the VOA characters 
including the $\mathcal{N}=2$ $\widehat{\Gamma}(SU(N))$ SCFTs \cite{DelZotto:2015rca,Xie:2016evu,Buican:2016arp,Closset:2020scj,Closset:2020afy}
whose Schur line defect correlators are derived from our formulae by specializing the fugacity. 

\item 
Interestingly, the multiple Kronecker theta series which we introduce has 
a close relationship to the multiple $q$-zeta values ($q$-MVZs) 
\cite{schlesinger2001some,MR2069738,MR2111222,MR1992130,MR2341851,MR2322731,
MR2843304,MR3141529,okounkov2014hilbert,MR3338962,MR3473421,MR3522085,milas2022generalized}
and $q$-multiple polylogarithms ($q$-MPLs) 
\cite{schlesinger2001some,MR2341851,MR3687119},
which can enjoy $q$-shuffle relations. 
Since the algebra of the line operators in our setup of $\mathcal{N}=2^*$ SYM theory would coincide with the spherical DAHA \cite{MR3018956} (also see \cite{MR2133033,Cirafici:2020qlf,Gaiotto:2020vqj,Gukov:2022gei}) as the non-commutative deformation of the coordinate ring of the Coulomb branch \cite{Gaiotto:2010be,Drukker:2009id,Alday:2009fs,Gomis:2011pf}, 
it would be interesting to address it by presenting more general $q,t$-shuffle relations. 
More detailed investigation would be an interesting future work. 

\item 
The grand canonical line defect correlation functions are conjectured to enjoy a hidden symmetry 
\cite{Gaiotto:2020vqj}. It takes a similar form as the  triality symmetry of the grand canonical correlation function of the Coulomb branch operators in 
the 3d $\mathcal{N}=4$ $U(N)$ ADHM theory on $S^3$ \cite{Gaiotto:2020vqj,Hatsuda:2021oxa}.  
It would be intriguing to show the hidden symmetry analytically by further analyzing our exact formulae. 

\end{itemize}

\section{Line defect Schur correlators}
\label{sec_setup}

\subsection{Wilson line operators}
\label{sec_wilson}
\begin{figure}
\begin{center}
\includegraphics[width=14.0cm]{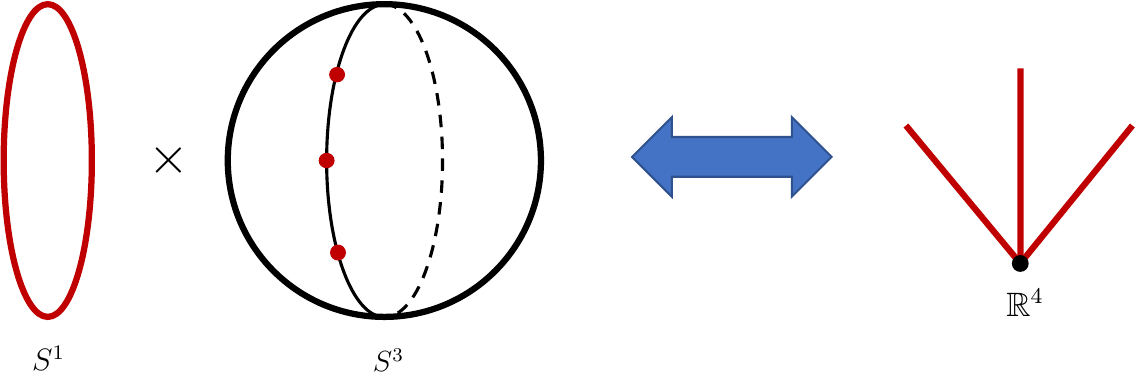}
\caption{
The line operators wrapping $S^1$ and inserted along a great circle in $S^3$ (left). 
The rays emanating from the origin in $\mathbb{R}^4$ (right). 
Since they map to one another under the conformal map, 
the Schur line defect correlators will count the BPS local operators living at a junction of lines. 
}
\label{fig_line1}
\end{center}
\end{figure}

A Wilson line operator is a non-local operator 
which is defined as a trace ${\Tr}_{\mathcal{R}} U$ in a representation $\mathcal{R}$ of a gauge or flavor group 
of the path-ordered exponential (i.e. holonomy matrix) $U$ 
for a given curve $L$. 
Let us consider a four-dimensional $\mathcal{N}=4$ $U(N)$ SYM theory on $S^1\times S^3$. 
We introduce 
the half-BPS Wilson line operators which wrap the $S^1$ and localize at points in the $S^3$. 
The supersymmetry can be preserved when the line operators sit along a great circle in the $S^3$ \cite{Cordova:2016uwk}.  
Upon a decompactification of the $S^1$ and a conformal map, they map to rays emanating from the origin in $\mathbb{R}^4$ 
(see Figure \ref{fig_line1}). 
\footnote{Unlike straight lines along $\mathbb{R}$ in $\mathbb{R}^4$ they have endpoint. 
They are also called the \textit{half line defects} \cite{Cordova:2016uwk}. 
}
When the two line operators are inserted at the north and its anti-counterpart at the south poles on the $S^3$, 
they map to the straight line in $\mathbb{R}^4$. 
The origin can preserve two supercharges 
and support local operators sitting at a junction of multiple rays. 

This setup can decorate the Schur index \cite{Gadde:2011ik,Gadde:2011uv} 
which can be regarded as a certain supersymmetric partition function of four-dimensional $\mathcal{N}\ge 2$ theories on $S^1\times S^3$. 
In the presence of the BPS line operators localized along a great circle in the $S^3$ it is interpreted as a correlation function of the line operators. 
We refer to it as the \textit{Schur line defect correlators}. 
The Schur line defect correlators are topological in that 
they do not depend on the distance between the inserted line operators. 
While without any insertion of the line operators the Schur index counts the BPS local operators annihilated by four supercharges, 
in the presence of a collection of the BPS line operators along a great circle in the $S^3$, 
the Schur line defect correlators would count the BPS local operators living at the junction of the rays annihilated by two supercharges. 
For the Schur indices and Schur line defect correlators of $\mathcal{N}=4$ SYM theory, 
one can introduce a fugacity $t$ associated with the difference of the Cartan generators of 
the $SU(2)_C$ and $SU(2)_H$ subgroups of the R-symmetry group $SU(4)_R$. 
We call them the flavored Schur indices and flavored Schur line defect correlators. 
They reduce to the unflavored ones by setting $t$ to unity. 
Alternatively, introducing the fugacity $\xi$ $=$ $q^{-1/2}t^2$ $=$ $e^{2\pi i\zeta}$ this is interpreted as the Schur index of $\mathcal{N}=2^*$ SYM theory 
whose mass parameter of the $\mathcal{N}=2$ adjoint hypermultiplet is $\zeta$. 

%
%
%
%
%

\subsection{Line defect Schur correlators}
\label{sec_wilson}
For $\mathcal{N}=4$ $U(N)$ SYM theory the flavored Schur correlation function of the $k$ half-BPS Wilson line operators $W_{\mathcal{R}_j}$, $j=1,\cdots, k$ 
\footnote{
While the Schur line defect 2-point functions (i.e. $k=2$) for $\mathcal{N}=4$ SYM theory have been studied in \cite{Gang:2012yr,Drukker:2015spa},  
we consider more general Schur line defect correlation functions. 
}
transforming as the representation $\mathcal{R}_j$ under the gauge group $U(N)$ 
can be evaluated from a matrix integral \cite{Gang:2012yr} 
\begin{align}
\label{W_integral}
&\langle W_{\mathcal{R}_1}\cdots W_{\mathcal{R}_k}\rangle^{U(N)} (t;q)
\nonumber\\
&=
\frac{1}{N!}
\frac{(q;q)_{\infty}^{2N}}{(q^{\frac12}t^{\pm2};q)_{\infty}^{N}}
\oint \prod_{i=1}^{N}
\frac{d\sigma_i}{2\pi i \sigma_i} 
\frac{
\prod_{i\neq j}
\left(\frac{\sigma_i}{\sigma_j};q\right)_{\infty} \left(q \frac{\sigma_i}{\sigma_j};q\right)_{\infty}}
{\prod_{i\neq j}\left(q^{\frac12} t^2 \frac{\sigma_i}{\sigma_j};q\right)_{\infty} \left(q^{\frac12} t^{-2} \frac{\sigma_i}{\sigma_j};q\right)_{\infty}}
\prod_{j=1}^{k}\chi_{\mathcal{R}_{j}}(\sigma), 
\end{align}
where the integration contour is chosen as a unit torus $\mathbb{T}^N$. 
It is a formal Taylor series in $q^{1/2}$ and 
its coefficients are Laurent polynomial in $t$ with integer coefficients. 
\footnote{
We follow the same notation and definition in \cite{Gaiotto:2019jvo,Hatsuda:2022xdv} for the flavored Schur index of $\mathcal{N}=4$ SYM theoy. 
}
Here $\chi_{\mathcal{R}_j}(\sigma)$ is a character of the representation $\mathcal{R}_j$. 
Physically, it corresponds to the classical value of the BPS Wilson line operator 
whose holonomy matrix is specified by gauge fields along the $S^1$. 
We have used a shorthand notation $(q^{\frac12}t^{\pm2};q)_{\infty}=(q^{\frac12}t^{2};q)_{\infty}(q^{\frac12}t^{-2};q)_{\infty}$.
The correlation function (\ref{W_integral}) is obviously invariant under the transformation 
\begin{align}
\label{t_transf}
t&\rightarrow t^{-1}, 
\end{align}
under which two $SU(2)$ subgroups of the $SU(4)$ R-symmetry are swapped. 
In the absence of the line operators, 
the flavored Schur line defect correlator (\ref{W_integral}) reduces the flavored Schur index $\mathcal{I}^{U(N)}$. 
The exact closed-form of the flavored Schur index is explored in \cite{Hatsuda:2022xdv}. 

\subsection{Symmetric functions}
The characters of the representations under the gauge group appearing in the matrix integral (\ref{W_integral}) is presented 
as certain symmetric functions in gauge fugacities $\sigma_i$. 

The Wilson line operator $W_{n}$ with charge $n\in \mathbb{Z}$ in $U(N)$ SYM theory is specified by the character given by the $n$-th power sum symmetric function in $N$ variables
\begin{align}
p_n(\sigma)&=\sum_{i=1}^{N}\sigma_i^n. 
\end{align}
The generating function for the Wilson line operators $W_{n}$ with charge $n$ is 
\begin{align}
P(s;\sigma)&=
\sum_{n=1}^{\infty}p_n(\sigma) s^n
=\sum_{i=1}^{N}\frac{s}{1-s\sigma_i}
=s \frac{\partial}{\partial s} \log \frac{1}{\prod_{i=1}^N (1-s\sigma_i)}. 
\end{align}

The Wilson line operator $W_{(1^k)}$ in the rank-$k$ antisymmetric representation is described by the character given by the $k$-th elementary symmetric function  
\begin{align}
e_k(\sigma)&=\sum_{1\le i_1< i_2 < \cdots < i_k \le N} \sigma_{i_1}\sigma_{i_2}\cdots \sigma_{i_k}. 
\end{align}
The generating function for the Wilson line operators $W_{(1^k)}$ in the antisymmetric representation reads
\begin{align}
\label{E_gene}
E(s;\sigma)&=\sum_{k=0}^{\infty} e_k(\sigma) s^k
=\prod_{i=1}^{N}(1+s\sigma_i). 
\end{align}
The elementary symmetric function can be expressed as a specialization of the Schur function $s_{\lambda}(\sigma)$
\begin{align}
e_k(\sigma)&=s_{(1^k)}(\sigma). 
\end{align}

The Wilson line operator $W_{(k)}$ in the rank-$k$ symmetric representation for $U(N)$ SYM theory is characterized by the complete homogeneous symmetric polynomial of degree $k$ in $N$ variables 
\begin{align}
h_k(\sigma)&=\sum_{1\le i_1 \le i_2 \le \cdots \le i_k \le N} \sigma_{i_1}\sigma_{i_2}\cdots \sigma_{i_k}. 
\end{align}
The generating function for the Wilson line operators $W_{(k)}$ in the symmetric representation is
\begin{align}
\label{H_gene}
H(s;\sigma)&=\sum_{k=0}^{\infty} h_k(\sigma) s^k
=\prod_{i=1}^{N}\frac{1}{1-s\sigma_i}. 
\end{align}
The complete homogeneous symmetric function can be expressed as a specialization of the Schur function $s_{\lambda}(\sigma)$
\begin{align}
h_k(\sigma)&=s_{(k)}(\sigma). 
\end{align}

\subsubsection{Newton's identities}
Newton's identities state that 
\begin{align}
ke_k(\sigma)&=\sum_{i=1}^k (-1)^{i-1}e_{k-i}(\sigma)p_i(\sigma). 
\end{align}
It implies that 
\begin{align}
\label{e_p}
e_k(\sigma)&=
\sum_{\lambda}(-1)^{k-r}\prod_{i=1}^{r}\frac{1}{\lambda_i^{m_i}(m_i !)}p_{\lambda_i}(\sigma)^{m_i}, 
\end{align}
where the sum is taken over all possible partitions $\lambda$ of $k=\sum_{i=1}^{r}\lambda_i m_i$ with $\lambda_1>\lambda_2>\cdots>\lambda_r$. 
Similarly, it follows that
\begin{align}
kh_k(\sigma)&=\sum_{i=1}^kh_{k-i}(\sigma)p_i(\sigma). 
\end{align}
Hence we have
\begin{align}
\label{h_p}
h_k(\sigma)&=
\sum_{\lambda}\prod_{i=1}^{r}\frac{1}{\lambda_i^{m_i}(m_i !)}p_{\lambda_i}(\sigma)^{m_i}. 
\end{align}
As each of families $\{p_k(\sigma)\}$, $\{e_k(\sigma)\}$ and $\{h_k(\sigma)\}$ generates the ring of symmetric polynomials as a polynomial ring. 
According to the relations (\ref{e_p}) and (\ref{h_p}), 
the correlation functions of the Wilson line operators in the rank-$k$ antisymmetric and symmetric representations can be expressed 
as linear combinations of those of the Wilson lines with fixed charges $n\le k$. 

For example, let us consider the 2-point function of the Wilson line operators in the conjugate representations $\mathcal{R}$ and $\overline{\mathcal{R}}$
\begin{align}
\label{asym2_int}
&
\langle W_{\mathcal{R}}W_{\overline{\mathcal{R}}}\rangle^{U(N)}
\nonumber\\
&=
\frac{1}{N!}
\frac{(q;q)_{\infty}^{2N}}{(q^{\frac12}t^{\pm2};q)_{\infty}^{N}}
\oint \prod_{i=1}^{N}
\frac{d\sigma_i}{2\pi i \sigma_i} 
\frac{
\prod_{i\neq j}
\left(\frac{\sigma_i}{\sigma_j};q\right)_{\infty} \left(q \frac{\sigma_i}{\sigma_j};q\right)_{\infty}}
{\prod_{i\neq j}\left(q^{\frac12} t^2 \frac{\sigma_i}{\sigma_j};q\right)_{\infty} \left(q^{\frac12} t^{-2} \frac{\sigma_i}{\sigma_j};q\right)_{\infty}}
\chi_{\mathcal{R}}(\sigma) \chi_{\overline{\mathcal{R}}}(\sigma), 
\end{align}
where $\chi_{\overline{\mathcal{R}}}(\sigma)$ $=$ $\chi_{\mathcal{R}}(\sigma^{-1})$. 
According to the relations  (\ref{e_p}) and (\ref{h_p})
we have 
\begin{align}
\label{asym2_power}
&
\langle W_{(1^2)}W_{\overline{(1^2)}}\rangle^{U(N)}
\nonumber\\
&=\frac14 \Bigl[
\langle W_1W_1W_{-1}W_{-1}\rangle^{U(N)}
-2\langle W_{1}W_{1}W_{-2}\rangle^{U(N)}
+\langle W_2W_{-2}\rangle^{U(N)}
\Bigr], \\
\label{sym2_power}
&
\langle W_{(2)}W_{\overline{(2)}}\rangle^{U(N)}
\nonumber\\
&=\frac14 \Bigl[
\langle W_1W_1W_{-1}W_{-1}\rangle^{U(N)}
+2\langle W_{1}W_{1}W_{-2}\rangle^{U(N)}
+\langle W_2W_{-2}\rangle^{U(N)}
\Bigr]
\end{align}
and 
\begin{align}
\label{asym3_power}
&
\langle W_{(1^3)}W_{\overline{(1^3)}}\rangle^{U(N)}
\nonumber\\
&=\frac{1}{36}
\Biggl[
\langle W_1W_1W_1W_{-1}W_{-1}W_{-1}\rangle^{U(N)}
-6\langle W_1W_1W_1W_{-1}W_{-2}\rangle^{U(N)}
+4\langle W_1W_1W_1W_{-3}\rangle^{U(N)}
\nonumber\\
&
+9\langle W_1W_2W_{-1}W_{-2}\rangle^{U(N)}
-12\langle W_1W_1W_2W_{-3}\rangle^{U(N)}
+\langle W_3W_{-3}\rangle^{U(N)}
\Biggr], 
\\
\label{asym3_power}
&
\langle W_{(3)}W_{\overline{(3)}}\rangle^{U(N)}
\nonumber\\
&=\frac{1}{36}
\Biggl[
\langle W_1W_1W_1W_{-1}W_{-1}W_{-1}\rangle^{U(N)}
+6\langle W_1W_1W_1W_{-1}W_{-2}\rangle^{U(N)}
+4\langle W_1W_1W_1W_{-3}\rangle^{U(N)}
\nonumber\\
&
+9\langle W_1W_2W_{-1}W_{-2}\rangle^{U(N)}
+12\langle W_1W_1W_2W_{-3}\rangle^{U(N)}
+\langle W_3W_{-3}\rangle^{U(N)}
\Biggr]. 
\end{align}

\subsubsection{Irreducible power sum symmetric functions}
Let $\lambda=(\lambda_1,\lambda_2,\cdots,\lambda_r)$ be a partition of weight $|\lambda|$ and $I=\{1,2,\cdots,k-1\}$ a set of integers. 
Given the partition $\lambda$ and the set $I$ we consider a decomposition $I=\bigoplus_{i=1}^{r}I_i$ with the conditions $I_i\cap I_j=\emptyset$ and $|I_i|=\lambda_i$. 
We then recursively define irreducible elements of products of $k$ power sum symmetric functions 
$p_{n_1},\cdots,p_{n_{k-1}}, p_{-n_1-\cdots-n_{k-1}}$ by
\begin{align}
&
\mathfrak{p}_{\{n_1,\cdots, n_{k-1}, -n_1-\cdots-n_{k-1}\}}
\nonumber\\
&:=
p_{n_1}p_{n_2}\cdots p_{n_{k-1}}p_{-n_1-n_2-\cdots-n_{k-1}}
\nonumber\\
&-
\sum_{j=1}^{k-1} 
\sum_{\begin{smallmatrix}
\lambda=(\lambda_1,\cdots,\lambda_r)\\
|\lambda|=j,\\
r\le k-2\\
\end{smallmatrix}}
\sum_{\{I_1,\cdots,I_r\}} 
\mathfrak{p}_{
\left\{
\sum_{i^{(1)}\in I_1} n_{i^{(1)}}, 
\cdots
\sum_{i^{(r)}\in I_r} n_{i^{(r)}}, 
-\sum_{\alpha=1}^{r}\sum_{i^{(\alpha)}\in I_\alpha} n_{i^{(\alpha)}}
\right\}
}. 
\end{align}
Here the sum $\sum_{\{I_1,\cdots,I_r\}}$ is taken over all the possible combinations of subsets of integers. 
For example, we have 
\begin{align}
&
\mathfrak{p}_{\{n_1,-n_1\}}
=p_{n_1}p_{-n_1},\\
&
\mathfrak{p}_{\{n_1,n_2,-n_1-n_2\}}
=
p_{n_1}p_{n_2}p_{-n_1-n_2}
-\mathfrak{p}_{\{n_1,-n_1\}}-
\mathfrak{p}_{\{n_2,-n_2\}}
-\mathfrak{p}_{\{n_1+n_2,-n_1-n_2\}}, \\
%
&
\mathfrak{p}_{\{n_1,n_2,n_3,-n_1-n_2-n_3\}}
=
p_{n_1}p_{n_2}p_{n_3}p_{-n_1-n_2-n_3}
-\mathfrak{p}_{\{n_1,-n_1\}}
-\mathfrak{p}_{\{n_2,-n_2\}}
-\mathfrak{p}_{\{n_3,-n_3\}}
\nonumber\\
&-\mathfrak{p}_{\{n_1+n_2,-n_1-n_2\}}
-\mathfrak{p}_{\{n_1+n_3,-n_1-n_3\}}
-\mathfrak{p}_{\{n_2+n_3,-n_2-n_3\}}
-\mathfrak{p}_{\{n_1+n_2+n_3,-n_1-n_2-n_3\}}
\nonumber\\
&-\mathfrak{p}_{\{n_1,n_2,-n_1-n_2\}}
-\mathfrak{p}_{\{n_1,n_3,-n_1-n_3\}}
-\mathfrak{p}_{\{n_2,n_3,-n_2-n_3\}}
\nonumber\\
&
-\mathfrak{p}_{\{n_1,n_2+n_3,-n_1-n_2-n_3\}}
-\mathfrak{p}_{\{n_2,n_1+n_3,-n_1-n_2-n_3\}}
-\mathfrak{p}_{\{n_3,n_1+n_2,-n_1-n_2-n_3\}}. 
\end{align}

For $k>N$ the products of power sum symmetric functions 
corresponding to the $k$-point functions 
can be decomposed into a sum of products which have at most $N$ power sum symmetric functions 
corresponding to the 2-, 3-, $\cdots$, $N$-point functions and a constant term. 
It follows that 
\begin{align}
\label{k_psum}
&
p_{n_1} p_{n_2}\cdots p_{n_{k-1}}p_{-n_1-n_2-\cdots-n_{k-1}}
\nonumber\\
&=
N\sum_{i=1}^{N} (-1)^{N-i} (N-i)! S(k,N-i+1)
\nonumber\\
&+
\sum_{j=1}^{k-1} 
\sum_{\begin{smallmatrix}
\lambda=(\lambda_1,\cdots,\lambda_r)\\
|\lambda|=j,\\
r\le N-1\
\end{smallmatrix}}
\sum_{\{I_1,\cdots,I_r\}} 
\mathfrak{p}_{
\{
\sum_{i^{(1)}\in I_1} n_{i^{(1)}}, 
\cdots, 
\sum_{i^{(r)}\in I_r} n_{i^{(r)}}, 
-\sum_{\alpha=1}^{r}\sum_{i^{(\alpha)}\in I_\alpha} n_{i^{(\alpha)}}
\}
}, 
\end{align}
where $S(n,k)$ are the Stirling numbers of the second kind. 
According to the relation (\ref{k_psum}), 
the $k$-point functions of the Wilson line operators in $U(N)$ SYM theory for $k>N$
can be built up from the $2$-, $3$-, $\cdots$, $N$-point functions. 

For example, for $N=2$ the partitions with a single row only contribute in the sum. 
They are $\lambda=(\lambda_1)=(j)$ with $1\le j\le k-1$ and correspond to the 2-point functions. 
Hence the $k$-point function of the charged Wilson line operators in $U(2)$ SYM theory can be simply decomposed into a sum of the 2-point functions 
according to the following relation: 
\begin{align}
\label{k_psum2}
&
p_{n_1}p_{n_2}\cdots p_{n_{k-1}}p_{-n_1-n_2-\cdots-n_{k-1}}
\nonumber\\
&=
2(-S(k,2)+S(k,1))+
\sum_{j=1}^{k-1}p_{n_j}p_{-n_j}
+\sum_{j_1<j_2} p_{n_{j_1}+n_{j_2}} p_{-n_{j_1}-n_{j_2}}
\nonumber\\
&+\cdots
+\sum_{j_1<j_2<\cdots<j_{k-2}} p_{n_{j_1}+n_{j_2}+\cdots+n_{j_{k-2}}} p_{-n_{j_1}-n_{j_2}-\cdots-n_{j_{k-2}}}
\nonumber\\
&+p_{n_{j_1}+n_{j_2}+\cdots+n_{j_{k-1}}} p_{-n_{j_1}-n_{j_2}-\cdots-n_{j_{k-1}}}. 
\end{align}
The $k=3,4$ and $5$-point functions of the charged Wilson line operators read 
\begin{align}
\label{u2_3pt_red}
&
\langle W_{n_1} W_{n_2} W_{-n_1-n_2}\rangle^{U(2)}
\nonumber\\
&=-4\mathcal{I}^{U(2)}
+\langle W_{n_1}W_{-n_1}\rangle^{U(2)}
+\langle W_{n_2}W_{-n_2}\rangle^{U(2)}
+\langle W_{n_1+n_2}W_{-n_1-n_2}\rangle^{U(2)}, 
\\
\label{u2_4pt_red}
&\langle W_{n_1} W_{n_2} W_{n_3} W_{-n_1-n_2-n_3}\rangle^{U(2)}
=-12\mathcal{I}^{U(2)}
+\sum_{i=1}^3 \langle W_{n_i}W_{-n_i}\rangle^{U(2)}
\nonumber\\
&
+\sum_{i<j}\langle W_{n_i+n_j}W_{-n_i-n_j}\rangle^{U(2)}
+\langle W_{n_1+n_2+n_3}W_{-n_1-n_2-n_3}\rangle^{U(2)}, 
\end{align}
\begin{align}
\label{u2_5pt_red}
&\langle W_{n_1} W_{n_2} W_{n_3}W_{n_4} W_{-n_1-n_2-n_3-n_4}\rangle^{U(2)}
=-28\mathcal{I}^{U(2)}
+\sum_{i=1}^{4}
\langle W_{n_i}W_{-n_i}\rangle^{U(2)}
\nonumber\\
&+\sum_{i_1<i_2}\langle W_{n_{i_1}+n_{i_2}}W_{-n_{i_1}-n_{i_2}}\rangle^{U(2)}
+\sum_{i_1<i_2<i_3}\langle W_{n_{i_1}+n_{i_2}+n_{i_3}}W_{-n_{i_1}-n_{i_2}-n_{i_3}}\rangle^{U(2)}
\nonumber\\
&+\langle W_{n_{i_1}+n_{i_2}+n_{i_3}+n_{i_4}}W_{-n_{i_1}-n_{i_2}-n_{i_3}-n_{i_4}}\rangle^{U(2)}. 
\end{align}

For $N=3$ the sum is taken over the two types of partitions with $r=1$ and $r=2$, 
which are $\lambda$ $=$ $(\lambda_1)$ and $(\lambda_1,\lambda_2)$ 
corresponding to the 2- and 3-point functions respectively. 
The $k$-point function can be written as a sum of the 2- and 3-point functions 
by using the following relation: 
\begin{align}
\label{k_psum3}
&
p_{n_1}p_{n_2}\cdots p_{n_{k-1}}p_{-n_1-n_2-\cdots-n_{k-1}}
\nonumber\\
&=3(2S(k,3)-S(k,2)+S(k,1))
\nonumber\\
&+\sum_{\lambda_1=1}^{k-1}\sum_{i_1^{(1)}<\cdots<i^{(1)}_{\lambda_1}}
p_{n_{i_1^{(1)}}+n_{i_2^{(1)}}+\cdots n_{i_{\lambda_{1}}^{(1)}}}
p_{-n_{i_1^{(1)}}-n_{i_2^{(1)}}-\cdots n_{i_{\lambda_{1}}^{(1)}}}
\nonumber\\
&
+\sum_{j=1}^{k-1}
\sum_{
\begin{smallmatrix}
0<\lambda_{1}\le \lambda_{2}\\
\lambda_1+\lambda_2=j\\
\end{smallmatrix}
}
\sum_{i_1^{(1)}<\cdots<i_{\lambda_{1}}^{(1)}}
\sum_{i_1^{(2)}<\cdots<i_{\lambda_{2}}^{(2)}}
\mathfrak{p}_{
\{
\sum_{a=1}^{\lambda_1} n_{i_a^{(1)}}, 
\sum_{a=1}^{\lambda_2} n_{i_a^{(2)}}, 
-\sum_{\alpha=1}^{2}\sum_{a=1}^{\lambda_\alpha} n_{i_a^{(\alpha)}}, 
\}
}. 
\end{align}
For $k=4$ one finds that 
\begin{align}
\label{u3_4pt_red}
&
\langle W_{n_1} W_{n_2} W_{n_3} W_{-n_1-n_2-n_3}\rangle^{U(3)}
\nonumber\\
&=18\mathcal{I}^{U(3)}
+\sum_{i=1}^2 \langle W_{n_i}W_{-n_i}\rangle^{U(3)}
+\sum_{i<j}\langle W_{n_i+n_j}W_{-n_i-n_j}\rangle^{U(3)}
+\langle W_{n_1+n_2+n_3}W_{-n_1-n_2-n_3}\rangle^{U(3)}
\nonumber\\
&+\sum_{i<j} \langle \mathfrak{W}_{n_i}\mathfrak{W}_{n_j}\mathfrak{W}_{-n_i-n_j} \rangle^{U(3)}
+\sum_{i=1}^{3}\sum_{
\begin{smallmatrix}
j_1\neq i, j_2\neq i\\
j_1<j_2\\
\end{smallmatrix}
}
\langle \mathfrak{W}_{n_i}\mathfrak{W}_{n_{j_1}+n_{j_2}}\mathfrak{W}_{-n_1-n_2-n_3} \rangle^{U(3)}, 
\end{align}
where 
\begin{align}
\label{3pt_irr}
&
\langle \mathfrak{W}_{n_1}\mathfrak{W}_{n_2}\mathfrak{W}_{-n_1-n_2} \rangle^{U(N)}
\nonumber\\
&:=\langle W_{n_1}W_{n_2}W_{-n_1-n_2}\rangle^{U(N)}
-\sum_{i=1}^2\langle W_{n_i}W_{-n_i}\rangle^{U(N)}
-\langle W_{n_1+n_2}W_{-n_1-n_2}\rangle^{U(N)}
\end{align}
is the irreducible part of the 3-point function. 
The numerical coefficient of the $U(3)$ Schur index in (\ref{u3_4pt_red}) is computed from the relation 
(\ref{k_psum3}) as $3(2S(4,3)-S(4,2)+S(4,1))=18$. 
For $k=5$ we have
\begin{align}
\label{u3_5pt_red}
&
\langle W_{n_1} W_{n_2} W_{n_3}W_{n_4} W_{-n_1-n_2-n_3-n_4}\rangle^{U(3)}
\nonumber\\
&=108\mathcal{I}^{U(3)}
+\sum_{i=1}^{4}\langle W_{n_i}W_{-n_i}\rangle^{U(3)}
+\sum_{i_1<i_2} \langle W_{n_{i_1}+n_{i_2}}W_{-n_{i_1}-n_{i_2}}\rangle^{U(3)}
\nonumber\\
&+\sum_{i_1<i_2<i_3}
\langle W_{n_{i_1}+n_{i_2}+n_{i_3}}W_{-n_{i_1}-n_{i_2}-n_{i_3}}\rangle^{U(3)}
+\langle W_{n_{i_1}+n_{i_2}+n_{i_3}+n_{i_4}}W_{-n_{i_1}-n_{i_2}-n_{i_3}-n_{i_4}}\rangle^{U(3)}
\nonumber\\
&
+\sum_{i_1<i_2}\langle \mathfrak{W}_{n_{i_1}}\mathfrak{W}_{n_{i_2}}\mathfrak{W}_{-n_{i_1}-n_{i_2}}\rangle^{U(3)}
+\sum_{i=1}^{4}\sum_{\begin{smallmatrix}
i_2<i_3\\
i_2\neq i_1\\
i_3\neq i_1\\
\end{smallmatrix}
}\langle \mathfrak{W}_{n_{i_1}}\mathfrak{W}_{n_{i_2}+n_{i_3}}\mathfrak{W}_{-n_{i_1}-n_{i_2}-n_{i_3}}\rangle^{U(3)}
\nonumber\\
&+\sum_{i=1}^{4}\langle \mathfrak{W}_{n_{i}}\mathfrak{W}_{n_1+n_2+n_3+n_4-n_i}\mathfrak{W}_{-n_2-n_2-n_3-n_4}\rangle^{U(3)}
\nonumber\\
&+\sum_{i_1<i_2}\langle \mathfrak{W}_{n_{i_1}+n_{i_2}}\mathfrak{W}_{n_1+n_2+n_3+n_4-n_{i_1}-n_{i_2}}\mathfrak{W}_{-n_2-n_2-n_3-n_4}\rangle^{U(3)}. 
\end{align}
Again the numerical coefficient of the $U(3)$ Schur index is fixed from (\ref{k_psum3}) as 
$3(2S(5,3)-S(5,2)+S(5,1))$ $=$ $108$. 

\subsection{Half-BPS limits}
When we keep $\mathfrak{q}$ $:=$ $q^{1/2}t^2$ fixed and take $q$ to zero, the Schur index reduces to the half-BPS index. 
In this limit the matrix integral (\ref{W_integral}) reduces to 
\begin{align}
\label{halfBPS_W}
&\langle W_{\mathcal{R}_1}\cdots W_{\mathcal{R}_k}\rangle^{U(N)}_{\textrm{$\frac12$BPS}} (\mathfrak{q})
=
\frac{1}{N!}\oint 
\prod_{i=1}^{N} \frac{d\sigma_i}{2\pi i\sigma_i}
\frac{\prod_{i\neq j} 
\left(1-\frac{\sigma_i}{\sigma_j}\right)}
{\prod_{i,j} 
\left(1-\mathfrak{q} \frac{\sigma_i}{\sigma_j} \right)
}
\prod_{j=1}^{k} \chi_{\mathcal{R}_j}(\sigma). 
\end{align}
The resulting integral (\ref{halfBPS_W}) defines an inner product of the symmetric functions
\begin{align}
\label{inner_P}
\langle f,g\rangle&:=
\frac{1}{N!}\oint 
\prod_{i=1}^{N} \frac{d\sigma_i}{2\pi i\sigma_i}
\frac{\prod_{i\neq j} 
\left(1-\frac{\sigma_i}{\sigma_j}\right)}
{\prod_{i,j} 
\left(1-\mathfrak{q} \frac{\sigma_i}{\sigma_j} \right)
}
f(\sigma) g(\sigma^{-1}). 
\end{align}
It can be viewed as a $q$-deformation of the Hall inner product. 
With respect to the inner product (\ref{inner_P}) the Hall-Littlewood functions $P_{\lambda}(\sigma;\mathfrak{q})$ are orthogonal 
\begin{align}
\label{ortho_P}
\langle P_{\lambda} (\sigma;\mathfrak{q}),P_{\mu}(\sigma^{-1},\mathfrak{q})\rangle 
&=\frac{1}{v_{\lambda}} \delta_{\lambda\mu}, 
\end{align}
where
\begin{align}
\label{v_lambda}
v_{\lambda}&=
\frac{
(\mathfrak{q};\mathfrak{q})_{N-l(\lambda)}\prod_{j\ge1}(\mathfrak{q};\mathfrak{q})_{m_j(\lambda)}
}{(1-\mathfrak{q})^{N}}
\end{align}
and $m_i(\lambda)$ is the multiplicity of the integer $i$ in the partition $\lambda$. 
In the absence of line defects the matrix integral (\ref{halfBPS_W}) reduces to the half-BPS index 
\begin{align}
\mathcal{I}_{\textrm{$\frac12$BPS}}^{U(N)}
&=\frac{1}{(\mathfrak{q};\mathfrak{q})_{N}}.
\end{align}
Consider the 2-point function of the Wilson line operators 
where the characters are given by the Schur functions 
\begin{align}
\label{halfBPS_2pt_Schur}
\langle W_{\lambda} W_{\bar{\mu}}\rangle_{\textrm{$\frac12$BPS}}^{U(N)}
&=
\frac{1}{N!}\oint 
\prod_{i=1}^{N} \frac{d\sigma_i}{2\pi i\sigma_i}
\frac{\prod_{i\neq j} 
\left(1-\frac{\sigma_i}{\sigma_j}\right)}
{\prod_{i,j} 
\left(1-\mathfrak{q} \frac{\sigma_i}{\sigma_j} \right)
}s_{\lambda}(\sigma) s_{\mu}(\sigma^{-1}). 
\end{align}
Since the Schur functions can be decomposed in terms of the Hall-Littlewood functions
\begin{align}
s_{\lambda}(\sigma)&=\sum_{\nu}K_{\lambda \nu}(\mathfrak{q})P_{\nu}(\sigma;\mathfrak{q}), 
\end{align}
where $K_{\lambda\nu}(\mathfrak{q})$ is the Kostka-Foulkes polynomial \cite{MR1354144}, 
the matrix integral can be formally evaluated from (\ref{ortho_P}) as
\begin{align}
\langle W_{\lambda} W_{\bar{\mu}}\rangle_{\textrm{$\frac12$BPS}}^{U(N)}
&=
\sum_{\nu}\frac{K_{\lambda\nu}(\mathfrak{q}) K_{\mu\nu}(\mathfrak{q})}
{\prod_{n=1}^{N-l(\nu)} (1-\mathfrak{q}^n)
\prod_{j\ge1}\prod_{n=1}^{m_j(\nu)}(1-\mathfrak{q}^n)
}. 
\end{align}

\section{Fermi-gas formulation}
\label{sec_Fermigas}
In \cite{Hatsuda:2022xdv} the closed-form expressions for the Schur index of $\mathcal{N}=2^*$ $U(N)$ SYM theory are presented by means of the Fermi-gas method. 
In this section we extend the analysis to the Schur line defect correlation functions. 

We redefine the flavor fugacity by replacing $t$ with $\xi=q^{-1/2}t^2=e^{2\pi i\zeta}$ 
which is associated with the $\mathcal{N}=2^*$ deformation due to the mass parameter for the adjoint hypermultiplet. 
We define a new function
\begin{align}
\theta(x;q)&:=\sum_{n\in \mathbb{Z}}(-1)^n x^{n+\frac12}q^{\frac{n^2+n}{2}}
\nonumber\\
&=(x^{\frac12}-x^{-\frac12})\prod_{n=1}^{\infty} (1-q^n)(1-xq^n)(1-x^{-1}q^n). 
\end{align}
Then the matrix integral (\ref{W_integral}) can be rewritten as 
\begin{align}
\label{W_integral2}
&\langle W_{\mathcal{R}_1}\cdots W_{\mathcal{R}_k}\rangle^{U(N)} (\xi;q)
\nonumber\\
&=
\frac{(-1)^N\xi^{N^2/2}}{N!}
\oint_{|\sigma_i|=1} \prod_{i=1}^{N}
\frac{d\sigma_i}{2\pi i\sigma_i}
\frac{\theta'(1;q)^N \prod_{i<j} \theta(\frac{\sigma_i}{\sigma_j};q)\theta(\frac{\sigma_i}{\sigma_j}:q)}
{\prod_{i,j}\theta(\frac{\sigma_i}{\sigma_j}\xi^{-2};q)}
\prod_{j=1}^{k}\chi_{\mathcal{R}_j}(\sigma). 
\end{align}
Corresponding to (\ref{t_transf}), 
the integral (\ref{W_integral2}) is invariant under
\begin{align}
\label{xi_transf}
\xi&\rightarrow q^{-1}\xi^{-1}. 
\end{align}

According to the Frobenius determinant formula \cite{Frobenius:1882uber,MR0335789,Mason:2008zzb}
\begin{align}
\label{Frobdet}
\frac{
(q;q)_{\infty}^{3N}\prod_{i<j}\theta(v_i v_j^{-1};q) \theta(w_j w_i^{-1};q)
}
{
\prod_{i,j}\theta(v_i w_j^{-1};q)
}
&=\frac{\theta(u;q)}{\theta(u\prod_i v_i w_i^{-1};q)}
\det_{i,j} F(v_i w_j^{-1},u;q), 
\end{align}
where 
\begin{align}
\label{Kronecker_fcn}
F(x,y;q)&:=\frac{\theta(xy;q) (q;q)_{\infty}^3}{\theta(x;q) \theta(y;q)}
\end{align}
is the Kronecker theta function \cite{zbMATH02706826,MR1723749,MR1106744,MR2796409}, 
one can express (\ref{W_integral2}) as
\begin{align}
\label{W_integral3}
&\langle W_{\mathcal{R}_1}\cdots W_{\mathcal{R}_k}\rangle^{U(N)} (\xi;q)
\nonumber\\
&=\frac{(-1)^N \xi^{N^2/2}}{N!}\frac{\theta(u;q)}{\theta(u\xi^{-N};q)}
\oint_{|\sigma_i|=1} \prod_{i=1}^{N}\frac{d\sigma_i}{2\pi i\sigma_i} \det_{i,j} F\left(\frac{\sigma_i}{\sigma_j}\xi^{-1},u;q\right)
\prod_{j=1}^{k}\chi_{\mathcal{R}_j}(\sigma). 
\end{align}
In the absence of the characters $\chi_{\mathcal{R}_j}$ in the integrand of (\ref{W_integral}) 
it reduces to the Schur index and the normalized function 
\begin{align}
\label{pfn_fermi0}
\mathcal{Z}(N;u;\xi;q)&=
\frac{1}{N!}\oint_{|\sigma_i|=1} \prod_{i=1}^{N}\frac{d\sigma_i}{2\pi i\sigma_i} \det_{i,j} F\left(\frac{\sigma_i}{\sigma_j}\xi^{-1},u;q\right)
\end{align}
can be regarded as a partition function of free Fermi-gas with $N$ particles on a circle 
which is characterized by a one-particle density matrix 
\begin{align}
\label{density_0}
\rho_{0}(\alpha,\alpha';u;\xi;q)
&=F(e^{2\pi i(\alpha-\alpha')}\xi^{-1},u;\xi;q)
\nonumber\\
&=-\sum_{p\in \mathbb{Z}}\frac{e^{2\pi ip(\alpha-\alpha')}\xi^{-p}}{1-uq^p}, 
\end{align}
where $0\le$ $\alpha$ $=\frac{1}{2\pi i}\log \sigma$ $\le1$ is 
the periodic position operator and $p$ is the discrete momentum operator. 

In order to generalize the Fermi-gas method to the Schur line defect correlation functions,
we use the idea in \cite{Hatsuda:2013yua, Drukker:2015spa}. 
We consider matrix integrals
\begin{align}
\label{geneW_E}
\mathcal{Z}_E^{\{n_j\}}(N;\{s_j\};u;\xi;q)&=
\frac{1}{N!}\oint_{|\sigma_i|=1} \prod_{i=1}^{N}\frac{d\sigma_i}{2\pi i\sigma_i} \det_{i,j} F\left(\frac{\sigma_i}{\sigma_j}\xi^{-1},u;q\right)
\prod_{j=1}^{k} E(s_j;\sigma^{n_j}) 
\end{align}
and 
\begin{align}
\label{geneW_H}
\mathcal{Z}_H^{\{n_j\}}(N;\{s_j\};u;\xi;q)&=
\frac{1}{N!}\oint_{|\sigma_i|=1} \prod_{i=1}^{N}\frac{d\sigma_i}{2\pi i\sigma_i} \det_{i,j} F\left(\frac{\sigma_i}{\sigma_j}\xi^{-1},u;q\right)
\prod_{j=1}^{k} H(s_j;\sigma^{n_j}),  
\end{align}
where $E(s_j;\sigma)$ and $H(s_j;\sigma)$ are the generating functions (\ref{E_gene}) and (\ref{H_gene}) for the characters 
of the antisymmetric representations and those of the symmetric representations. 
These matrix integrals play a role of generating functions for the correlation functions of the Wilson line operators. 
For example, the correlation functions of the Wilson line operators $W_{n_j}$ of charges $\{n_j\}_{j=1}^{k}$ can be obtained 
from the coefficient of the term with $\prod_{j=1}^{k}s_j$ in either of (\ref{geneW_E}) or (\ref{geneW_H}). 
Besides, for $k=2$ and $(n_1,n_2)=(1,-1)$ one can extract the 2-point functions of the Wilson line operators 
$W_{(1^l)}$ (resp. $W_{(l,0)}$) 
in the rank-$l$ antisymmetric representations (resp. symmetric representations) 
by reading off the term including $s_1^{l}s_2^l$ in (\ref{geneW_E}) (resp. in (\ref{geneW_H})). 

We observe that the introductions of the products $\prod_{j=1}^k$ $\prod_{i=1}^N$ $E(s_j; \sigma^{n_j})$ in (\ref{geneW_E}) 
and $\prod_{j=1}^k$ $\prod_{i=1}^N$ $H(s_j;\sigma^{n_j})$ in (\ref{geneW_H}) 
replace the density matrix (\ref{density_0}) with 
\begin{align}
\label{density_E}
\rho_{E}^{(n_1,n_2,\cdots,n_k)}(\{s_j\};\alpha,\alpha';u,\xi;q)
&=X_E^{(n_1,n_2,\cdots,n_k)} (\{s_j\};\alpha) \rho_0(\alpha,\alpha';u,\xi;q)
\end{align}
and 
\begin{align}
\label{density_H}
\rho_{H}^{(n_1,n_2,\cdots,n_k)}(\{s_j\};\alpha,\alpha';u,\xi;q)
&=X_H^{(n_1,n_2,\cdots,n_k)} (\{s_j\};\alpha) \rho_0(\alpha,\alpha';u,\xi;q), 
\end{align}
where we have defined position-dependent matrices 
\begin{align}
X_E^{(n_1,n_2,\cdots,n_k)} (\{s_j\};\alpha) &=\prod_{j=1}^{k} E(s_j;e^{2\pi i n_j \alpha}). 
\end{align}
and 
\begin{align}
X_H^{(n_1,n_2,\cdots,n_k)} (\{s_j\};\alpha) &=\prod_{j=1}^{k} H(s_j;e^{2\pi i n_j \alpha}). 
\end{align}
In other words, the matrix integrals (\ref{geneW_E}) and (\ref{geneW_H}) are now identified with 
the canonical partition functions of free Fermi-gas with $N$ particles whose density matrices are given by 
(\ref{density_E}) and (\ref{density_H}). 

\subsection{Spectral zeta functions}
\label{sec_spectralzeta}

\subsubsection{Multiple Kronecker theta series}
\label{sec_kronecker}
We define a function
\begin{align}
\label{s_zeta}
Q(\{l_i\};\{n_i\};u;\xi;q)
&:=\sum_{p\in \mathbb{Z}}
\prod_{i=0}^{k}
\frac{(-1)^{l_i} \xi^{-l_i p-l_i n_i}}
{(1-uq^{p+n_i})^{l_i}}, 
\end{align}
where $n_0=0$. 
As this function generalizes the Fourier series (\ref{density_0}) of the Kronecker theta function  
by including mult-index obeying certain condition, 
we call the function (\ref{s_zeta}) \textit{multiple Kronecker theta series}. 

The multiple Kronecker theta series (\ref{s_zeta}) plays a role of elementary blocks of the Schur indices and line defect correlators. 
If $k=0$ and $l_0=l$, the multiple Kronecker theta series (\ref{s_zeta}) becomes the spectral zeta function 
associated with the density matrix $\rho_0$ of the Fermi-gas for the Schur index of $\mathcal{N}=2^*$ $U(N)$ SYM theory \cite{Hatsuda:2022xdv}
\begin{align}
\label{szeta_0}
Z_{l}(u;\xi;q)
&:=\Tr(\rho_0^l)
\nonumber\\
&=Q(l;0;u;\xi;q)
=\sum_{p\in \mathbb{Z}}
\left(
\frac{-\xi^{-p}}{1-uq^p}
\right)^l. 
\end{align}
We can write the spectral zeta function for $l=1$ as
\begin{align}
\label{q1_tP1}
Z_{1}(u;\xi;q)
&=
\frac{(q)_{\infty}^3 \theta(\xi^{-1}u)}{\theta(\xi^{-1})\theta(u)}
=P_1\left[
\begin{matrix}
\xi\\
1\\
\end{matrix}
\right](\nu,\tau)
\end{align}
and 
\begin{align}
\label{qk_tPk}
Z_{l}(u;\xi;q)
&=\frac{u^{-(l-1)}}{(l-1)!}\sum_{k=1}^{l-1}
k! |s(l-1,k)|P_{k+1}
\left[
\begin{matrix}
q^{l-1}\xi^l\\
1\\
\end{matrix}
\right]
(\nu,\tau)
\end{align}
for $l\ge 2$ with $u=e^{2\pi i\nu}$ and $q=e^{2\pi i\tau}$. 
Here $s(n,k)$ are the Stirling numbers of the first kind and
\begin{align}
P_k\left[\begin{matrix}
\theta\\\phi\\
\end{matrix}
\right](z,\tau)
&=\frac{(-1)^k}{(k-1)!}
{\sum_{n\in \mathbb{Z}}}' 
\frac{(n+\lambda)^{k-1}x^{n+\lambda}}
{1-\theta^{-1}q^{n+\lambda}}
\end{align}
is the twisted Weierstrass function \cite{Mason:2008zzb} 
where $\sum'$ stands for the sum that omits $n=0$ if $(\theta,\phi)=(1,1)$. 

In \cite{Hatsuda:2022xdv} it is shown that the Schur indices of $\mathcal{N}=2^*$ $U(N)$ SYM theories can be expressible 
in terms of the spectral zeta functions (\ref{szeta_0}). 

More generally, 
the multiple Kronecker theta series (\ref{s_zeta}) can be written in terms of the twisted Weierstrass function 
by means of the partial expansion into the function (\ref{szeta_0}). 
It is convenient to define functions 
\begin{align}
\label{charge_fcn}
(n)_{q,\xi}&:=
-\frac{\xi^{-n}}{1-q^n}
=\frac{q^{-\frac{n}{2}} \xi^{-n}}{q^{\frac{n}{2}}-q^{-\frac{n}{2}}}, 
\\
\label{charge_fcn2}
[n]_{q}&:=
\frac{1-q^n}{1-q}. 
\end{align}
The function (\ref{charge_fcn}) transforms as
\begin{align}
(n)_{q,q^{-1}\xi^{-1}}&=-(-n)_{q,\xi}
\nonumber\\
&=q^n \xi^{2n} (n)_{q,\xi}
\end{align}
under (\ref{xi_transf}). 
It follows that
\begin{align}
\label{chargefcn_prop1}
[n]_{q}&=\frac{(1)_{q,1}}{(n)_{q,1}}, \\
\label{chargefcn_prop2}
(n)_{q,\xi}+(-n)_{q,\xi}&=
-\frac{(n)_{q,\xi}}{(n)_{q\xi^2,1}}
=-\frac{q^{\frac{n}{2}}\xi^{n}-q^{-\frac{n}{2}}\xi^{-n}}
{q^{\frac{n}{2}}-q^{-\frac{n}{2}}}
,\\
\label{chargefcn_prop3}
\frac{(n)_{q,\xi}^2}{(n)_{q\xi,1}}
+\frac{(-n)_{q,\xi}^2}{(-n)_{q\xi,1}}
&=-\frac{(n)_{q,\xi}^2}{(n)_{q\xi,1}(n)_{q\xi^3,1}}, 
\end{align}
where 
\begin{align}
\frac{1}{(n)_{q\xi^2,1}}
&=q^{\frac{n}{2}}\xi^{n}
(q^{\frac{n}{2}}\xi^n - q^{-\frac{n}{2}}\xi^{-n}). 
\end{align}
These relations are useful to describe the correlation functions. 

Let $l_{\max}$ be the maximal value of the integers $\{l_i\}$, $i=0,\cdots, k$ for the function $Q(\{l_i\};\{n_i\};u;\xi;q)$. 
For $k>0$ the function $Q(\{l_i\};\{n_i\};u;\xi;q)$ can be decomposed into 
$\sum_{i} l_i$ parts, each of which is expressed in terms of the spectral zeta function $Q(l;0;u;\xi;q)=Z_{l}(u;\xi;q)$ 
with $1\le l\le l_{\max}$ and the function (\ref{charge_fcn}). 

For example, for $k=1$ we have 
\begin{align}
\label{ql0l1}
&
Q(l_0,l_1;0,n;u;\xi;q)
\nonumber\\
&=\sum_{k=0}^{l_0-1}
\left(
\begin{matrix}
k+l_1-1\\
l_1-1\\
\end{matrix}
\right)
(-1)^k 
q^{kn}\xi^{kn}
(n)_{q,\xi}^{k+l_1}
Q(l_0-k;0;\xi^{\frac{l_0+l_1}{l_0-k}};q)
\nonumber\\
&+\sum_{k=0}^{l_1-1}
\left(
\begin{matrix}
k+l_0-1\\
k\\
\end{matrix}
\right)
(-1)^{k}
q^{-kn}\xi^{-kn}
(-n)_{q,\xi}^{k+l_0}
Q(l_1-k;0;u;\xi^{\frac{l_0+l_1}{l_1-k}};q). 
\end{align}
Clearly, it follows that 
\begin{align}
Q(l_1,l_0;0,n;u;\xi;q)&=Q(l_0,l_1;0,-n,u;\xi;q). 
\end{align}
Several examples are shown in Appendix \ref{app_qxx}. 

For $k=2$ with $l_0=l$, $l_1=l_2=1$ one finds that 
\begin{align}
&
Q(l,1,1;0,n_1,n_2;u;\xi;q)
\nonumber\\
&=\sum_{m=1}^{l} \sum_{k=0}^{m-1} \sum_{k_1+k_2=k}
\left(
\begin{matrix}
m\\
k+1\\
\end{matrix}
\right)
(-1)^{k+m-1}
q^{(m-1)(n_1+n_2)-k_1n_1-k_2n_2}
\xi^{(m-1)(n_1+n_2)}
\nonumber\\
&\times (n_1)_{q,\xi}^{m} (n_2)_{q,\xi}^{m} 
Q(l-m+1;0;u;\xi^{\frac{l+2}{l-m+1}};q)
\nonumber\\
&+(-n_1)_{q,\xi}^{l} (-n_1+n_2)_{q,\xi} Q(1;0;u;\xi^{l+2};q)
\nonumber\\
&+(-n_2)_{q,\xi}^{l} (-n_2+n_1)_{q,\xi} Q(1;0;u;\xi^{l+2};q). 
\end{align}
We have
\begin{align}
\label{ql11_sym}
Q(1,l,1;0,n_1,n_2;u;\xi;q)
&=Q(l,1,1;0,-n_1,n_2-n_1;u;\xi;q),\\
Q(1,1,l;0,n_1,n_2;u;\xi;q)
&=Q(l,1,1;0,n_1-n_2,-n_2;u;\xi;q). 
\end{align}
Further examples for $k=2$ are found in Appendix \ref{app_qxxx1}. 

For general $k$ with $l_0=l_1=\cdots =l_{k}=1$ 
the multiple Kronecker theta series (\ref{s_zeta}) can be decomposed as
\begin{align}
\label{q1x1}
&
Q(1,\cdots,1;0,n_1,\cdots,n_k;u;\xi;q)
\nonumber\\
&=\Bigl[
\prod_{i=1}^{k}(n_i)_{q,\xi}
+\sum_{i=1}^{k} \sum_{j\neq i} (-n_i)_{q,\xi}(-n_i+n_j)_{q,\xi}
\Bigr]Q(1;0;u;\xi^{k+1};q). 
\end{align}

Likewise, the spectral zeta functions for the modified density matrices (\ref{density_E}) and (\ref{density_H}) can be expressed in terms of 
the multiple Kronecker theta series (\ref{s_zeta}). 
Noting that $\sigma=e^{2\pi i\alpha}$ is a translation operator 
\begin{align}
\label{translate1}
\sigma^{-n} \mathcal{O}(p) \sigma^{n}
&=e^{-2\pi in\alpha} \mathcal{O}(p) e^{2\pi in\alpha}
=\mathcal{O}(p+n), 
\end{align}
where $\mathcal{O}(p)$ is some $p$-dependent operator, 
the spectral zeta functions can be calculated by taking traces of normal ordered operators $(X_{E/H}^{(n_1,n_2,\cdots,n_k)}(\alpha) \rho(p))^l$. 

We obtain the spectral zeta function associated with the modified density matrix (\ref{density_E}) of the form
\begin{align}
\label{szeta_E}
&Z_{l}^{E}(n_1,n_2,\cdots,n_{k-1})
=\Tr \left({\rho_E^{(n_1,n_2,\cdots,n_k)}}^l\right)
\nonumber\\
&=Z_l(u;\xi;q)
+Z_{l;1}^{(k-1)}(\{n_i\};u;\xi;q)\prod_{i=1}^{k}s_i 
+\prod_{m\ge2} Z_{l;m}^{E;(k-1)}(\{n_i\};u;\xi;q) \prod_{i=1}^k s_i^m, 
\end{align}
where $Z_{l}(u;\xi;q)$ $=$ $Q(l;0;u;\xi;q)$ which is independent of the fugacities $\{s_i\}$ encodes the Schur index without any insertion of the line operators. 
It is nothing but the spectral zeta function (\ref{szeta_0}) for the density matrix $\rho_0$. 
The function $Z_{l;1}^{(k-1)}$ which appears in the terms with $\prod_i s_i$ captures the $k$-point functions of the charged Wilson line operators. 
It is given by 
\begin{align}
\label{szeta_1}
&Z^{(k-1)}_{l;1}(\{n_i\};u;\xi;q)
=l\sum_{j=1}^k \sum_{l_1+\cdots l_{j}=l}
\sum_{\{ N_{J_i}\}_{i=1}^j}
Q\left(
\{l_i\}_{i=1}^{j}; \{N_{J_i}\}_{i=1}^j; u;\xi;q
\right), 
\end{align}
where 
\begin{align}
N_{J_i}&=\sum_{a_* \in J_i} n_{a_*}
\end{align}
and each $J_i$ is a subset of integers labeling the charged Wilson line operators obeying the condition
\begin{align}
\emptyset=J_1\subset J_2\subset \cdots \subset J_{j}\subseteq I=\{1,2,\cdots,k-1\}. 
\end{align}
Here $A\subseteq B$ allows the case $A=B$ while $A\subset B$ excludes it. 
Since $J_1$ is empty, $N_{J_1}$ is $0$. 
For example, when $k>5$ the subsets
\begin{align}
J_1=\emptyset,\qquad 
J_2=\{1,2\}, \qquad 
J_3=\{1,2,3,4,5\} 
\end{align}
are allowed so that we have 
\begin{align}
N_{J_1}=0, \qquad
N_{J_2}=n_1+n_2, \qquad 
N_{J_3}=n_1+n_2+n_3+n_4+n_5. 
\end{align}

The other terms $Z_{l;m}^{E;(k-1)}(\{n_j\};u;\xi;q)$ in (\ref{szeta_E}) encode the 2-point functions of the Wilson line operators transforming in the rank-$m$ antisymmetric representations. 
The $k$-point function of $\mathcal{N}=2^*$ $U(N)$ SYM theory can be constructed from  
the spectral zeta functions $Z_l^E(n_1,\cdots,n_{k-1})$ with $l=1,\cdots, N$. 

Similarly, the spectral zeta function specified by the other modified density matrix (\ref{density_H}) takes the form
\begin{align}
\label{szeta_H}
&Z_{l}^{H}(n_1,n_2,\cdots,n_{k-1})
=\Tr \left({\rho_H^{(n_1,n_2,\cdots,n_k)}}^l\right)
\nonumber\\
&=Z_l(u;\xi;q)
+Z_{l;1}^{(k-1)}(\{n_i\};u;\xi;q)\prod_{i=1}^{k}s_i
+\prod_{m\ge2} Z_{l;m}^{H;(k-1)}(\{n_i\};u;\xi;q) \prod_{i=1}^k s_i^m. 
\end{align}
Again whereas the function $Z_{l;1}^{(k-1)}$ appears as a coefficient of the terms with $\prod_j s_j$, 
the terms $Z_{l;m}^{H;(k-1)}(\{n_j\};u;\xi;q)$ encode the 2-point functions of the Wilson line operators transforming in the rank-$m$ symmetric representations. 

\subsubsection{$Z_l^E$}
We show several examples of the spectral zeta functions. 
For simplicity we abbreviate $Q(\{l_i\};\{n_i\};u;\xi;q)=Q(\{l_i\};\{n_i\})$. 

For $k=2$ the spectral zeta functions for the modified density matrix (\ref{density_E}) are 
\begin{align}
\label{ZE2_1}
Z_{1}^{E}(n)
&=(1+s_1s_2)Q(1;0),\\
\label{ZE2_2}
Z_{2}^{E}(n)
&=(1+s_1^2s_2^2)Q(2;0)
+2s_1s_2\left[
Q(2;0)+Q(1,1;0,n)
\right],\\
\label{ZE2_3}
Z_{3}^{E}(n)
&=(1+s_1^3s_2^3)Q(3;0)
\nonumber\\
&+3(s_1s_2+s_1^2s_2^2)\Bigl[
Q(3;0)
+Q(2,1;0,n)
+Q(1,2;0,n)
\Bigr]
, \\
\label{ZE2_4}
Z_{4}^{E}(n)
&=(1+s_1^4s_2^4)
Q(4;0)
\nonumber\\
&+4(s_1s_2+s_1^3s_2^3)\Bigl[
Q(4;0)
+Q(3,1;0,n)
+Q(2,2;0,n)
+Q(1,3;0,n)
\Bigr]
\nonumber\\
&+s_1^2s_2^2\Bigl[
6Q(4;0)
+8Q(3,1;0,n)
+10Q(2,2;0,n)
\nonumber\\
&+8Q(1,3;0,n)
+4Q(1,2,1;0,n,2n)
\Bigr],
\end{align}
\begin{align}
\label{ZE2_5}
Z_{5}^{E}(n)
&=(1+s_1^5s_2^5)
Q(5;0)
\nonumber\\
&+5(s_1s_2+s_1^4s_2^4)
\Bigl[
Q(5;0)+Q(4,1;0,n)+Q(3,2;0,n)
\nonumber\\
&+Q(2,3;0,n)+Q(1,4;0,n)
\Bigr]
\nonumber\\
&+(s_1^2s_2^2+s_1^3s_2^3)
\Bigl[
10Q(5;0)+15Q(4,1;0,n)+20Q(3,2;0,n)+20Q(2,3;0,n)
\nonumber\\
&+15Q(1,4;0,n)+5Q(2,2,1;0,n,2n)
+10Q(1,3,1;0,n,2n)
+5Q(1,2,2;0,n,2n)
\Bigr]. 
\end{align}
These spectral zeta functions with $k=2$ are the blocks of the 2-point functions. 
In particular, we have
\begin{align}
Z_{l;1}^{(1)}(n)&=lQ(l;0)+l\sum_{k=1}^{l-1}Q(l-k,k;0,n). 
\end{align}

For $k=3$ we get
\begin{align}
\label{ZE3_1}
Z_{1}^{E}(n_1,n_2)
&=(1+s_1s_2s_3)Q(1;0),\\
\label{ZE3_2}
Z_{2}^{E}(n_1,n_2)
&=(1+s_1^2s_2^2s_3^2)Q(2;0)
+2s_1s_2s_3\Bigl[
Q(2;0)+Q(1,1;0,n_1)
\nonumber\\
&
+Q(1,1;0,n_2)
+Q(1,1;0,n_1+n_2)
\Bigr]
,\\
\label{ZE3_3}
Z_{3}^{E}(n_1,n_2)
&=(1+s_1^3s_2^3s_3^3)Q(3;0)
\nonumber\\
&+3(s_1s_2s_3+s_1^2s_2^2s_3^2)\Bigl[
Q(3;0)
+\sum_{i=1}^2Q(2,1;0,n_i)
+Q(2,1;0,n_1+n_2)
\nonumber\\
&+\sum_{i=1}^2Q(1,2;0,n_i)
+Q(1,2;0,n_1+n_2)
+\sum_{i=1}^2Q(1,1,1;0,n_i,n_1+n_2)
\Bigr]. 
\end{align}
These spectral zeta functions are associated to the 3-point functions. 
The terms which are associated with $s_1s_2s_3$ describe the 3-point functions 
of the charged Wilson line operators. 
They are given by
\begin{align}
Z_{l;1}^{(2)}&=lQ(l;0)
+l\Bigl[
\sum_{k=1}^{l-1}
\sum_{i=1}^{2}
Q(l-k,k;0,n_i)
+Q(l-k,k;0,n_1+n_2)
\Bigr] 
\nonumber\\
&+l
\sum_{
\begin{smallmatrix}
0<k_1,k_2\\
2\le k_1+k_2\le l-1\\
\end{smallmatrix}
}
\sum_{i=1}^2 
Q(l-k_1-k_2,k_1,k_2;0,n_i,n_1+n_2). 
\end{align}

For $k=4$ we find
\begin{align}
\label{ZE4_1}
Z_{1}^{E}(n_1,n_2,n_3)
&=(1+s_1s_2s_3s_4)Q(1;0),\\
\label{ZE4_2}
Z_{2}^{E}(n_1,n_2,n_3)
&=(1+s_1^2s_2^2s_3^2s_4^2)Q(2;0)
\nonumber\\
&+2s_1s_2s_3s_4\Bigl[
Q(2;0)
+Q(1,1;0,n_1)
+Q(1,1;0,n_2)
+Q(1,1;0,n_3)
\nonumber\\
&
+Q(1,1;0,n_1+n_2)
+Q(1,1;0,n_1+n_3)
+Q(1,1;0,n_2+n_3)
\nonumber\\
&
+Q(1,1;0,n_1+n_2+n_3)
\Bigr]. 
\end{align}
\begin{align}
\label{ZE4_3}
Z_{3}^{E}(n_1,n_2,n_3)
&=(1+s_1^3s_2^3s_3^3s_4^3)Q(3;0)
\nonumber\\
&+3(s_1s_2s_3s_4+s_1^2s_2^2s_3^2s_4^2)\Bigl[
Q(3;0)
+\sum_{i=1}^{3}Q(2,1;0,n_i)
\nonumber\\
&
+\sum_{i=1}^3 \sum_{i<j} Q(2,1;0,n_i+n_j)
+Q(2,1;0,n_1+n_2+n_3)
\nonumber\\
&
+\sum_{i=1}^3 Q(1,2;0,n_i)
+\sum_{i<j} Q(1,2;0,n_i+n_j)
+Q(1,2;0,n_1+n_2+n_3)
\nonumber\\
&+\sum_{i=1}^3 \sum_{j\neq i}Q(1,1,1;0,n_i,n_i+n_j)
+\sum_{i=1}^{3}Q(1,1,1;0,n_i,n_1+n_2+n_3)
\nonumber\\
&+\sum_{i<j}Q(1,1,1;0,n_i+n_j,n_1+n_2+n_3)
\Bigr]. 
\end{align}
The 4-point functions of the charged Wilson line operators are captured by 
\begin{align}
Z_{l;1}^{(3)}&=
lQ(l;0)
+l\Bigl[
\sum_{k=1}^{l-1}
\Bigl\{
\sum_{i=1}^{3}
Q(l-k,k;0,n_i)
+\sum_{i<j}Q(l-k,k;0,n_i+n_j)
\Bigr\}
\Bigr]
\nonumber\\
&+l\Bigl[
\sum_{
\begin{smallmatrix}
0<k_1,k_2\\
2\le k_1+k_2\le l-1\\
\end{smallmatrix}
}
\Bigl\{
\sum_{i=1}^{3}\sum_{j\neq i}
Q(l-k_1-k_2;k_1,k_2;0,n_i,n_i+n_j)
\nonumber\\
&+\sum_{i=1}^3 Q(l-k_1-k_2;k_1,k_2;0,n_i,n_1+n_2+n_3)
\nonumber\\
&+\sum_{i<j}Q(l-k_1-k_2,k_1,k_2;0,n_i+n_j,n_1+n_2+n_3)
\Bigr]
\nonumber\\
&+l
\sum_{
\begin{smallmatrix}
0<k_1,k_2,k_3\\
3\le k_1+k_2+k_3\le l-1\\
\end{smallmatrix}
}\sum_{i=1}^{3}\sum_{j\neq i}
Q(l-k_1-k_2-k_3,k_1,k_2,k_3;0,n_i,n_i+n_j,n_1+n_2+n_3), 
\end{align}
which appears in the terms associated with $s_1s_2s_3s_4$. 

\subsubsection{$Z_l^H$}
The 2-point function of the Wilson line operator in the symmetric representation and that in its conjugate representation 
can be obtained from $Z_l^H$ with $k=2$. 
We find
\begin{align}
\label{ZH2_1}
Z_{1}^{H}(n)
&=\sum_{k=0}^{\infty}s_1^ks_2^kQ(1;0)
,\\
\label{ZH2_2}
Z_{2}^{H}(n)
&=\sum_{k=0}^{\infty}s_1^ks_2^k 
\left[
(k+1)Q(2;0)
+\sum_{l=1}^k 2(k-l+1)
Q(1,1;0,ln)
\right],\\
\label{ZH2_3}
Z_{3}^{H}(n)
&=\sum_{k=0}^{\infty}s_1^ks_2^k\Bigl[
\frac{(k+1)(k+2)}{2}Q(3;0)
\nonumber\\
&+\sum_{l=1}^k \frac{3(k-l+1)(k-l+2)}{2}
\{
Q(2,1;0,ln)
+Q(1,2;0,ln)
\}
\nonumber\\
&+\sum_{l_1=2}\sum_{0<l_2<l_1}
3(k-l_1+1)(k-l_1+2)
Q(1,1,1;0,l_2,l_1)
\Bigr]. 
\end{align}
See Appendix \ref{app_spectralZ} for more examples. 

\subsection{Closed-form formula}
Let $\lambda=(\lambda_1^{m_1}\lambda_{2}^{m_2}\cdots\lambda_{r}^{m_r})$ be a partition of integer $N$ 
with $\sum_{i=1}^{r}m_i\lambda_i=N$ and $\lambda_1>\lambda_2$ $>\cdots>$ $\lambda_r>\lambda_{r+1}=0$. 
Then we have
\begin{align}
\label{geneW_E/H}
\mathcal{Z}_{E/H}^{\{n_j\}}(N;\{s_j\};u;\xi;q)
&=\sum_{\lambda}(-1)^{N-r}\prod_{i=1}^{r}\frac{1}{\lambda_i^{m_i}(m_i!)}
{Z_{\lambda_i}^{E/H}(n_1,\cdots,n_{k-1})}^{m_i}. 
\end{align}
Now we can obtain the closed-form expressions for the Schur line defect correlation functions. 
Since the Schur line defect correlation functions are independent of the variable $u$, 
it is convenient to fix $u$ to some special value. 

When we set $u$ to $\xi^{N/2}$, 
the canonical partition function of the Fermi-gas is identical to the Schur index up to the overall factor $(-1)^{N+1}\xi^{N^2/2}$. 
Besides, this specialization yields the closed-form expressions for the Schur line defect correlators 
as multiple series which generalize the nested sum of the Schur index obtained in \cite{Hatsuda:2022xdv}
\begin{align}
\mathcal{I}^{U(N)}
&=
-
\sum_{
\begin{smallmatrix}
p_1,\cdots,p_N \in \mathbb{Z}\\
p_1<\cdots<p_N\\
\end{smallmatrix}
}
\frac{\xi^{\frac{N^2}{2}-\sum_{i=1}p_i}}
{\prod_{i=1}^{N}(1-\xi^{\frac{N}{2}} q^{p_i})}. 
\end{align}
It is closely related to the multiple $q$-zeta values ($q$-MVZs) 
\cite{schlesinger2001some,MR2069738,MR2111222,MR1992130,MR2341851,MR2322731,
MR2843304,MR3141529,okounkov2014hilbert,MR3338962,MR3473421,MR3522085,milas2022generalized}
and $q$-multiple polylogarithms ($q$-MPLs) 
\cite{schlesinger2001some,MR2341851,MR3687119}. 
We leave more detailed investigation of the relation to these functions to future work. 

When we choose $u$ as $\xi$, the multiple Kronecker theta series $Q(1;0;u;\xi;q)$ vanishes
\begin{align}
Q(1;0;u=\xi)&=0, 
\end{align}
which can lead to the expression with fewer terms. 
For simplicity, here and in the following we omit the dependence on $\xi$ and $q$ to write 
$Q(\{l_i\};$ $\{n_i\};$ $u;$ $\xi;$ $q)$ as $Q(\{l_i\}; \{n_i\}; u)$. 

Plugging the expression (\ref{szeta_E}) or (\ref{szeta_H}) for the spectral zeta function 
into (\ref{geneW_E/H}) with $u=\xi^{N/2}$ and reading off the coefficients of the terms with $\prod_{j=1}^k s_j$, 
we find that the $k$-point function of the Wilson line operators of charges $\{n_i\}_{i=1}^{k}$ is given by 
\begin{align}
\label{closed_W}
&\left\langle W_{n_1}W_{n_2}\cdots W_{n_{k}} \right\rangle^{U(N)}
\nonumber\\
&=\xi^{N^2/2}\sum_{\lambda} (-1)^{r+1}\prod_{i=1}^{r} 
\frac{1}{\lambda_i^{m_i} (m_i) !}
\nonumber\\
&\times \Biggl[
\sum_{i=1}^r 
m_i \lambda_i 
Q(\lambda_i;0;\xi^{\frac{N}{2}})^{m_i-1}
\left\{
\sum_{p=1}^k \sum_{l_1+\cdots l_{p}=\lambda_i}
\sum_{\{ N_{J_i}\}_{i=1}^p}
Q\left(
\{l_i\}_{i=1}^{p}; \{N_{J_i}\}_{i=1}^p; \xi^{\frac{N}{2}}
\right)
\right\}
\nonumber\\
&\times \prod_{j\neq i}^r Q(\lambda_j;0;\xi^{\frac{N}{2}})^{m_j}
\Biggr], 
\end{align}
where $\sum_{i=1}^{k}n_i=0$. 
The terms for $p=1$ in the third sum yield $N \mathcal{I}^{U(N)}$. 
Thus we find an exact closed-form expression for the $k$-point function 
of the charged Wilson line operators in terms of the Kronecker theta series 
\begin{align}
\label{closed_W1}
&\left\langle W_{n_1}W_{n_2}\cdots W_{n_{k}} \right\rangle^{U(N)}
\nonumber\\
&=
N\mathcal{I}^{U(N)}
+
\xi^{N^2/2}\sum_{\lambda} (-1)^{r+1}\prod_{i=1}^{r} 
\frac{1}{\lambda_i^{m_i} (m_i) !}
\nonumber\\
&\times \Biggl[
\sum_{i=1}^r 
m_i \lambda_i 
Q(\lambda_i;0;\xi^{\frac{N}{2}})^{m_i-1}
\left\{
\sum_{p>1}^k \sum_{l_1+\cdots l_{p}=\lambda_i}
\sum_{\{ N_{J_i}\}_{i=1}^p}
Q\left(
\{l_i\}_{i=1}^{p}; \{N_{J_i}\}_{i=1}^p; \xi^{\frac{N}{2}}
\right)
\right\}
\nonumber\\
&\times \prod_{j\neq i}^r Q(\lambda_j;0;\xi^{\frac{N}{2}})^{m_j}
\Biggr]. 
\end{align}
The multiple Kronecker theta series (\ref{s_zeta}) can be decomposed into the spectral zeta functions (\ref{szeta_0}) 
which are given by the twisted Weierstrass functions from the relations (\ref{q1_tP1}) and (\ref{qk_tPk}). 
This implies that the Schur line defect correlation functions can be expressed in terms of the twisted Weierstrass functions. 
Since the general expression is quite complicated,
we give several examples in the following. 

\subsection{Charged Wilson line correlators}
\subsubsection{$U(2)$ 2-point functions}
Consider the 2-point functions of the Wilson line operators with charge $+n$ and with $-n$. 
For $\mathcal{N}=2^*$ $U(2)$ SYM theory the 2-point function can be constructed from $Z_1^{E/H}(n)$ and $Z_2^{E/H}(n)$. 
It is given by
\begin{align}
\label{u2_2pt_exact1}
\langle W_n W_{-n}\rangle^{U(2)}
&=-\xi^2 \left[
Q(1;0;\xi)^2-Q(2;0;\xi)-Q(1,1;0,n;\xi)
\right]
\nonumber\\
&=\xi^2 \left[
Q(2;0;\xi)+Q(1,1;0,n;\xi)
\right], 
\end{align}
where we have used $Q(1;0;\xi)=0$. 
Since the $\mathcal{N}=2^*$ $U(2)$ Schur index is given by \cite{Hatsuda:2022xdv}
\begin{align}
\label{u2_exact1}
\mathcal{I}^{U(2)}&=\frac{\xi^2}{2}Q(2;0;\xi)
=\frac{\xi}{2}
P_2\left[
\begin{matrix}
q \xi^2 \\
1\\
\end{matrix}
\right](\zeta,\tau)
\nonumber\\
&=-\sum_{\begin{smallmatrix}
p_1,p_2\in \mathbb{Z}\\
p_1<p_2\\
\end{smallmatrix}}
\frac{\xi^{-p_1-p_2+2}}
{(1-\xi q^{p_1}) (1-\xi q^{p_2})}, 
\end{align}
we have
\begin{align}
\label{u2_2pt_prop1}
\langle W_n W_{-n}\rangle^{U(2)}
&=2\mathcal{I}^{U(2)}+\xi^2Q(1,1;0,n;\xi)
\nonumber\\
&=
\left(
-2\sum_{\begin{smallmatrix}
p_1,p_2\in \mathbb{Z}\\
p_1<p_2\\
\end{smallmatrix}}
+\sum_{\begin{smallmatrix}
p_1,p_2\in \mathbb{Z}\\
p_2=p_1+n\\
\end{smallmatrix}}
\right)
\frac{\xi^{-p_1-p_2+2}}
{(1-\xi q^{p_1}) (1-\xi q^{p_2})}
. 
\end{align}

When $n$ $=$ $0$, $Q(1,1;0,n)$ reduces to $Q(2;0)$ so that the 2-point function (\ref{u2_2pt_prop1}) becomes $4\mathcal{I}^{U(2)}$. 
From (\ref{q1_tP1}) and (\ref{q11_expand}) we have 
\footnote{
The expression (\ref{q11_tW}) is valid for $n\neq 0$. 
}
\begin{align}
\label{q11_tW}
\xi^2
Q(1,1;0,n;u;\xi;q)&=
\xi^2
\Bigl[
(n)_{q,\xi}+(-n)_{q,\xi}
\Bigr]P_1\left[
\begin{matrix}
\xi^2\\
1\\
\end{matrix}
\right]
(\nu,\tau), 
\end{align}
where $u=e^{2\pi i\nu}$. 
The expression (\ref{q11_tW}) which captures the 2-point function of the charged Wilson line operators is invariant under the transformation (\ref{xi_transf}). 

It follows from (\ref{u2_exact1}), (\ref{u2_2pt_prop1}) and (\ref{q11_tW}) 
that the 2-point function (\ref{u2_2pt_prop1}) is expressed in terms of the twisted Weierstrass functions
\begin{align}
\label{u2_2pt_exact2}
\langle W_n W_{-n}\rangle^{U(2)}
&=\xi P_2\left[
\begin{matrix}
q\xi^2\\
1
\end{matrix}
\right](\zeta,\tau)
-\xi^2
\frac{(n)_{q,\xi}}{(n)_{q\xi^2,1}}
P_1\left[
\begin{matrix}
\xi^2\\
1\\
\end{matrix}
\right](\zeta,\tau)
\nonumber\\
&=
\xi P_2\left[
\begin{matrix}
q\xi^2\\
1
\end{matrix}
\right](\zeta,\tau)
-\xi^2
\frac{q^{\frac{n}{2}} \xi^{n}-q^{-\frac{n}{2}}\xi^{-n}}
{q^{\frac{n}{2}}-q^{-\frac{n}{2}}}
P_1\left[
\begin{matrix}
\xi^2\\
1\\
\end{matrix}
\right](\zeta,\tau). 
\end{align}
where $\xi=e^{2\pi i\zeta}$. 
Here we have used the relation (\ref{chargefcn_prop2}). 

A simple calculation also leads to another closed-form of the $U(2)$ 2-point function 
\begin{align}
\label{u2_2pt_exact2}
\langle W_nW_{-n}\rangle^{U(2)}
&=
\frac{(q;q)_{\infty}^2}{(\xi^{-2};q)_{\infty} (q^2\xi^{2};q)_{\infty}}
\sum_{m\in \mathbb{Z}\setminus \{0,n\}}
\frac{q^{\frac{m}{2}}\xi^{m}-q^{-\frac{m}{2}}\xi^{-m}}{q^{\frac12}\xi-q^{-\frac12}\xi^{-1}}
\frac{q^{\frac{m-1}{2}}}{1-q^m}. 
\end{align}
This can be simply obtained from the Schur index of $\mathcal{N}=2^*$ $U(2)$ SYM theory of the form 
\begin{align}
\label{u2_exact2}
\mathcal{I}^{U(2)}&=
\frac{(q;q)_{\infty}^2}{(\xi^{-2};q)_{\infty} (q^2\xi^{2};q)_{\infty}}
\sum_{m>0}
\frac{q^{\frac{m}{2}}\xi^{m}-q^{-\frac{m}{2}}\xi^{-m}}{q^{\frac12}\xi-q^{-\frac12}\xi^{-1}}
\frac{q^{\frac{m-1}{2}}}{1-q^m} 
\end{align}
by modifying the domain of integers in the sum. 
By setting $\xi$ to $q^{-\frac12}$ we get the unflavored 2-point function 
\begin{align}
\langle W_nW_{-n}\rangle^{U(2)}
&\xrightarrow{\xi\rightarrow q^{-1/2}}
\sum_{m\in \mathbb{Z}\setminus \{0,n\}}
\frac{mq^{\frac{m-1}{2}}}{1-q^m}
\nonumber\\
&=2\sum_{m>0}
\frac{mq^{\frac{m-1}{2}}}{1-q^m}
-\frac{nq^{\frac{n-1}{2}}}{1-q^n}. 
\end{align}

\subsubsection{$U(3)$ 2-point functions}
The 2-point function for $\mathcal{N}=2^*$ $U(3)$ SYM theory 
can be obtained from the three spectral zeta functions, $Z_1^{E/H}(n)$, $Z_2^{E/H}(n)$ and $Z_3^{E/H}(n)$. 
We first set $u=\xi^{3/2}$. 
It is then given by 
\begin{align}
\label{u3_2pt_exact1}
&
\langle W_n W_{-n}\rangle^{U(3)}
\nonumber\\
&=\frac{\xi^{\frac92}}{2}
\Bigl[
Q(1;0;\xi^{\frac32})^3-3Q(1;0;\xi^{\frac32})Q(2;0;\xi^{\frac32})+2Q(3;0;\xi^{\frac32})
\nonumber\\
&-2Q(1;0;\xi^{\frac32})Q(1,1;0,n;\xi^{\frac32})
+2Q(2,1;0,n;\xi^{\frac32})+2Q(1,2;0,n;\xi^{\frac32})
\Bigr]. 
\end{align}
Since the $U(3)$ Schur index is given by \cite{Hatsuda:2022xdv}
\begin{align}
\label{u3_exact1}
\mathcal{I}^{U(3)}&=\frac{\xi^{\frac92}}{6} 
\Bigl[
Q(1;0;\xi^{\frac32})^3-3Q(1;0;\xi^{\frac32})Q(2;0;\xi^{\frac32})+2Q(3;0;\xi^{\frac32})
\Bigr]
\nonumber\\
&=
-\sum_{\begin{smallmatrix}
p_1,p_2,p_3\in \mathbb{Z}\\
p_1<p_2<p_3\\
\end{smallmatrix}}
\frac{\xi^{-p_1-p_2-p_3+\frac92}}
{(1-\xi^{\frac32}q^{p_1}) (1-\xi^{\frac32}q^{p_2}) (1-\xi^{\frac32}q^{p_3})}
,
\end{align}
we can rewrite the correlation function (\ref{u3_2pt_exact1}) as 
\begin{align}
\label{u3_2pt_exact1a}
&
\langle W_n W_{-n}\rangle^{U(3)}
\nonumber\\
&=3\mathcal{I}^{U(3)}
+\xi^{\frac92}
\Bigl[
-Q(1;0;\xi^{\frac32})Q(1,1;0,n;\xi^{\frac32})+Q(2,1;0,n;\xi^{\frac32})+Q(1,2;0,n;\xi^{\frac32})
\Bigr]
\nonumber\\
&=
\left(
-3\sum_{\begin{smallmatrix}
p_1,p_2,p_3\in \mathbb{Z}\\
p_1<p_2<p_3\\
\end{smallmatrix}}
+\sum_{\begin{smallmatrix}
p_1,p_2,p_3\in \mathbb{Z}\\
p_3=p_2+n\\
\end{smallmatrix}}
-\sum_{\begin{smallmatrix}
p_1,p_2,p_3\in \mathbb{Z}\\
p_2=p_1,p_3=p_1+n\\
\end{smallmatrix}}
-\sum_{\begin{smallmatrix}
p_1,p_2,p_3\in \mathbb{Z}\\
p_2=p_1+n,p_3=p_1+n\\
\end{smallmatrix}}
\right)
\nonumber\\
&\times 
\frac{\xi^{-p_1-p_2-p_3+\frac92}}
{(1-\xi^{\frac32}q^{p_1}) (1-\xi^{\frac32}q^{p_2}) (1-\xi^{\frac32}q^{p_3})}
. 
\end{align}
The charge dependent term 
\begin{align}
\xi^{\frac92}
\Bigl[
-Q(1;0;\xi^{\frac32})Q(1,1;0,n;\xi^{\frac32})+Q(2,1;0,n;\xi^{\frac32})+Q(1,2;0,n;\xi^{\frac32})
\Bigr]
\end{align}
is invariant under the transformation (\ref{xi_transf}). 

Setting the fugacity $u$ to $\xi$, we can find another expression with fewer terms. 
In this case, $Q(1;0;\xi)$ vanishes so that the Schur index can be simply written as \cite{Hatsuda:2022xdv}
\begin{align}
\mathcal{I}^{U(3)}
&=-\frac{\xi^{\frac92}}{3}\frac{\theta(\xi)}{\theta(\xi^{-2})}Q(3;0;\xi), 
\end{align}
where $\theta(x):=\theta(x;q)$  
and the 2-point function is given by
\begin{align}
\label{u3_2pt_exact2}
\langle W_n W_{-n}\rangle^{U(3)}
&=-\xi^{\frac92}
\frac{\theta(\xi)}{\theta(\xi^{-2})}
\Bigl[
Q(3;0;\xi)+Q(2,1;0,n;\xi)+Q(1,2;0,n;\xi)
\Bigr]
\nonumber\\
&=3\mathcal{I}^{U(3)}-\xi^{\frac92}\frac{\theta(\xi)}{\theta(\xi^{-2})}
\Bigl[
Q(2,1;0,n;\xi)+Q(1,2;0,n;\xi)
\Bigr]. 
\end{align}

When $n=0$, both $Q(2,1;0,n;\xi)$ and $Q(1,2;0,n;\xi)$ reduce to $Q(3;0;\xi)$ 
so that the $U(3)$ 2-point function (\ref{u3_2pt_exact1a}) becomes $9 \mathcal{I}^{U(3)}$. 

From (\ref{q1_tP1}), (\ref{qk_tPk}), (\ref{u3_2pt_exact2}) and (\ref{q21_expand}) 
it can be expressed in terms of the twisted Weierstrass functions
\begin{align}
\label{u3_2pt_exact3}
&
\langle W_n W_{-n}\rangle^{U(3)}
\nonumber\\
&=
\frac{\xi^{\frac52}}{2}
\frac{\theta(\xi)}{\theta(\xi^2)}
\Biggl[
P_2\left[
\begin{matrix}
q^2\xi^3\\
1\\
\end{matrix}
\right](\zeta,\tau)
+2P_3\left[
\begin{matrix}
q^2\xi^3\\
1\\
\end{matrix}
\right](\zeta,\tau)
\nonumber\\
&-2\xi 
\frac{(n)_{q,\xi}}{(n)_{q\xi^2,1}}
P_2\left[
\begin{matrix}
q\xi^3\\
1\\
\end{matrix}
\right](\zeta,\tau)
+2\xi^2 
\frac{(n)_{q,\xi}^2}{(n)_{q\xi,1}(n)_{q\xi^3,1}}
P_1\left[
\begin{matrix}
\xi^3\\
1\\
\end{matrix}
\right](\zeta,\tau)
\Biggr]. 
\end{align}

\subsubsection{$U(4)$ 2-point functions}
Next consider the 2-point function for $\mathcal{N}=2^*$ $U(4)$ SYM theory. 
In this case there are four spectral zeta functions which contribute to the correlator. 
If we set $u$ to $\xi^2$, we find
\begin{align}
\label{u4_2pt_exact1}
\langle W_n W_{-n}\rangle^{U(4)}
&=-\frac{\xi^8}{6}
\Bigl[
Q(1;0;\xi^2)^4-6Q(1;0;\xi^2)^2Q(2;0;\xi^2)
\nonumber\\
&
+3Q(2;0;\xi^2)^2+8Q(1;0;\xi^2)Q(3;0;\xi^2)-6Q(4;0;\xi^2)
\nonumber\\
&
-3Q(1;0;\xi^2)^2Q(1,1;0,n;\xi^2)
+3Q(2;0)Q(1,1;0,n;\xi^2)
\nonumber\\
&+6Q(1;0;\xi^2)Q(2,1;0,n;\xi^2)+6Q(1;0;\xi^2)Q(1,2;0,n;\xi^2)
\nonumber\\
&-6Q(3,1;0,n;\xi^2)-6Q(2,2;0,n;\xi^2)-6Q(1,3;0,n;\xi^2)
\Bigr]. 
\end{align}
As the $\mathcal{N}=2^*$ $U(4)$ Schur index is given by \cite{Hatsuda:2022xdv}
\begin{align}
\label{u4_exact1}
\mathcal{I}^{U(4)}&=
-\frac{\xi^8}{24}
\Bigl[
Q(1;0;\xi^2)-6Q(1;0;\xi^2)^2Q(2;0;\xi^2)
\nonumber\\
&+3Q(2;0;\xi^2)^2+8Q(1;0;\xi^2)Q(3;0;\xi^2)-6Q(4;0;\xi^2)
\Bigr]
\nonumber\\
&=
-\sum_{
\begin{smallmatrix}
p_1,p_2,p_3,p_4\in \mathbb{Z}\\
p_1<p_2<p_3<p_4\\
\end{smallmatrix}
}
\frac{\xi^{-p_1-p_2-p_3-p_4+8}}
{(1-\xi^2 q^{p_1}) (1-\xi^2 q^{p_2}) (1-\xi^2 q^{p_3}) (1-\xi^2 q^{p_4}) }
, 
\end{align}
it can be expressed as
\begin{align}
\label{u4_2pt_exact1a}
&\langle W_n W_{-n}\rangle^{U(4)}
\nonumber\\
&=4\mathcal{I}^{U(4)}
-\frac{\xi^8}{2}
\Bigl[
-Q(1;0;\xi^2)^2Q(1,1;0,n;\xi^2)
+Q(2;0;\xi^2)Q(1,1;0,n;\xi^2)
\nonumber\\
&+2Q(1;0;\xi^2)Q(2,1;0,n;\xi^2)+2Q(1;0;\xi^2)Q(1,2;0,n;\xi^2)
\nonumber\\
&-2Q(3,1;0,n;\xi^2)-2Q(2,2;0,n;\xi^2)-2Q(1,3;0,n;\xi^2)
\Bigr]
\nonumber\\
&=
\Biggl(
-4\sum_{
\begin{smallmatrix}
p_1,p_2,p_3,p_4\in \mathbb{Z}\\
p_1<p_2<p_3<p_4\\
\end{smallmatrix}
}
+\frac12 
\sum_{
\begin{smallmatrix}
p_1,p_2,p_3,p_4\in \mathbb{Z}\\
p_4=p_3+n\\
\end{smallmatrix}
}
-\frac12 
\sum_{
\begin{smallmatrix}
p_1,p_2,p_3,p_4\in \mathbb{Z}\\
p_2=p_1, p_4=p_3+n\\
\end{smallmatrix}
}
-\sum_{
\begin{smallmatrix}
p_1,p_2,p_3,p_4\in \mathbb{Z}\\
p_3=p_2, p_4=p_2+n\\
\end{smallmatrix}
}
-\sum_{
\begin{smallmatrix}
p_1,p_2,p_3,p_4\in \mathbb{Z}\\
p_3=p_2+n, p_4=p_2+n\\
\end{smallmatrix}
}
\nonumber\\
&
+\sum_{
\begin{smallmatrix}
p_1,p_2,p_3,p_4\in \mathbb{Z}\\
p_2=p_1,p_3=p_1, p_4=p_1+n\\
\end{smallmatrix}
}
+\sum_{
\begin{smallmatrix}
p_1,p_2,p_3,p_4\in \mathbb{Z}\\
p_2=p_1,p_3=p_1+n, p_4=p_1+n\\
\end{smallmatrix}
}
+\sum_{
\begin{smallmatrix}
p_1,p_2,p_3,p_4\in \mathbb{Z}\\
p_2=p_1+n,p_3=p_1+n, p_4=p_1+n\\
\end{smallmatrix}
}
\Biggr)
\nonumber\\
&\times 
\frac{\xi^{-p_1-p_2-p_3-p_4+8}}
{(1-\xi^2 q^{p_1}) (1-\xi^2 q^{p_2}) (1-\xi^2 q^{p_3}) (1-\xi^2 q^{p_4}) }
. 
\end{align}
Again the charge dependent terms in (\ref{u4_2pt_exact1}) are invariant under the transformation (\ref{xi_transf}). 

Specializing the fugacity $u$ to $\xi$, we have alternative expression
\begin{align}
\label{u4_2pt_exact2}
\langle W_n W_{-n}\rangle^{U(4)}
&=4\mathcal{I}^{U(4)}
+\frac{\xi^8}{2}
\frac{\theta(\xi)}{\theta(\xi^{-3})}
\Bigl[
Q(2;0;\xi)Q(1,1;0,n;\xi)
\nonumber\\
&-2Q(3,1;0,n;\xi)-2Q(2,2;0,n;\xi)-2Q(1,3;0,n;\xi)
\Bigr]. 
\end{align}
From (\ref{q1_tP1}), (\ref{qk_tPk}), (\ref{u4_exact1}) (\ref{u4_2pt_exact2}), (\ref{q31_expand}) and (\ref{q22_expand}) 
we can write the correlation function in terms of the twisted Weierstrass function
\begin{align}
\label{u4_2pt_exact3}
&
\langle W_n W_{-n}\rangle^{U(4)}
\nonumber\\
&=-\frac{\xi^5}{6}
\frac{\theta(\xi)}{\theta(\xi^3)}
\Biggl[
3\xi P_2\left[
\begin{matrix}
q\xi^2\\
1\\
\end{matrix}
\right]^2
(\zeta,\tau)
-2P_2\left[
\begin{matrix}
q^3\xi^4\\
1\\
\end{matrix}
\right]
(\zeta,\tau)
-6P_3\left[
\begin{matrix}
q^3\xi^4\\
1\\
\end{matrix}
\right]
(\zeta,\tau)
-6P_4\left[
\begin{matrix}
q^3\xi^4\\
1\\
\end{matrix}
\right](\zeta,\tau)
\nonumber\\
&- 3\xi^2
\frac{(n)_{q,\xi}}{(n)_{q\xi^2,1}}
P_2\left[
\begin{matrix}
q\xi^2\\
1\\
\end{matrix}
\right]
(\zeta,\tau)
P_1\left[
\begin{matrix}
\xi^2\\
1\\
\end{matrix}
\right]
(\zeta,\tau)
+3\xi 
\frac{(n)_{q,\xi}}{(n)_{q\xi^2,1}}
\Bigl(
P_2\left[
\begin{matrix}
q^2\xi^4\\
1\\
\end{matrix}
\right]
(\zeta,\tau)
+
2P_3\left[\begin{matrix}q^2\xi^4\\1\\\end{matrix}\right]
(\zeta,\tau)
\Bigr)
\nonumber\\
&
-6\xi^2 
\frac{(n)_{q,\xi}^2}{(n)_{q\xi,1} (n)_{q\xi^3,1}}
P_2\left[
\begin{matrix}
q\xi^4\\
1\\
\end{matrix}
\right]
(\zeta,\tau)
+6\xi^3
\frac{(n)_{q,\xi}^3}{(n)_{q\xi^4,1} (n)_{q\xi,1}^2}
P_1\left[
\begin{matrix}
\xi^4\\
1\\
\end{matrix}
\right]
(\zeta,\tau)
\Biggr]. 
\end{align}

\subsubsection{$U(2)$ 3-point functions}
Next consider the 3-point functions of the Wilson line operators which carry charges $n_1$, $n_2$ and $n_3$ 
obeying the Gauss law condition $n_1+n_2+n_3=0$. 

For $\mathcal{N}=2^*$ $U(2)$ SYM theory the 3-point function can be obtained from the spectral zeta functions $Z_{1}^{E/H}(n_1,n_2)$ and $Z_{2}^{E/H}(n_1,n_2)$. 
With the specialization $u=\xi$, we find
\begin{align}
\label{u2_3pt_exact1}
&
\langle W_{n_1} W_{n_2} W_{-n_1-n_2}\rangle^{U(2)}
\nonumber\\
&=-\xi^2 \Bigl[
Q(1;0;\xi)^2-Q(2;0;\xi)
\nonumber\\
&-Q(1,1;0,n_1;\xi)-Q(1,1;0,n_2;\xi)-Q(1,1;0,n_1+n_2;\xi)
\Bigr], 
\end{align}
where $Q(1;0;\xi)$ $=$ $0$. 
This is consistent with the relation (\ref{u2_3pt_red}) and the expression (\ref{u2_2pt_exact1}) of the $U(2)$ 2-point function. 
According to the closed-form expression (\ref{u2_exact1}) of the $U(2)$ Schur index, 
we can write it as
\begin{align}
\label{u2_3pt_exact1a}
&
\langle W_{n_1} W_{n_2} W_{-n_1-n_2}\rangle^{U(2)}
\nonumber\\
&=
\left(
-2\sum_{
\begin{smallmatrix}
p_1,p_2\in \mathbb{Z}\\
p_1<p_2\\
\end{smallmatrix}
}
+\sum_{
\begin{smallmatrix}
p_1,p_2\in \mathbb{Z}\\
p_2=p_1+n_1\\
\end{smallmatrix}
}
+\sum_{
\begin{smallmatrix}
p_1,p_2\in \mathbb{Z}\\
p_2=p_1+n_2\\
\end{smallmatrix}
}
+\sum_{
\begin{smallmatrix}
p_1,p_2\in \mathbb{Z}\\
p_2=p_1+n_1+n_2\\
\end{smallmatrix}
}
\right)
\frac{\xi^{-p_1-p_2+2}}
{(1-\xi q^{p_1}) (1-\xi q^{p_2})}. 
\end{align}

\subsubsection{$U(3)$ 3-point functions}
Consider the 3-point function for $\mathcal{N}=2^*$ $U(3)$ SYM theory. 
It is produced by three spectral zeta functions 
$Z_{1}^{E/H}(n_1,n_2)$, $Z_{2}^{E/H}(n_1,n_2)$ and $Z_{3}^{E/H}(n_1,n_2)$. 
By taking $u=\xi^{\frac32}$, we obtain 
\begin{align}
\label{u3_3pt_exact1}
&
\langle W_{n_1}W_{n_2}W_{-n_1-n_2}\rangle^{U(3)}
\nonumber\\
&=\frac{\xi^{\frac92}}{2}
\Bigl[
Q(1;0;\xi^{\frac32})^3-3Q(1;0;\xi^{\frac32})Q(2;0;\xi^{\frac32})+2Q(3;0;\xi^{\frac32})
\nonumber\\
&-2Q(1;0;\xi^{\frac32})\sum_{i=1}^2Q(1,1;0,n_i;\xi^{\frac32})
-2Q(1;0;\xi^{\frac32})Q(1,1;0,n_1+n_2;\xi^{\frac32})
\nonumber\\
&+2\sum_{i=1}^2Q(2,1;0,n_i;\xi^{\frac32})
+2Q(2,1;0,n_1+n_2;\xi^{\frac32})
\nonumber\\
&+2\sum_{i=1}^2Q(1,2;0,n_i;\xi^{\frac32})
+2Q(1,2;0,n_1+n_2;\xi^{\frac32})
+2\sum_{i=1}^2Q(1,1,1;0,n_i,n_1+n_2;\xi^{\frac32})
\Bigr]. 
\end{align}
We can rewrite this as
\begin{align}
\label{u3_3pt_exact1a}
&
\langle W_{n_1}W_{n_2}W_{-n_1-n_2}\rangle^{U(3)}
\nonumber\\
&=
\Biggl(
-3\sum_{\begin{smallmatrix}
p_1,p_2,p_3\in \mathbb{Z}\\
p_1<p_2<p_3\\
\end{smallmatrix}
}
+
\sum_{i=1}^2
\sum_{\begin{smallmatrix}
p_1,p_2,p_3\in \mathbb{Z}\\
p_3=p_2+n_i\\
\end{smallmatrix}
}
+\sum_{\begin{smallmatrix}
p_1,p_2,p_3\in \mathbb{Z}\\
p_3=p_2+n_1+n_2\\
\end{smallmatrix}
}
-\sum_{i=1}^2
\sum_{\begin{smallmatrix}
p_1,p_2,p_3\in \mathbb{Z}\\
p_2=p_1,p_3=p_1+n_i\\
\end{smallmatrix}
}
-\sum_{\begin{smallmatrix}
p_1,p_2,p_3\in \mathbb{Z}\\
p_2=p_1,p_3=p_1+n_1+n_2\\
\end{smallmatrix}
}
\nonumber\\
&
-\sum_{i=1}^2
\sum_{\begin{smallmatrix}
p_1,p_2,p_3\in \mathbb{Z}\\
p_2=p_1+n_i,p_3=p_1+n_i\\
\end{smallmatrix}
}
-\sum_{\begin{smallmatrix}
p_1,p_2,p_3\in \mathbb{Z}\\
p_2=p_1+n_1+n_2,p_3=p_1+n_1+n_2\\
\end{smallmatrix}
}
-
\sum_{i=1}^2
\sum_{\begin{smallmatrix}
p_1,p_2,p_3\in \mathbb{Z}\\
p_2=p_1+n_i,p_3=p_1+n_1+n_2\\
\end{smallmatrix}
}
\Biggr)
\nonumber\\
&\times 
\frac{\xi^{-p_1-p_2-p_3+\frac92}}
{(1-\xi^{\frac32}q^{p_1}) (1-\xi^{\frac32}q^{p_2}) (1-\xi^{\frac32}q^{p_3})}. 
\end{align}

According to the closed-form expression (\ref{u3_exact1}) of the $U(3)$ Schur index, 
we have
\begin{align}
\label{u3_3pt_exact2}
\langle W_{n_1}W_{n_2}W_{-n_1-n_2}\rangle^{U(3)}
&=-6\mathcal{I}^{U(3)}
+\sum_{i=1}^2 \langle W_{n_i}W_{-n_i}\rangle^{U(3)}
+\langle W_{n_1+n_2}W_{-n_1-n_2}\rangle^{U(3)}
\nonumber\\
&+\xi^{\frac92} \sum_{i=1}^2 Q(1,1,1;0,n_i,n_1+n_2;\xi^{\frac32}). 
\end{align}
Unlike the $U(2)$ case, the 3-point function is not only given by the $U(3)$ Schur index and the $U(3)$ 2-point functions. 
The remaining term is 
\begin{align}
\label{u3_3pt_remain}
\xi^{\frac92} \sum_{i=1}^2 Q(1,1,1;0,n_i,n_1+n_2;\xi^{\frac32}). 
\end{align}
From (\ref{q111_expand1}) the term (\ref{u3_3pt_remain}) can be rewritten in terms of the twisted Weierstrass function as
\begin{align}
&\xi^{\frac92} \sum_{i=1}^2 Q(1,1,1;0,n_i,n_1+n_2;\xi^{\frac32})
\nonumber\\
&
=\xi^{\frac92} 
\Biggl[
\sum_{\pm }
\sum_{i=1}^2 
(\pm n_i)_{q,\xi}(\pm n_1\pm n_2)_{q,\xi}
+\sum_{i\neq j}(-n_i)_{q,\xi}(n_j)_{q,\xi}
\Biggr]
P_1\left[
\begin{matrix}
\xi^3\\
1\\
\end{matrix}
\right]
( \frac32\zeta ,\tau ). 
\end{align}

Note that this term is equal to 
\begin{align}
&\xi^{\frac92} \frac{\theta(\xi)}{\theta(\xi^2)} \sum_{i=1}^2 Q(1,1,1;0,n_i,n_1+n_2;\xi)
\nonumber\\
&=\xi^{\frac92} 
\Biggl[
\sum_{\pm}
\sum_{i=1}^2 
(\pm n_i)_{q,\xi}(\pm n_1\pm n_2)_{q,\xi}
+\sum_{i\neq j}(-n_i)_{q,\xi}(n_j)_{q,\xi}
\Biggr]
P_1\left[
\begin{matrix}
\xi^3\\
1\\
\end{matrix}
\right]
(\zeta ,\tau ), 
\end{align}
which is obtained by setting $u=\xi$ since $P_1\left[ \begin{smallmatrix}
\xi^3\\
1\\
\end{smallmatrix}
\right]
(\frac32\zeta,\tau)$ $=$ $\frac{(q)_{\infty}^3}{\theta(\xi^{-3})}$ 
and $P_1\left[ \begin{smallmatrix}
\xi^3\\
1\\
\end{smallmatrix}
\right]
(\zeta,\tau)$ $=$ $\frac{(q)_{\infty}^3 \theta(\xi^{-2})}{\theta(\xi^{-3}) \theta(\xi)}$. 
Also the term (\ref{u3_3pt_remain}) is invariant under the transformation (\ref{xi_transf}). 

\subsubsection{$U(4)$ 3-point functions}
For $\mathcal{N}=2^*$ $U(4)$ SYM theory 
the 3-point function can be built up from the four spectral zeta functions 
$Z_1^{E/H}(n_1,n_2)$, $Z_2^{E/H}(n_1,n_2)$, $Z_3^{E/H}(n_1,n_2)$ and $Z_4^{E/H}(n_1,n_2)$. 
With $u=\xi^2$, it is given by
\begin{align}
\label{u4_3pt_exact1}
&
\langle W_{n_1}W_{n_2}W_{-n_1-n_2}\rangle^{U(4)}
=-\frac{\xi^8}{6}
\Biggl[
Q(1;0;\xi^2)^4-6Q(1;0;\xi^2)^2Q(2;0;\xi^2)
\nonumber\\
&+3Q(2;0;\xi^2)^2
+8Q(1;0;\xi^2)Q(3;0;\xi^2)-6Q(4;0;\xi^2)
\nonumber\\
&-3Q(1;0;\xi^2)^2 
\sum_{n=n_1, n_2, n_1+n_2} 
Q(1,1;0,n;\xi^2)
+3Q(2;0;\xi^2)\sum_{n=n_1, n_2, n_1+n_2}
Q(1,1;0,n;\xi^2)
\nonumber\\
&+6Q(1;0;\xi^2)
\sum_{n=n_1, n_2, n_1+n_2}  
Q(2,1;0,n;\xi^2)
+6Q(1;0;\xi^2)
\sum_{n=n_1, n_2, n_1+n_2} 
Q(1,2;0,n;\xi^2)
\nonumber\\
&-6\sum_{n=n_1, n_2, n_1+n_2} 
( Q(3,1;0,n;\xi^2)
+Q(2,2;0,n;\xi^2)
+Q(1,3;0,n;\xi^2) )
\nonumber\\
&+6Q(1;0)\sum_{i=1}^2Q(1,1,1;0,n_i,n_1+n_2;\xi^2)
-6\sum_{i=1}^2Q(2,1,1;0,n_i,n_1+n_2;\xi^2)
\nonumber\\
&
-6\sum_{i=1}^2Q(1,2,1;0,n_i,n_1+n_2;\xi^2)
-6\sum_{i=1}^2Q(1,1,2;0,n_i,n_1+n_2;\xi^2)
\Biggr]. 
\end{align}
While the first five lines contain generalized terms 
appearing in the $U(4)$ 2-point function (\ref{u4_2pt_exact1}), 
the last two lines are particular terms for the $U(4)$ 3-point function. 
The correlator (\ref{u4_3pt_exact1}) also can be written as multiple series 
\begin{align}
\label{u4_3pt_exact1a}
\langle W_{n_1}W_{n_2}W_{-n_1-n_2}\rangle^{U(4)}
&=
\left(
{\sum_{p_1,p_2,p_3,p_4\in \mathbb{Z}}}^{(1)}
+{\sum_{p_1,p_2,p_3,p_4\in \mathbb{Z}}}^{(2)}
+{\sum_{p_1,p_2,p_3,p_4\in \mathbb{Z}}}^{(3)}
\right)
\nonumber\\
&\times 
\frac{\xi^{-p_1-p_2-p_3-p_4+8}}
{(1-\xi^2 q^{p_1}) (1-\xi^2 q^{p_2}) (1-\xi^2 q^{p_3}) (1-\xi^2 q^{p_4})}, 
\end{align}
where 
\begin{align}
&
{\sum_{p_1,p_2,p_3,p_4\in \mathbb{Z}}}^{(1)}
=-4\sum_{\begin{smallmatrix}
p_1,p_2,p_3,p_4\in \mathbb{Z}\\
p_1<p_2<p_3<p_4\\
\end{smallmatrix}}
\end{align}
is the sum producing a scalar multiple of the $U(4)$ Schur index, 
\begin{align}
&{\sum_{p_1,p_2,p_3,p_4\in \mathbb{Z}}}^{(2)}
=
\frac12 
\sum_{i=1}^2 
\sum_{\begin{smallmatrix}
p_1,p_2,p_3,p_4\in \mathbb{Z}\\
p_4=p_3+n_i\\
\end{smallmatrix}}
+\frac12
\sum_{\begin{smallmatrix}
p_1,p_2,p_3,p_4\in \mathbb{Z}\\
p_4=p_3+n_1+n_2\\
\end{smallmatrix}}
-\frac12 
\sum_{i=1}^2 
\sum_{\begin{smallmatrix}
p_1,p_2,p_3,p_4\in \mathbb{Z}\\
p_2=p_1,p_4=p_3+n_i\\
\end{smallmatrix}}
-\frac12 
\sum_{\begin{smallmatrix}
p_1,p_2,p_3,p_4\in \mathbb{Z}\\
p_2=p_1,p_4=p_3+n_1+n_2\\
\end{smallmatrix}}
\nonumber\\
&-\sum_{i=1}^2 
\sum_{\begin{smallmatrix}
p_1,p_2,p_3,p_4\in \mathbb{Z}\\
p_3=p_2,p_4=p_2+n_i\\
\end{smallmatrix}}
-\sum_{\begin{smallmatrix}
p_1,p_2,p_3,p_4\in \mathbb{Z}\\
p_3=p_2,p_4=p_2+n_1+n_2\\
\end{smallmatrix}}
-\sum_{i=1}^2 
\sum_{\begin{smallmatrix}
p_1,p_2,p_3,p_4\in \mathbb{Z}\\
p_3=p_2+n_i,p_4=p_2+n_i\\
\end{smallmatrix}}
-\sum_{\begin{smallmatrix}
p_1,p_2,p_3,p_4\in \mathbb{Z}\\
p_3=p_2+n_1+n_2,p_4=p_2+n_1+n_2\\
\end{smallmatrix}}
\nonumber\\
&
+\sum_{i=1}^2
\sum_{\begin{smallmatrix}
p_1,p_2,p_3,p_4\in \mathbb{Z}\\
p_2=p_1,p_3=p_1,p_4=p_1+n_i\\
\end{smallmatrix}}
+\sum_{\begin{smallmatrix}
p_1,p_2,p_3,p_4\in \mathbb{Z}\\
p_2=p_1,p_3=p_1,p_4=p_1+n_1+n_2\\
\end{smallmatrix}}
\nonumber\\
&
+\sum_{i=1}^2
\sum_{\begin{smallmatrix}
p_1,p_2,p_3,p_4\in \mathbb{Z}\\
p_2=p_1,p_3=p_1+n_i,p_4=p_1+n_i\\
\end{smallmatrix}}
+\sum_{\begin{smallmatrix}
p_1,p_2,p_3,p_4\in \mathbb{Z}\\
p_2=p_1,p_3=p_1+n_1+n_2,p_4=p_1+n_1+n_2\\
\end{smallmatrix}}
\nonumber\\
&+\sum_{i=1}^2
\sum_{\begin{smallmatrix}
p_1,p_2,p_3,p_4\in \mathbb{Z}\\
p_2=p_1+n_i,p_3=p_1+n_i,p_4=p_1+n_i\\
\end{smallmatrix}}
+\sum_{\begin{smallmatrix}
p_1,p_2,p_3,p_4\in \mathbb{Z}\\
p_2=p_1+n_1+n_2,p_3=p_1+n_1+n_2,p_4=p_1+n_1+n_2\\
\end{smallmatrix}}
\end{align}
is the sum appearing in the $U(4)$ 2-point function and 
\begin{align}
&{\sum_{p_1,p_2,p_3,p_4\in \mathbb{Z}}}^{(3)}
=
\Biggl[
-\sum_{i=1}^2
\sum_{\begin{smallmatrix}
p_1,p_2,p_3,p_4\in \mathbb{Z}\\
p_3=p_2+n_i,p_4=p_2+n_1+n_2\\
\end{smallmatrix}}
+\sum_{i=1}^2
\Bigl(
\sum_{\begin{smallmatrix}
p_1,p_2,p_3,p_4\in \mathbb{Z}\\
p_2=p_1,p_3=p_1+n_i,p_4=p_1+n_1+n_2\\
\end{smallmatrix}}
\nonumber\\
&
+
\sum_{\begin{smallmatrix}
p_1,p_2,p_3,p_4\in \mathbb{Z}\\
p_2=p_1+n_i,p_3=p_1+n_i,p_4=p_1+n_1+n_2\\
\end{smallmatrix}}
+
\sum_{\begin{smallmatrix}
p_1,p_2,p_3,p_4\in \mathbb{Z}\\
p_2=p_1+n_i,p_3=p_1+n_1+n_2,p_4=p_1+n_1+n_2\\
\end{smallmatrix}}
\Bigr)
\Biggr]
\end{align}
is the sum characterizing the $U(4)$ 3-point function. 

The expression (\ref{u4_3pt_exact1}) is reducible 
in that it contains the terms as a scalar multiple of the $U(4)$ Schur index and that of the $U(4)$ 2-point function. 
We find that  
\begin{align}
\label{u4_3pt_exact2}
&
\langle W_{n_1}W_{n_2}W_{-n_1-n_2}\rangle^{U(4)}
\nonumber\\
&=-8\mathcal{I}^{U(4)}
+\sum_{i=1}^{2} \langle W_{n_i}W_{-n_i}\rangle^{U(4)}
+\langle W_{n_1+n_2}W_{-n_1-n_2}\rangle^{U(4)}
\nonumber\\
&+\xi^{8}\sum_{i=1}^2 \Bigl[
-Q(1;0;\xi^2)Q(1,1,1;0,n_i,n_1+n_2;\xi^2)+
Q(2,1,1;0,n_i,n_1+n_2;\xi^2)
\nonumber\\
&
+Q(1,2,1;0,n_i,n_1+n_2;\xi^2)
+Q(1,1,2;0,n_i,n_1+n_2;\xi^2)
\Bigr]. 
\end{align}

By setting $u=\xi$, we get
\begin{align}
\label{u4_3pt_exact3}
&
\langle W_{n_1}W_{n_2}W_{-n_1-n_2}\rangle^{U(4)}
\nonumber\\
&=-8\mathcal{I}^{U(4)}
+\sum_{i=1}^{2} \langle W_{n_i}W_{-n_i}\rangle^{U(4)}
+\langle W_{n_1+n_2}W_{-n_1-n_2}\rangle^{U(4)}
\nonumber\\
&+\xi^{8} \frac{\theta(\xi)}{\theta(\xi^{3})}\sum_{i=1}^2 \Bigl[
Q(2,1,1;0,n_i,n_1+n_2;\xi)
+Q(1,2,1;0,n_i,n_1+n_2;\xi)
\nonumber\\
&
+Q(1,1,2;0,n_i,n_1+n_2;\xi)
\Bigr]. 
\end{align}
To express the 3-point function (\ref{u4_3pt_exact3}) in terms of the twisted Weierstrass function, 
it suffices to rewrite the irreducible part as
\begin{align}
&\xi^{8} \frac{\theta(\xi)}{\theta(\xi^{3})}\sum_{i=1}^2 
\Bigl[
Q(2,1,1;0,n_i,n_1+n_2;\xi)
+Q(1,2,1;0,n_i,n_1+n_2;\xi)
\nonumber\\
&+Q(1,1,2;0,n_i,n_1+n_2;\xi)
\Bigr]
\nonumber\\
&=
\Bigl[
\sum_{\pm}
\sum_{i=1}^2
(\pm n_i)_{q,\xi}(\pm n_1\pm n_2)_{q,\xi}
+
\sum_{i\neq j}
(-n_i)_{q,\xi}(n_j)_{q,\xi}
\Bigr]
P_2\left[
\begin{matrix}
q\xi^4\\
1\\
\end{matrix}
\right](\zeta,\tau)
\nonumber\\
&+\Bigl[
\sum_{\pm}
\sum_{i=1}^2 (\pm n_i)_{q,\xi}(\pm n_1\pm n_2)_{q,\xi}
\Bigl\{
-(q\xi)^{\pm n_i\pm (n_1+n_2)}
c_{1,\pm}^{(3)}
(\pm n_i)_{q,\xi}(\pm n_1\pm n_2)_{q,\xi}
\nonumber\\
&+(\pm n_i)_{q,\xi}
+(\pm n_1\pm n_2)_{q,\xi}
\Bigr\}
+
\sum_{i\neq j}(-n_i)_{q,\xi} (n_j)_{q,\xi}
\Bigl\{
-(q\xi)^{-n_i+n_j}
c_{2}^{(3)}
(-n_i)_{q,\xi}(n_j)_{q,\xi}
\nonumber\\
&+(-n_i)_{q,\xi}
+(n_j)_{q,\xi}
\Bigr\}
\Bigr]
P_1\left[
\begin{matrix}
\xi^4\\
1\\
\end{matrix}
\right](\zeta,\tau),
\end{align}
where
\begin{align}
c_{1,\pm}^{(3)}&=(2-q^{\mp n_i}-q^{\mp n_1\mp n_2}), \qquad 
c_{2}^{(3)}=(2-q^{n_i}-q^{-n_i}). 
\end{align}

\subsubsection{$U(2)$ 4-point functions}
The 4-point function of the Wilson line operators of charges 
$n_1$, $n_2$, $n_3$ and $n_4$ is allowed when the condition $n_1+n_2+n_3+n_4=0$ holds. 
So we write $n_4=-n_1-n_2-n_3$. 

The 4-point function of the charged Wilson line operators for $\mathcal{N}=2^*$ $U(2)$ SYM theory 
can be obtained from the two spectral zeta functions $Z_1^{E/H}(n_1,n_2,n_3)$ and $Z_{2}^{E/H}(n_1,n_2,n_3)$. 
With $u=\xi$ it is given by
\begin{align}
\label{u2_4pt_exact1}
&
\langle W_{n_1} W_{n_2} W_{n_3} W_{-n_1-n_2-n_3}\rangle^{U(2)}
=-\xi^2 \Bigl[
Q(1;0;\xi)^2-Q(2;0;\xi)
\nonumber\\
&-\sum_{i=1}^3Q(1,1;0,n_i;\xi)
-\sum_{i<j}Q(1,1;0,n_i+n_j;\xi)
-Q(1,1;0,n_1+n_2+n_3;\xi)
\Bigr]
\nonumber\\
&=\xi^2 \Bigl[
Q(2;0;\xi)+\sum_{i=1}^3 Q(1,1;0,n_i;\xi)
+\sum_{i<j}Q(1,1;0,n_i+n_j;\xi)
\nonumber\\
&+Q(1,1;0,n_1+n_2+n_3;\xi)
\Bigr]. 
\end{align}
This is consistent with the relation (\ref{u2_4pt_red})  
where it is expressible in terms of the $U(2)$ Schur index and the 2-point functions. 
We can also write it as 
\begin{align}
\label{u2_4pt_exacta}
&
\langle W_{n_1} W_{n_2} W_{n_3} W_{-n_1-n_2-n_3}\rangle^{U(2)}
\nonumber\\
&=
\Biggl(
-2\sum_{\begin{smallmatrix}
p_1,p_2\in \mathbb{Z}\\
p_1<p_2\\
\end{smallmatrix}}
+
\sum_{i=1}^2
\sum_{\begin{smallmatrix}
p_1,p_2\in \mathbb{Z}\\
p_2=p_1+n_i\\
\end{smallmatrix}}
+
\sum_{i>j}
\sum_{\begin{smallmatrix}
p_1,p_2\in \mathbb{Z}\\
p_2=p_1+n_i+n_j\\
\end{smallmatrix}}
+
\sum_{\begin{smallmatrix}
p_1,p_2\in \mathbb{Z}\\
p_2=p_1+n_1+n_2+n_3\\
\end{smallmatrix}}
\Biggr)
\nonumber\\
&\times 
\frac{\xi^{-p_1-p_2+2}}{(1-\xi q^{p_1}) (1-\xi q^{p_2}) }, 
\end{align}
where the first sum leads to $2\mathcal{I}^{U(2)}$ 
and the others produce the $U(2)$ 2-point functions. 

\subsubsection{$U(3)$ 4-point functions}
For $\mathcal{N}=2^*$ $U(3)$ SYM theory, 
the 4-point function is given by the three spectral zeta functions 
$Z_1^{E/H}(n_1,n_2,n_3)$, $Z_{2}^{E/H}(n_1,n_2,n_3)$ and $Z_{3}^{E/H}(n_1,n_2,n_3)$. 
Setting $u=\xi^{\frac32}$, 
we get the 4-point function 
for $\mathcal{N}=2^*$ $U(3)$ SYM \begin{align}
\label{u3_4pt_exact1}
&
\langle W_{n_1} W_{n_2} W_{n_3} W_{-n_1-n_2-n_3}\rangle^{U(3)}
\nonumber\\
&=\frac{\xi^{\frac92}}{2} \Bigl[
Q(1;0;\xi^{\frac32})^3-3Q(1;0;\xi^{\frac32})Q(2;0;\xi^{\frac32})+2Q(3;0;\xi^{\frac32})
\nonumber\\
&-2Q(1;0;\xi^{\frac32})\sum_{i=1}^3Q(1,1;0,n_i;\xi^{\frac32})
-2Q(1;0;\xi^{\frac32})\sum_{i<j}Q(1,1;0,n_i+n_j;\xi^{\frac32})
\nonumber\\
&-2Q(1;0;\xi^{\frac32})Q(1,1;0,n_1+n_2+n_3;\xi^{\frac32})
+2\sum_{i=1}^3Q(2,1;0,n_i;\xi^{\frac32})
+2\sum_{i<j}Q(2,1;0,n_i+n_j;\xi^{\frac32})
\nonumber\\
&+2Q(2,1;0,n_1+n_2+n_3;\xi^{\frac32})
+2\sum_{i=1}^{3}Q(1,2;0,n_i;\xi^{\frac32})
+2\sum_{i<j}Q(1,2;0,n_i+n_j;\xi^{\frac32})
\nonumber\\
&+2Q(1,2;0,n_1+n_2+n_3;\xi^{\frac32})
+2\sum_{i=1}^3\sum_{j\neq i}Q(1,1,1;0,n_i,n_i+n_j;\xi^{\frac32})
\nonumber\\
&
+2\sum_{i=1}^3Q(1,1,1;0,n_i,n_1+n_2+n_3;\xi^{\frac32})
+2\sum_{i<j}Q(1,1,1;0,n_i+n_j,n_1+n_2+n_3;\xi^{\frac32})
\Bigr]. 
\end{align}
This can be rewritten in terms of the $U(3)$ Schur index, 
the 2- and 3-point functions as (\ref{u3_4pt_red}). 
It is given by the multiple series 
\begin{align}
\label{u3_4pt_exact1a}
&
\langle W_{n_1} W_{n_2} W_{n_3} W_{-n_1-n_2-n_3}\rangle^{U(3)}
\nonumber\\
&=\left(
{\sum_{p_1,p_2,p_3\in \mathbb{Z}}}^{(1)}
+
{\sum_{p_1,p_2,p_3\in \mathbb{Z}}}^{(2)}
+
{\sum_{p_1,p_2,p_3\in \mathbb{Z}}}^{(3)}
\right)
\frac{\xi^{-p_1-p_2-p_3+\frac{9}{2}}}
{(1-\xi^{\frac32} q^{p_1}) (1-\xi^{\frac32} q^{p_2}) (1-\xi^{\frac32} q^{p_3})}, 
\end{align}
where the first sum 
\begin{align}
{\sum_{p_1,p_2,p_3\in \mathbb{Z}}}^{(1)}
&=-3\sum_{\begin{smallmatrix}
p_1,p_2,p_3\in \mathbb{Z}\\
p_1<p_2<p_3\\
\end{smallmatrix}}
\end{align}
generates a scalar multiple of the $U(3)$ Schur index, 
the second
\begin{align}
&
{\sum_{p_1,p_2,p_3\in \mathbb{Z}}}^{(2)}
=
\Biggl(
\sum_{i=1}^{3}
\sum_{\begin{smallmatrix}
p_1,p_2,p_3\in \mathbb{Z}\\
p_3=p_2=n_i\\
\end{smallmatrix}}
+
\sum_{i<j}
\sum_{\begin{smallmatrix}
p_1,p_2,p_3\in \mathbb{Z}\\
p_3=p_2+n_i+n_j\\
\end{smallmatrix}}
+
\sum_{\begin{smallmatrix}
p_1,p_2,p_3\in \mathbb{Z}\\
p_3=p_2+n_1+n_2+n_3\\
\end{smallmatrix}}
\nonumber\\
&-
\sum_{i=1}^3
\sum_{\begin{smallmatrix}
p_1,p_2,p_3\in \mathbb{Z}\\
p_2=p_1,p_3=p_1+n_i\\
\end{smallmatrix}}
-
\sum_{i<j}
\sum_{\begin{smallmatrix}
p_1,p_2,p_3\in \mathbb{Z}\\
p_2=p_1,p_3=p_1+n_i+n_j\\
\end{smallmatrix}}
-
\sum_{\begin{smallmatrix}
p_1,p_2,p_3\in \mathbb{Z}\\
p_2=p_1,p_3=p_1+n_1+n_2+n_3\\
\end{smallmatrix}}
\nonumber\\
&-
\sum_{i=1}^3
\sum_{\begin{smallmatrix}
p_1,p_2,p_3\in \mathbb{Z}\\
p_2=p_1+n_i,p_3=p_1+n_i\\
\end{smallmatrix}}
-
\sum_{i<j}
\sum_{\begin{smallmatrix}
p_1,p_2,p_3\in \mathbb{Z}\\
p_2=p_1+n_i+n_j,p_3=p_1+n_i+n_j\\
\end{smallmatrix}}
-
\sum_{\begin{smallmatrix}
p_1,p_2,p_3\in \mathbb{Z}\\
p_2=p_1+n_1+n_2+n_3,p_3=p_1+n_1+n_2+n_3\\
\end{smallmatrix}}
\Biggr)
\end{align}
yields the $U(3)$ 2-point functions and the third
\begin{align}
{\sum_{p_1,p_2,p_3\in \mathbb{Z}}}^{(3)}&=
\sum_{i=1}^3
\sum_{j\neq i}
\sum_{\begin{smallmatrix}
p_1,p_2,p_3\in \mathbb{Z}\\
p_2=p_1+n_i,p_3=p_1+n_i+n_j\\
\end{smallmatrix}}
+
\sum_{i<j}
\sum_{\begin{smallmatrix}
p_1,p_2,p_3\in \mathbb{Z}\\
p_2=p_1+n_i,p_3=p_1+n_1+n_2+n_3\\
\end{smallmatrix}}
\nonumber\\
&+
\sum_{i<j}
\sum_{\begin{smallmatrix}
p_1,p_2,p_3\in \mathbb{Z}\\
p_2=p_1+n_i+n_j,p_3=p_1+n_1+n_2+n_3\\
\end{smallmatrix}}
\end{align}
gives rise to the $U(3)$ 3-point functions. 

\subsubsection{$U(4)$ 4-point functions}
Similarly, the closed-form expression for the 4-point function for $\mathcal{N}=2^*$ $U(4)$ can be found 
by collecting the four spectral zeta functions. 
When we set $u$ to $\xi^2$, 
\begin{align}
\label{u4_4pt_exact1}
&
\langle W_{n_1} W_{n_2} W_{n_3} W_{-n_1-n_2-n_3}\rangle^{U(4)}
\nonumber\\
&=24 \mathcal{I}^{U(4)}
+\sum_{i=1}^3 \langle W_{n_i}W_{-n_i}\rangle^{U(4)}
+\sum_{i<j}\langle W_{n_i+n_j}W_{-n_i-n_j}\rangle^{U(4)}
+\langle W_{n_1+n_2+n_3}W_{-n_1-n_2-n_3}\rangle^{U(4)}
\nonumber\\
&+\sum_{i<j}\langle 
\mathfrak{W}_{n_i}\mathfrak{W}_{n_j}\mathfrak{W}_{-n_i-n_j}\rangle^{U(4)}
+\sum_{i=1}^{3} \sum_{\begin{smallmatrix}
j_1\neq i,j_2\neq i\\
j_1<j_2\\
\end{smallmatrix}
}
\langle \mathfrak{W}_{n_i}\mathfrak{W}_{n_{j_1}+n_{j_2}}\mathfrak{W}_{-n_1-n_2-n_3}\rangle^{U(4)}
\nonumber\\
&+\xi^{8}\sum_{i=1}^3 \sum_{j\neq i} 
Q(1,1,1,1;0,n_i,n_i+n_j,n_1+n_2+n_3;\xi^2), 
\end{align}
where the irreducible parts of the 3-point functions are defined by (\ref{3pt_irr}). 
The irreducible part of the $U(4)$ 4-point function 
\begin{align}
&
\xi^8\sum_{i=1}^3\sum_{j\neq i} 
Q(1,1,1,1;0,n_i,n_i+n_j,n_1+n_2+n_3;\xi^2)
\nonumber\\
&=
\sum_{\begin{smallmatrix}
p_1,p_2,p_3,p_4\in \mathbb{Z}\\
p_2=p_1+n_i, p_3=p_1+n_i+n_j, \\
p_4=p_1+n_1+n_2+n_3\\
\end{smallmatrix}}
\frac{\xi^{-p_1-p_2-p_3-p_4+8}}
{(1-\xi^2 q^{p_1}) (1-\xi^2 q^{p_2}) (1-\xi^2 q^{p_3}) (1-\xi^2 q^{p_4})}
,
\end{align}
which is given by the multiple Kronecker theta series 
can be written as 
\begin{align}
&
\sum_{i=1}^3 \sum_{j\neq i}
\Biggl[
-\frac{(n_i)_{q,\xi} (n_i+n_j)_{q,\xi} (n_1+n_2+n_3)_{q,\xi}}{(2n_i+n_j+n_1+n_2+n_3)_{q\xi^2,1}}
\nonumber\\
&+(q\xi^2)^{n_i}
\frac{(n_i)_{q,\xi} (n_j)_{q,\xi} (n_1+n_2+n_3-n_i)_{q,\xi}}
{(n_1+n_2+n_3-2n_i+n_j)_{q\xi^2,1}}
\Biggr]
P_1\left[
\begin{matrix}
\xi^4\\
1\\
\end{matrix}
\right](2\zeta,\tau)
\end{align}
in terms of the twisted Weierstrass function. 

\subsection{Antisymmetric Wilson line correlators}
While the correlation functions of the Wilson line operators in the antisymmetric and symmetric representations 
are given by those of the charged Wilson line operators by using Newton's identities, 
we can also obtain them from the spectral zeta functions. 
From the generating function (\ref{geneW_E}) 
the 2-point functions of the Wilson line operators in the rank-$m$ antisymmetric representation 
and in its conjugate representation can be also obtained by reading off the coefficients of the terms with $s_1^ms_2^m$. 

For $\mathcal{N}=2^*$ $U(1)$ and $U(2)$ SYM theory there is no non-trivial 2-point functions of 
the Wilson line operators in the antisymmetric representation. 
We have 
\begin{align}
\langle W_{(1^2)}W_{\overline{(1^2)}}\rangle^{U(2)}&=\mathcal{I}^{U(2)}. 
\end{align}
and 
\begin{align}
\langle W_{(1^2)}W_{\overline{(1^2)}}\rangle^{U(3)}&=
\langle W_{1}W_{-1}\rangle^{U(3)}. 
\end{align}

\subsubsection{$U(4)$ 2-point function}
The non-trivial 2-point function of the Wilson line operators in the rank-$2$ antisymmetric representation 
appears for $\mathcal{N}=2^*$ $U(4)$ SYM. 
Substituting the spectral zeta functions $Z_1^{E}(n)$, $Z_2^{E}(n)$, $Z_3^{E}(n)$ and $Z_4^{E}(n)$ 
into (\ref{geneW_E/H}), reading off the coefficients of the terms with $s_1^2 s_2^2$ 
and setting $u$ to $\xi^2$, 
we obtain 
\begin{align}
\label{u4_2asym_exact1}
&\langle W_{(1^2)} W_{\overline{(1^2)}} \rangle^{U(4)}
=\frac{-\xi^{8}}{4} 
\Bigl[
Q(1;0;\xi^2)^4-6Q(1;0;\xi^2)^2Q(2;0;\xi^2)
\nonumber\\
&+3Q(2;0;\xi^2)^2+8Q(1;0;\xi^2)Q(3;0;\xi^2)-6Q(4;0;\xi^2)
\nonumber\\
&-4Q(1;0;\xi^2)^2Q(1,1;0,1;\xi^2)
+4Q(2;0;\xi^2)Q(1,1;0,1;\xi^2)
+2Q(1,1;0,1;\xi^2)^2
\nonumber\\
&+8Q(1;0;\xi^2)Q(2,1;0,1;\xi^2)
+8Q(1;0;\xi^2)Q(1,2;0,1;\xi^2)
\nonumber\\
&-8Q(1,3;0,1;\xi^2)-8Q(3,1;0,1;\xi^2)-10Q(2,2;0,1;\xi^2)-4Q(1,2,1;0,1,2;\xi^2)
\Bigr]. 
\end{align}
The expression (\ref{u4_2asym_exact1}) contains 
the $U(4)$ Schur index and the $U(4)$ 2-point function of the Wilson line operators in the fundamental representation. 
It can be rewritten as 
\begin{align}
\label{u4_2asym_exact2}
&
\langle W_{(1^2)} W_{\overline{(1^2)}} \rangle^{U(4)}
=-2\mathcal{I}^{U(4)}+2\langle W_1W_{-1}\rangle^{U(4)}
\nonumber\\
&-\frac{\xi^{8}}{2}
\Bigl[
-Q(2,2;0,1;\xi^2)
-2Q(1,2,1;0,1,2;\xi^2)
\Bigr], 
\end{align}
where we have eliminated the term involving $Q(1,1;0,1;\xi^2)$ 
as it vanishes for $u=\xi^2$.

We eventually get
\begin{align}
\label{u4_2asym_exact3}
&
\langle W_{(1^2)} W_{\overline{(1^2)}} \rangle^{U(4)}
=-2\mathcal{I}^{U(4)}+2\langle W_1W_{-1}\rangle^{U(4)}
\nonumber\\
&+\frac{\xi^{8}}{2}
\Biggl[
\xi^{-2}
\frac{(1)_{q,\xi}^2}{(1)_{q\xi^2,1}^2}
P_2\left[
\begin{matrix}
q\xi^4\\
1\\
\end{matrix}
\right](2\zeta,\tau)
+2\Bigl\{
q\xi \frac{(1)_{q,\xi}^3}{(1)_{q\xi^4,1}}
+q(1)_{q,1}^2
-\frac{(1)_{q,\xi}^2(2)_{q,\xi}}{(4)_{q\xi^2,1}}
\Bigr\}
P_1\left[
\begin{matrix}
\xi^4\\
1\\
\end{matrix}
\right](2\zeta,\tau)
\Biggr]. 
\end{align}
Equivalently, 
\begin{align}
\label{u4_2asym_exact4}
&
\langle W_{(1^2)} W_{\overline{(1^2)}} \rangle^{U(4)}
=-2\mathcal{I}^{U(4)}+2\langle W_1W_{-1}\rangle^{U(4)}
\nonumber\\
&+\frac{\xi^{4}}{2}
\Biggl[
\frac{(1-q\xi^2)^2}{(1-q)^2}
P_2\left[
\begin{matrix}
q\xi^4\\
1\\
\end{matrix}
\right](2\zeta,\tau)
-2
\frac{(1-q^2\xi^2)(1-q\xi^2)(1-q\xi^4)}{(1-q)^3 (1+q)}
P_1\left[
\begin{matrix}
\xi^4\\
1\\
\end{matrix}
\right](2\zeta,\tau)
\Biggr]. 
\end{align}
Note that this can be also obtained from the relation (\ref{asym2_power}) 
and the previous results for the 2-, 3- and 4-point functions. 
Under S-duality the $U(4)$ Wilson line operator in the rank-$2$ antisymmetric representation 
is expected to map to the $U(4)$ 't Hooft line operator $T_{B}$ of magnetic charge $B=(1^2,0,0)$. 
So the expression (\ref{u4_2asym_exact3}) or (\ref{u4_2asym_exact4}) should also be equal to 
the dual 2-point function \cite{Gang:2012yr}
\begin{align}
\label{u4_tHooft_asym2}
&\langle T_{(1^2,0,0)}T_{\overline{(1^2,0,0)}} \rangle^{U(4)}(\xi;q)
\nonumber\\
&=\frac{1}{2\cdot 2}
\frac{(q)_{\infty}^8}{(\xi^{-1};q)_{\infty}^4(q\xi;q)_{\infty}^4}
\oint \prod_{i=1}^4 \frac{d\sigma_i}{2\pi i\sigma_i} 
\frac{(\sigma_1^{\pm}\sigma_2^{\mp};q)_{\infty} (q\sigma_1^{\pm}\sigma_2^{\mp};q)_{\infty}}
{(q\xi \sigma_1^{\pm}\sigma_2^{\mp})_{\infty} (\xi^{-1}\sigma_1^{\pm}\sigma_2^{\mp})_{\infty}}
\nonumber\\
&\times 
\frac{(\sigma_3^{\pm}\sigma_4^{\mp};q)_{\infty} (q\sigma_3^{\pm}\sigma_4^{\mp};q)_{\infty}}
{(q\xi \sigma_3^{\pm}\sigma_4^{\mp})_{\infty} (\xi^{-1}\sigma_3^{\pm}\sigma_4^{\mp})_{\infty}}
\prod_{i=1}^{2}\prod_{j=3}^{4}
\frac{(q^{\frac12}\sigma_i^{\pm}\sigma_j^{\mp};q)_{\infty} (q^{\frac32}\sigma_i^{\pm}\sigma_j^{\mp};q)_{\infty}}
{(q^{\frac32}\xi\sigma_i^{\pm}\sigma_j^{\mp})_{\infty} (q^{\frac12}\xi^{-1} \sigma_i^{\pm}\sigma_j^{\mp})_{\infty}}. 
\end{align}
While we have checked that they coincide by expanding the two expressions, 
it would be interesting to analytically prove the equality. 

\subsubsection{$U(5)$ 2-point function}
For $\mathcal{N}=2^*$ $U(5)$ SYM 
the 2-point function of the Wilson line operators in the rank-$2$ representation and its conjugate representation 
can be obtained from the five spectral zeta functions (\ref{ZE2_1})-(\ref{ZE2_5}). 
With $u=\xi^{\frac52}$ we get
\begin{align}
\label{u5_2asym_exact1}
&\langle W_{(1^2)} W_{\overline{(1^2)}} \rangle^{U(5)}
\nonumber\\
&=-5\mathcal{I}^{U(5)}
+3\langle W_{1}W_{-1}\rangle^{U(5)}
+\frac{\xi^{\frac52}}{2}
\Biggl[
Q(1;0;\xi^{\frac52})Q(1,1;0,1;\xi^{\frac52})^2
\nonumber\\
&-2Q(1,1;0,1;\xi^{\frac52})Q(1,2;0,1;\xi^{\frac52})
-2Q(1;0;\xi^{\frac52})Q(1,2,1;0,1,2;\xi^{\frac52})
\nonumber\\
&
+2Q(1,2,2;0,1,2;\xi^{\frac52})
+4Q(1,3,1;0,1,2;\xi^{\frac52})
\nonumber\\
&-2Q(1,1;0,1;\xi^{\frac52})Q(2,1;0,1;\xi^{\frac52})
-Q(1,0;\xi^{\frac52})Q(2,2;0,1;\xi^{\frac52})
\nonumber\\
&
+2Q(2,2,1;0,1,2;\xi^{\frac52})
+2Q(2,3;0,1;\xi^{\frac52})
+2Q(3,2;0,1;\xi^{\frac52})
\Biggr], 
\end{align}
where 
\begin{align}
&
\mathcal{I}^{U(5)}
=\frac{\xi^{\frac{25}{2}}}{12}
\Biggl[
Q(1;0;\xi^{\frac52})^5
-10Q(1;0;\xi^{\frac52})^3Q(2;0;\xi^{\frac52})
\nonumber\\
&
+15Q(1;0;\xi^{\frac52})Q(2;0;\xi^{\frac52})^2
+20Q(1;0;\xi^{\frac52})^2Q(3;0;\xi^{\frac52})
\nonumber\\
&
-20Q(2;0;\xi^{\frac52})Q(3;0;\xi^{\frac52})
-30Q(1;0;\xi^{\frac52})Q(4;0;\xi^{\frac52})
+24Q(5;0;\xi^{\frac52})
\Biggr]
\end{align}
is the $U(5)$ Schur index 
and 
\begin{align}
&
\langle W_{1}W_{-1}\rangle^{U(5)}
=5\mathcal{I}^{U(5)}
+\frac{\xi^{\frac52}}{2}
\Biggl[
-Q(1;0;\xi^{\frac52})^3Q(1,1;0,1;\xi^{\frac52})
+3Q(1;0;\xi^{\frac52})^2Q(1,2;0,1;\xi^{\frac52})
\nonumber\\
&-6Q(1;0;\xi^{\frac52})Q(1,3;0,1;\xi^{\frac52})
+4Q(1,3,1;0,1,2;\xi^{\frac52})
+6Q(1,4;0,1;\xi^{\frac52})
\nonumber\\
&+3Q(1;0;\xi^{\frac52})Q(1,1;0,1;\xi^{\frac52})Q(2;0;\xi^{\frac52})
-3Q(1,2;0,1;\xi^{\frac52})Q(2;0;\xi^{\frac52})
\nonumber\\
&+3Q(1;0;\xi^{\frac52})^2Q(2,1;0,1;\xi^{\frac52})
\Biggr]
\end{align}
is the $U(5)$ 2-point function of the fundamental Wilson line operators. 

\subsection{Symmetric Wilson line correlators}
One can also find the 2-point functions of the Wilson line operators in the rank-$m$ symmetric representation and its conjugate representation 
by extracting the coefficients associated with the terms of $s_1^ms_2^m$ from the generating function (\ref{geneW_H}). 
As opposed to the antisymmetric Wilson line operators, the rank $m$ of the representation can be larger than the rank $N$ of the gauge group. 

\subsubsection{$U(2)$ 2-point function}
For $\mathcal{N}=2^*$ $U(2)$ SYM theory we set $u=\xi$. 
Substituting the spectral zeta functions $Z_1^H(n)$ and $Z_2^H(n)$ into (\ref{geneW_E/H}), 
we obtain from the coefficients of the terms with $s_1^m s_2^m$ the 2-point function of the $U(2)$ 2-point function 
of the Wilson line operators in the rank-$m$ symmetric representation
\begin{align}
\label{u2_sym_exact1}
&\langle W_{(m)} W_{\overline{(m)}} \rangle^{U(2)}
\nonumber\\
&=\frac{\xi^2}{2}\Bigl[
(m+1)
Q(2;0;\xi)
+\sum_{m_1=1}^{m}2 (m-m_1+1)Q(1,1;0,m_1;\xi)
\Bigr]. 
\end{align}
This can be rewritten as
\begin{align}
\label{u2_sym_exact2}
&\langle W_{(m)} W_{\overline{(m)}} \rangle^{U(2)}
\nonumber\\
&=(m+1)\mathcal{I}^{U(2)}
-\xi^2 \sum_{m_1=1}^m 
(m-m_1+1)
\frac{q^{\frac{m_1}{2}}\xi^{m_1}-q^{-\frac{m_1}{2}}\xi^{-m_1}}
{q^{\frac{m_1}{2}}-q^{-\frac{m_1}{2}}}
P_1\left[
\begin{matrix}
\xi^2\\
1\\
\end{matrix}
\right](\zeta,\tau). 
\end{align}
It follows from the relation of symmetric functions that 
\begin{align}
&\langle W_{(m)} W_{\overline{(m)}} \rangle^{U(2)}
\nonumber\\
&=(1-m^2)\mathcal{I}^{U(2)}
+\sum_{m_1=1}^{m}(m-m_1+1)\langle W_{m_1}W_{-m_1}\rangle^{U(2)}. 
\end{align}
This is consistent with the formula (\ref{q11_tW}) for the $U(2)$ 2-point function of the charged Wilson line operators. 

When we take the unflavored limit $\xi\rightarrow q^{-1/2}$, 
the 2-point function (\ref{u2_sym_exact1}) in the large $m$ limit coincides with 
\begin{align}
\langle W_{(m=\infty)} W_{\overline{(m=\infty)}} \rangle^{U(2)}
&=\sum_{n>0}\frac{n^2 q^{\frac{n-1}{2}}}{1-q^n}
\nonumber\\
&=1+4q^{1/2}+10q+16q^{3/2}+26q^2+40q^{5/2}+50q^3
\nonumber\\
&+64q^{7/2}+91q^4+104q^{9/2}+122q^5+\cdots,
\end{align}
which is the generating function for 
the sum of squares of divisors $d$ of $n$ for which $n/d$ is odd. 

\subsubsection{$U(3)$ 2-point function}
Next consider the 2-point function of the rank-$m$ symmetric Wilson line operators for $\mathcal{N}=2^*$ $U(3)$ SYM theory. 
It can be constructed from the three spectral zeta functions $Z_1^H(n)$, $Z_2^H(n)$ and $Z_3^H(n)$.  
If we set $u$ to $\xi^{\frac32}$, we obtain
\begin{align}
\label{u3_sym_exact1}
&\langle W_{(m)} W_{\overline{(m)}} \rangle^{U(3)}
\nonumber\\
&=\frac{\xi^{\frac92}}{6}\Bigl[
\frac{(m+1)(m+2)}{2}
\Bigl(
Q(1;0;\xi^{\frac32})^3-3Q(1;0;\xi^{\frac32})Q(2;0;\xi^{\frac32})+2Q(3;0;\xi^{\frac32})
\Bigr)
\nonumber\\
&+\sum_{m_1=1}^m 3(m-m_1+1)(m-m_1+2)
\nonumber\\
&\times \{
-Q(1;0;\xi^{\frac32})Q(1,1;0,m_1;\xi^{\frac32})+Q(2,1;0,m_1;\xi^{\frac32})+Q(1,2;0,m_1;\xi^{\frac32})
\}
\nonumber\\
&+
\sum_{m_1=2}^{m}
\sum_{0<m_2<m_1}
6(m-m_1+1)(m-m_1+2)Q(1,1,1;0,m_2,m_1;\xi^{\frac32})
\Bigr]. 
\end{align}
This can be expressed as
\begin{align}
\label{u3_sym_exact1}
&\langle W_{(m)} W_{\overline{(m)}} \rangle^{U(3)}
\nonumber\\
&=-\left(
\begin{matrix}
m+2\\
2\\
\end{matrix}
\right)\left(
\begin{matrix}
m+1\\
1\\
\end{matrix}
\right)\mathcal{I}^{U(3)}
+\frac12\sum_{m_1=1}^m (m-m_1+1)(m-m_1+2)
\langle W_{m_1}W_{-m_1}\rangle^{U(3)}
\nonumber\\
&+\xi^{\frac92}
\sum_{m_1=2}^{m}
\sum_{0<m_2<m_1}
(m-m_1+1)(m-m_1+2)Q(1,1,1;0,m_2,m_1;\xi^{\frac32}). 
\end{align}

\section{Grand canonical correlators}
\label{sec_GC}
We consider the Wilson line correlation functions in the grand canonical ensemble. 
We define the normalized grand canonical Schur correlation function of the Wilson line operators by 
\begin{align}
\label{W_GC}
&
\langle \mathcal{W}_{\mathcal{R}_1} \cdots \mathcal{W}_{\mathcal{R}_k}\rangle^{\mathrm{GC}}
(u;\mu;\xi;q)
\nonumber\\
&:=\frac{1}{\Xi(u;\mu;\xi;q)}
\sum_{N=1}^{\infty} (-1)^N \xi^{-N^2/2}
\frac{\theta(u\xi^{-N};q)}{\theta(u;q)}
\langle W_{\mathcal{R}_1} \cdots W_{\mathcal{R}_k}\rangle^{U(N)}(\xi;q) \mu^N
\nonumber\\
&=\frac{1}{\Xi(u;\mu;\xi;q)}
\sum_{N=1}^{\infty}
\left[
\oint_{|\sigma_i|=1}\prod_{i=1}^{N}\frac{d\sigma_i}{2\pi i\sigma_i}
\det_{i,j} F\left( \frac{\sigma_i}{\sigma_j}\xi^{-1},u;q \right)
\prod_{j=1}^{k}\chi_{\mathcal{R}_j}(\sigma)
\right]
\mu^N, 
\end{align}
where 
\begin{align}
\label{g_pfn}
\Xi(\mu;u;\xi;q)&=\sum_{N=0}^{\infty}\mathcal{Z}(N;u;\xi;q)\mu^N
\nonumber\\
&=\prod_{p\in \mathbb{Z}}
\frac{1-uq^p-\mu\xi^{-p}}{1-uq^p}
\end{align}
is the grand canonical partition function of the Fermi-gas. 

The grand canonical correlation function (\ref{W_GC}) and the partition function (\ref{g_pfn}) are invariant under the following transformation: 
\begin{align}
\label{ximuu_transf}
\xi&\rightarrow q^{-1}\xi^{-1},\nonumber\\
u&\rightarrow u^{-1},\nonumber\\
\mu&\rightarrow -u^{-1}\mu, 
\end{align}
which extends the transformation (\ref{xi_transf}). 
This transformation turns out to be useful to deform the expressions of the grand canonical correlators. 

For the modified density matrix (\ref{density_E}) or (\ref{density_H}), 
we introduce a function 
\begin{align}
\label{grand_G}
\mathcal{G}_{E/H}^{(n_1,n_2,\cdots,n_k)}(\mu;\{s_j\};u;\xi;q)
&=
\frac{\det (1+\mu\rho_{E/H}^{(n_1,n_2,\cdots,n_k)})}{\det(1+\mu\rho_0)}, 
\end{align}
where the functions
\begin{align}
\label{Xi_E/H}
\Xi_{E/H}^{(n_1,\cdots,n_k)}
&:=
\det (1+\mu\rho_{E/H}^{(n_1,\cdots,n_k)})
=\sum_{N=1}^{\infty}
\mathcal{Z}_{E/H}^{\{n_j\}} \mu^N
\end{align}
appearing in the numerator are the grand canonical partition functions 
which applies to the grand canonical ensembles of the Fermi-gas systems 
whose canonical partition functions are given by (\ref{geneW_E}) and (\ref{geneW_H}). 
Analogous to (\ref{geneW_E}) and (\ref{geneW_H}), 
the function (\ref{grand_G}) can be regarded as a generating function for the normalized grand canonical correlation functions (\ref{W_GC}) 
by reading off the coefficients of the terms with equal powers of $s_j$ with $j=1,\cdots,k$. 

We can write (\ref{grand_G}) as
\begin{align}
\label{grand_G2}
\mathcal{G}_{E/H}^{(n_1,n_2,\cdots,n_k)}(\mu;\{s_j\};u;\xi;q)
&=\det(1+\mathcal{X}_{E/H}^{(n_1,n_2,\cdots,n_k)}\varrho(p))
\nonumber\\
&=\exp\left[
\sum_{m=1}^{\infty} \frac{(-1)^{m+1}}{m}\Tr (\mathcal{X}_{E/H}^{(n_1,n_2,\cdots,n_k)}(\sigma) \varrho(p))^m
\right], 
\end{align}
where 
\begin{align}
\label{XEg}
\mathcal{X}_{E}^{(n_1,n_2,\cdots,n_k)}(\sigma)&=
\prod_{j=1}^{k}(1+s_j\sigma^{n_j})-1
\end{align}
and
\begin{align}
\label{XHg}
\mathcal{X}_{H}^{(n_1,n_2,\cdots,n_k)}(\sigma)&=
\prod_{j=1}^{k}
\frac{1}{1-s_j\sigma^{n_j}}-1
\end{align}
are the position-dependent operators and 
\begin{align}
\varrho(p)&=\frac{\mu\rho_0}{1+\mu\rho_0}
=\frac{-\mu\xi^{-p}}{1-uq^p-\mu\xi^{-p}}
\end{align}
is the momentum-dependent operator. 
Then a further analysis follow exactly the same line as the discussion in section \ref{sec_spectralzeta}. 
The normalized grand canonical correlation functions can be obtained by 
expanding (\ref{grand_G2}) 
and evaluating the normal ordered operators $(\mathcal{X}_{E/H}^{(n_1,n_2,\cdots,n_k)}(\sigma) \varrho(p))$ and their traces. 

\subsection{Generating functions for multiple Kronecker theta series}
Again it is useful to observe the relation (\ref{translate1}) 
and to define a function
\begin{align}
\label{g_zeta}
R(\{n_i\};\mu;u;\xi;q)&:=
\sum_{p\in \mathbb{Z}}\frac{(-\mu)^k \xi^{-kp-\sum_{i=1}^{k}n_i}}{\prod_{i=1}^k (1-uq^{p+n_i}-\mu\xi^{-p-n_i})}
\end{align}
in the calculation of the traces of the normal ordered operators. 
Under (\ref{ximuu_transf}) the function (\ref{g_zeta}) transforms as
\begin{align}
R(\{n_i\};-u^{-1}\mu;u^{-1};q^{-1}\xi^{-1};q)&=R(\{-n_i\};\mu;u;\xi;q). 
\end{align}

As we will see, 
the functions (\ref{g_zeta}) show up in the exact expression of the normalized grand canonical correlators as building blocks 
since they are generating functions for the multiple Kronecker theta series (\ref{s_zeta})
\begin{align}
\label{g_zeta_gene}
&
R(n_1,\cdots,n_k;\mu;u;\xi;q)
\nonumber\\
&=(-1)^k \sum_{m_1\ge 1}\cdots \sum_{m_k\ge 1}
Q(m_1,\cdots,m_k;n_1,\cdots,n_k;u;\xi;q)(-\mu)^{m_1+\cdots+m_k}. 
\end{align}

The simplest example is 
\begin{align}
\label{g_zeta_0}
R(0;\mu;u;\xi;q)&=\sum_{p\in \mathbb{Z}}
\frac{-\mu\xi^{-p}}
{1-uq^p-\mu\xi^{-p}}. 
\end{align}
The function (\ref{g_zeta_0}) is invariant under the transformation (\ref{ximuu_transf}). 
It is a generating function for the spectral zeta function $Z_{l}(u;\xi;q)$ or $Q(l;0;u;\xi;q)$ given by (\ref{szeta_0})
\begin{align}
R(0;\mu;u;\xi;q)&=-\sum_{l=1}^{\infty}Z_l(u;\xi;q) (-\mu)^l
\nonumber\\
&=-\sum_{l=1}^{\infty}Q(l;0;u;\xi;q) (-\mu)^l. 
\end{align}

\subsection{Closed-form formula}
The normalized grand canonical correlation functions of the Wilson line operators of fixed charges  
can be obtained from either $\mathcal{G}_{E}^{(n_1,\cdots,n_k)}$ or $\mathcal{G}_{H}^{(n_1,\cdots,n_k)}$ 
by finding the coefficients of the term with $\prod_{j}s_j$. 

\subsubsection{2-point functions}
In terms of the function (\ref{g_zeta}) 
we can express the traces of the normal ordered operators $(\mathcal{X}_E^{(n,-n)} \varrho)^m$. 
It is convenient to abbreviate (\ref{g_zeta}) as $R(\{n_i\})$ $=$ $R(\{n_i\};\mu;u;\xi;q)$. 
We have 
\begin{align}
\label{tr_2ptGE1}
\Tr(\mathcal{X}_E^{(n,-n)} \varrho)&=s_1s_2 R(0),\\
\label{tr_2ptGE2}
\Tr(\mathcal{X}_E^{(n,-n)} \varrho)^2&=2s_1s_2 R(0,n)+s_1^2s_2^2R(0,0),\\
\label{tr_2ptGE3}
\Tr(\mathcal{X}_E^{(n,-n)} \varrho)^3&=3s_1^2s_2^2 \Bigl(R(0,0,n)+R(0,n,n)\Bigr)
+s_1^3s_2^3 R(0,0,0), 
\end{align}
\begin{align}
\label{tr_2ptGE4}
\Tr(\mathcal{X}_E^{(n,-n)} \varrho)^4&=2s_1^2s_2^2\Bigl(R(0,0,n,n)+2R(0,n,n,2n) \Bigr)
\nonumber\\
&+4s_1^3s_2^3\Bigl(R(0,0,0,n)+R(0,0,n,n)+R(0,n,n,n) \Bigr)
\nonumber\\
&+s_1^4s_2^4 R(0,0,0,0),
\end{align}
\begin{align}
\label{tr_2ptGE5}
\Tr(\mathcal{X}_E^{(n,-n)} \varrho)^5&=
5s_1^3s_2^3\Bigl(
R(0,0,0,n,n)+R(0,0,n,n,n)
+R(0,0,n,n,2n)
\nonumber\\
&+2R(0,n,n,n,2n)+R(0,n,n,2n,2n)
\Bigr)
\nonumber\\
&+5s_1^4s_2^4\Bigl(
R(0,0,0,0,n)+R(0,0,0,n,n)
\nonumber\\
&+R(0,0,n,n,n)+R(0,n,n,n,n)
\Bigr)
\nonumber\\
&+s_1^5s_2^5 R(0,0,0,0,0),
\end{align}
\begin{align}
\label{tr_2ptGE6}
\Tr(\mathcal{X}_E^{(n,-n)} \varrho)^6&=
2s_1^3s_2^3\Bigl(
R(0,0,0,n,n,n)+3R(0,0,n,n,n,2n)
\nonumber\\
&+3R(0,n,n,n,2n,2n)+3R(0,n,n,2n,2n,3n)
\Bigr)
\nonumber\\
&+3s_1^4s_2^4 \Bigl(
3R(0,0,0,0,n,n)+4R(0,0,0,n,n,n)
\nonumber\\
&+3R(0,0,n,n,n,n)+2R(0,0,0,n,n,2n)
\nonumber\\
&+4R(0,0,n,n,n,2n)
+6R(0,n,n,n,n,2n)
\nonumber\\
&+2R(0,0,n,n,2n,2n)
+4R(0,n,n,n,2n,2n)
\nonumber\\
&+2R(0,n,n,2n,2n,2n)
\Bigr)
\nonumber\\
&+6s_1^5s_2^5\Bigl(
R(0,0,0,0,0,n)+
R(0,0,0,0,n,n)
\nonumber\\
&+R(0,0,0,n,n,n)
+R(0,0,n,n,n,n)
+R(0,n,n,n,n,n)
\Bigr). 
\end{align}
The normalized grand canonical 2-point function of the Wilson line operators of charges $\pm n$ is obtained from the terms with $s_1s_2$. 
These terms only appear from (\ref{tr_2ptGE1}) and (\ref{tr_2ptGE2}). 
Plugging them into (\ref{grand_G2}), we obtain the normalized grand canonical 2-point function of the Wilson line operators of charges $\pm n$
\begin{align}
\label{g2pt_1}
\langle \mathcal{W}_{n}\mathcal{W}_{-n}\rangle^{\mathrm{GC}}
&=R(0)-R(0,n)
\nonumber\\
&=-\sum_{p\in \mathbb{Z}}
\Biggl[
\frac{\mu\xi^{-p}}{1-uq^p-\mu\xi^{-p}}
+
\frac{\mu^2\xi^{-2p-n}}{(1-uq^p-\mu\xi^{-p})(1-uq^{p+n}-\mu\xi^{-p-n})}
\Biggr]
\nonumber\\
&=-\sum_{p\in \mathbb{Z}}
\frac{\mu\xi^{-p}(1-uq^{p+n})}{(1-uq^p-\mu\xi^{-p})(1-uq^{p+n}-\mu\xi^{-p-n})}. 
\end{align}
From (\ref{g_zeta_gene}) it can be also expressed as
\begin{align}
\label{g2pt_1a}
&
\langle \mathcal{W}_{n}\mathcal{W}_{-n}\rangle^{\mathrm{GC}}
\nonumber\\
&=-\sum_{m\ge1}Q(m;0)(-\mu)^m
-\sum_{m_1,m_2\ge1}Q(m_1,m_2;0,n)(-\mu)^{m_1+m_2}. 
\end{align}
By multiplying the normalized grand canonical 2-point function (\ref{g2pt_1a}) by the grand canonical partition function $\Xi(\mu;u;\xi;q)$ 
and expanding (\ref{g2pt_1}) in powers of $\mu$, 
we can rederive the previous exact expressions of the canonical 2-point functions of the charged Wilson line operators. 

By using the transformation (\ref{ximuu_transf}), 
the grand canonical 2-point function (\ref{g2pt_1}) can be also written as
\begin{align}
\label{g2pt_2}
\langle \mathcal{W}_{n}\mathcal{W}_{-n}\rangle^{\mathrm{GC}}
&=-\sum_{p\in \mathbb{Z}}
\frac{\mu\xi^{-p}(1-uq^{p-n})}{(1-uq^p-\mu\xi^{-p})(1-uq^{p-n}-\mu\xi^{-p+n})}
\nonumber\\
&=-\sum_{p\in \mathbb{Z}}
\frac{\mu\xi^{-p-n}(1-uq^{p})}{(1-uq^p-\mu\xi^{-p})(1-uq^{p+n}-\mu\xi^{-p-n})}, 
\end{align}
where in the second line we have shifted the integer $p$ $\rightarrow$ $p+n$. 
Multiplying (\ref{g2pt_1}) by $\xi^{-n}$ and subtracting it by (\ref{g2pt_2}), we find
\begin{align}
\label{g2pt_3}
\langle \mathcal{W}_{n}\mathcal{W}_{-n}\rangle^{\mathrm{GC}}
&=
-\frac{1-q^n}{1-\xi^{-n}}
\sum_{p\in \mathbb{Z}}
\frac{u\mu q^p \xi^{-p-n}}
{(1-uq^p-\mu\xi^{p})(1-uq^{p+n}-\mu\xi^{-p-n})}. 
\end{align}

\subsubsection{3-point functions}
While there are two relevant traces for the normalized grand canonical 2-point function of the charged Wilson line operators, 
there are three relevant traces for the 3-point functions. 
They are given by
\begin{align}
\label{tr_3ptGE1}
\Tr(\mathcal{X}_E^{(n_1,n_2,-n_1-n_2)} \varrho)
&=s_1s_2s_3 R(0),\\
\label{tr_3ptGE2}
\Tr(\mathcal{X}_E^{(n_1,n_2,-n_1-n_2)} \varrho)^2
&=2s_1s_2s_3 \Bigl(R(0,n_1,n_1+n_2)+R(0,n_2,n_1+n_2)\Bigr)
+s_1^2s_2^2s_3^2R(0,0),\\
\label{tr_3ptGE3}
\Tr(\mathcal{X}_E^{(n_1,n_2,-n_1-n_2)} \varrho)^3
&=3s_1s_2s_3 \Bigl(R(0,n_1,n_1+n_2)+R(0,n_2,n_1+n_2)\Bigr)
\nonumber\\
&+3s_1^2s_2^2s_3^2 \Bigl(
R(0,0,n_1)+R(0,n_1,n_1)
+R(0,0,n_2)+R(0,n_2,n_2)
\nonumber\\
&+R(0,0,n_1+n_2)+R(0,n_1+n_2,n_1+n_2)
\nonumber\\
&
+R(0,n_1,n_1+n_2)+R(0,n_2,n_1+n_2)
\Bigr)
+s_1^3s_2^3s_3^3 R(0,0,0). 
\end{align}
Substituting (\ref{tr_3ptGE1})-(\ref{tr_3ptGE3}) into (\ref{grand_G2}) and extracting the terms with $s_1s_2s_3$, 
we can get the normalized grand canonical 3-point function of the Wilson line operators with charges $n_1$, $n_2$ and $-n_1-n_2$. 

We find
\begin{align}
\label{g3pt_1}
&
\langle \mathcal{W}_{n_1}\mathcal{W}_{n_2}\mathcal{W}_{-n_1-n_2}\rangle^{\mathrm{GC}}
\nonumber\\
&=R(0)-R(0,n_1)-R(0,n_2)-R(0,n_1+n_2)
\nonumber\\
&+R(0,n_1,n_1+n_2)+R(0,n_2,n_1+n_2). 
\nonumber\\
&=
-\sum_{p\in \mathbb{Z}}\left[
\frac{\mu \xi^{-p}}{1-uq^{p}-\mu \xi^{-p}}
+\sum_{i=1}^3 \frac{\mu^2\xi^{-2p-n_{i}}}{(1-uq^{p}-\mu\xi^{-p})(1-uq^{p+n_i}-\mu\xi^{-p-n_i})}
\right.
\nonumber\\
&
\left.
+\sum_{i=1}^2 \frac{\mu^3\xi^{-3p-n_{i}-n_1-n_2}}
{(1-uq^{p}-\mu\xi^{-p})(1-uq^{p+n_i}-\mu\xi^{-p-n_i})(1-uq^{p+n_1+n_2}-\mu\xi^{-p-n_1-n_2})}
\right],
\end{align}
where $n_3=-n_1-n_2$. 
In terms of the multiple Kronecker theta series (\ref{s_zeta}) it is given by
\begin{align}
\label{g3pt_1a}
&
\langle \mathcal{W}_{n_1}\mathcal{W}_{n_2}\mathcal{W}_{-n_1-n_2}\rangle^{\mathrm{GC}}
\nonumber\\
&=-\sum_{m\ge1} Q(m;0)(-\mu)^m 
-\sum_{m_1,m_2\ge 0}
\Bigl[
\sum_{i=1}^2 Q(m_1,m_2;0,n_i)
+Q(m_1,m_2;0,n_1+n_2)
\Bigr](-\mu)^{m_1+m_2}
\nonumber\\
&-\sum_{m_1,m_2,m_3\ge0}
\sum_{i=1}^2 
Q(m_1,m_2,m_3;0,n_i,n_1+n_2)(-\mu)^{m_1+m_2+m_3}. 
\end{align}
All the canonical 3-point functions of the charged Wilson line operators can be obtained by 
multiplying the normalized grand canonical 3-point function (\ref{g3pt_1a}) by the grand canonical partition function (\ref{g_pfn})
and expanding (\ref{g3pt_1}) in powers of $\mu$. 

From (\ref{g3pt_1}) we get
\begin{align}
\label{g3pt_2}
&
\langle \mathcal{W}_{n_1}\mathcal{W}_{n_2}\mathcal{W}_{-n_1-n_2}\rangle^{\mathrm{GC}}
\nonumber\\
&=-\sum_{p\in \mathbb{Z}}
\frac{\mu\xi^{-p}
(1-uq^{p+n_1+n_2}) 
\Bigl[
(1-uq^{p+n_1}) (1-uq^{p+n_2}) -\mu^2 \xi^{-2p-n_1-n2}
\Bigr]
}
{
(1-uq^p-\mu\xi^{-p})
\prod_{i=1}^2 (1-uq^{p+n_i}-\mu\xi^{-p-n_i}) 
(1-uq^{p+n_1+n_2}-\mu\xi^{-p-n_1-n_2})
}. 
\end{align}
Using the transformation (\ref{ximuu_transf}), it can be written as
\begin{align}
\label{g3pt_3}
&
\langle \mathcal{W}_{n_1}\mathcal{W}_{n_2}\mathcal{W}_{-n_1-n_2}\rangle^{\mathrm{GC}}
\nonumber\\
&=-\sum_{p\in \mathbb{Z}}
\frac{\mu\xi^{-p-n_1-n_2}
(1-uq^{p}) 
\Bigl[
(1-uq^{p+n_1}) (1-uq^{p+n_2}) -\mu^2 \xi^{-2p-n_1-n2}
\Bigr]
}
{
(1-uq^p-\mu\xi^{-p})
\prod_{i=1}^2 (1-uq^{p+n_i}-\mu\xi^{-p-n_i}) 
(1-uq^{p+n_1+n_2}-\mu\xi^{-p-n_1-n_2})
}. 
\end{align}
Multiplying (\ref{g3pt_2}) by $\xi^{-n_1-n_2}$ and subtracting it by (\ref{g3pt_3}), 
we get
\begin{align}
\label{g3pt_4}
&
\langle \mathcal{W}_{n_1}\mathcal{W}_{n_2}\mathcal{W}_{-n_1-n_2}\rangle^{\mathrm{GC}}
=
\mu u\frac{(q^{n_1+n_2}-1)}{(\xi^{-n_1-n_2}-1)}
\nonumber\\
&\times \sum_{p\in \mathbb{Z}}
\frac{
q^{p+n_1+n_2}
\xi^{-p-n_1-n_2}
\Bigl[
(1-uq^{p+n_1}) (1-uq^{p+n_2}) -\mu^2 \xi^{-2p-n_1-n2}
\Bigr]
}
{
(1-uq^p-\mu\xi^{-p})
\prod_{i=1}^2 (1-uq^{p+n_i}-\mu\xi^{-p-n_i}) 
(1-uq^{p+n_1+n_2}-\mu\xi^{-p-n_1-n_2})
}. 
\end{align}

\subsubsection{4-point functions}
There are four traces which encode 
the normalized grand canonical 4-point function of the charged Wilson line operators. 
Since only the terms with $s_1s_2s_3s_4$ are required to find the exact expression of these correlation functions, we only show them for simplicity. 
We get
\begin{align}
\label{tr_4ptGE1}
\Tr(\mathcal{X}_E^{(n_1,n_2,n_3,-n_1-n_2-n_3)} \varrho)\Bigl|_{s_1s_2s_3s_4}
&=s_1s_2s_3s_4 R(0),\\
\label{tr_4ptGE2}
\Tr(\mathcal{X}_E^{(n_1,n_2,n_3,-n_1-n_2-n_3)} \varrho)^2\Bigl|_{s_1s_2s_3s_4}
&=2s_1s_2s_3s_4
\Bigl(
R(0,n_1)+R(0,n_2)+R(0,n_3)
\nonumber\\
&+R(0,n_1+n_2)+R(0,n_1+n_3)
\nonumber\\
&+R(0,n_2+n_3)+R(0,n_1+n_2+n_3)
\Bigr), 
\end{align}
\begin{align}
\label{tr_4ptGE3}
\Tr(\mathcal{X}_E^{(n_1,n_2,n_3,-n_1-n_2-n_3)} \varrho)^3\Bigl|_{s_1s_2s_3s_4}
&=3s_1s_2s_3s_4
\Bigl(
R(0,n_1,n_1+n_2)+R(0,n_1,n_1+n_3)
\nonumber\\
&+R(0,n_2,n_1+n_2)+R(0,n_2,n_2+n_3)
\nonumber\\
&+R(0,n_3,n_1+n_3)+R(0,n_3,n_2+n_3)
\nonumber\\
&+R(0,n_1,n_1+n_2+n_3)+R(0,n_2,n_1+n_2+n_3)
\nonumber\\
&+R(0,n_3,n_1+n_2+n_3)+R(0,n_1+n_2,n_1+n_2+n_3)
\nonumber\\
&+R(0,n_1+n_3,n_1+n_2+n_3)
\nonumber\\
&+R(0,n_2+n_3,n_1+n_2+n_3)
\Bigr),\\
\label{tr_4ptGE4}
\Tr(\mathcal{X}_E^{(n_1,n_2,n_3,-n_1-n_2-n_3)} \varrho)^4\Bigl|_{s_1s_2s_3s_4}
&=4s_1s_2s_3s_4
\Bigl(
R(0,n_1,n_1+n_2,n_1+n_2+n_3)
\nonumber\\
&+R(0,n_1,n_1+n_3,n_1+n_2+n_3)
\nonumber\\
&+R(0,n_2,n_1+n_2,n_1+n_2+n_3)
\nonumber\\
&+R(0,n_2,n_2+n_3,n_1+n_2+n_3)
\nonumber\\
&+R(0,n_3,n_1+n_3,n_1+n_2+n_3)
\nonumber\\
&+R(0,n_3,n_2+n_3,n_1+n_2+n_3)
\Bigr). 
\end{align}
Plugging these traces into (\ref{grand_G2}) and reading the terms with $s_1s_2s_3s_4$, 
one can find the normalized grand canonical 4-point function of the charged Wilson line operators. 
It is given by
\begin{align}
\label{g4pt_1}
&
\langle \mathcal{W}_{n_1}\mathcal{W}_{n_2}\mathcal{W}_{n_3}\mathcal{W}_{-n_1-n_2-n_3}\rangle^{\mathrm{GC}}
\nonumber\\
&=R(0)-\sum_{i=1}^3 R(0,n_i)-\sum_{i<j}R(0,n_i+n_j)-R(0,n_1+n_2+n_3)
\nonumber\\
&+\sum_{i=1}^3 \sum_{j\neq i}R(0,n_i,n_i+n_j)
+\sum_{i=1}^3 R(0,n_i,n_1+n_2+n_3)
\nonumber\\
&+\sum_{i<j}R(0,n_i+n_j,n_1+n_2+n_3)
-\sum_{i=1}^3 \sum_{j\neq i}R(0,n_i,n_i+n_j,n_1+n_2+n_3). 
\end{align}
In terms of the multiple Kronecker theta series (\ref{s_zeta}) we can also write it as
\begin{align}
\label{g4pt_2}
&
\langle \mathcal{W}_{n_1}\mathcal{W}_{n_2}\mathcal{W}_{n_3}\mathcal{W}_{-n_1-n_2-n_3}\rangle^{\mathrm{GC}}
\nonumber\\
&=\sum_{m\ge0}A_1 (-\mu)^m+\sum_{m_1,m_2\ge0}A_2  (-\mu)^{m_1+m_2}
\nonumber\\
&=
+\sum_{m_1,m_2,m_3\ge0}A_3  (-\mu)^{m_1+m_2+m_3}
+\sum_{m_1,m_2,m_3,m_4\ge0}A_4  (-\mu)^{m_1+m_2+m_3+m_4}, 
\end{align}
where
\begin{align}
A_1&=Q(m;0),\\
A_2&=\sum_{i=1}^3 Q(m_1,m_2;0,n_i)
+\sum_{i\neq j}Q(m_1,m_2;0,n_i+n_j)+Q(m_1,m_2;0,n_1+n_2+n_3),\\
A_3&=\sum_{i=1}^3 \sum_{j\neq i}Q(m_1,m_2,m_3;0,n_i,n_i+n_j)
+\sum_{i=1}^3 Q(m_1,m_2,m_3;0,n_i,n_1+n_2+n_3)
\nonumber\\
&+\sum_{i<j}Q(m_1,m_2,m_3;0,n_i+n_j,n_1+n_2+n_3),\\
A_4&=\sum_{i=1}^3\sum_{j\neq i}
Q(m_1,m_2,m_3,m_4;0,n_i,n_i+n_j,n_1+n_2+n_3). 
\end{align}

\subsubsection{$k$-point functions}
It is now straightforward to find the exact expression for the general normalized grand canonical $k$-point functions of the charged Wilson line operators 
by calculating the relevant traces of the normal ordered operators. 
We have 
\begin{align}
\label{gkpt_1}
&
\langle 
\mathcal{W}_{n_1}\mathcal{W}_{n_2}\cdots \mathcal{W}_{n_{k-1}}\mathcal{W}_{-n_1-\cdots-n_{k-1}}\rangle^{\mathrm{GC}}
\nonumber\\
&=R(0)
+\sum_{j=1}^{k-1}\sum_{
\begin{smallmatrix}
\lambda=(\lambda_1,\cdots,\lambda_r)\\
|\lambda|=j\\
\end{smallmatrix}}
\sum_{\{I_1,\cdots, I_r\}}
(-1)^r 
R\left( 0,
\bigoplus_{i^{(1)}\in I_1}n_{i^{(1)}}, 
\bigoplus_{i^{(2)}\in I_1,I_2}n_{i^{(2)}}, 
\cdots,
\bigoplus_{i^{(r)}\in I_1,\cdots,I_r}n_{i^{(r)}} 
\right). 
\end{align}
Again we have used the notation of the set $\{I_1,\cdots, I_r\}$ of integers 
with cardinality $|I_i|=|\lambda_i|$ for a given partition $\lambda$ $=$ $(\lambda_1,\lambda_2,\cdots,\lambda_r)$. 

\subsubsection{Antisymmetric representations}
The normalized grand canonical 2-point function of the Wilson line operators  
transforming in the rank-$2$ antisymmetric representation and its conjugate are associated to the terms with $s_1^2s_2^2$ in (\ref{grand_G2}). 
They are contained in the traces of $(\mathcal{X}_E^{(n,-n)}\varrho)^l$ with $l=1,2,3,4$, which are given by (\ref{tr_2ptGE1})-(\ref{tr_2ptGE4}). 
Inserting them into (\ref{grand_G2}) and setting $n=1$, we find 
\begin{align}
\label{g2pt_Asym2}
&
\langle \mathcal{W}_{(1,1)}\mathcal{W}_{\overline{(1,1)}}\rangle^{\mathrm{GC}}
\nonumber\\
&=-\frac12 R(0,0)+R(0,0,1)+R(0,1,1)
-\frac12 R(0,0,1,1)-R(0,1,1,2)
\nonumber\\
&+\frac12( R(0)-R(0,1))^2. 
\end{align}

For the grand canonical 2-point function of the Wilson line operators 
transforming in the rank-$3$ antisymmetric representation, 
one needs the traces of $(\mathcal{X}_E^{(n,-n)}\varrho)^l$ with $l=1,\cdots, 6$, which are given by (\ref{tr_2ptGE1})-(\ref{tr_2ptGE6}). 
We get
\begin{align}
\label{g2pt_Asym3}
&
\langle \mathcal{W}_{(1,1,1)}\mathcal{W}_{\overline{(1,1,1)}}\rangle^{\mathrm{GC}}
\nonumber\\
&=
-\frac13 R(0,0,0,1,1,1)+\frac13R(0,0,0)
+R(0,0,0,1,1)+R(0,0,1,1,1)
\nonumber\\
&-R(0,0,1,1)-R(0,0,0,1)-R(0,1,1,1)
-R(0,0,1,1,1,2)-R(0,1,1,1,2,2)
\nonumber\\
&+R(0,0,1,1,2)+R(0,1,1,2,2)
+2R(0,1,1,1,2)-R(0,1,1,2,2,3)
\nonumber\\
&
+\left( R(0)-R(0,1) \right)\left( -\frac12 R(0,0)+R(0,0,1)+R(0,1,1) \right)
+\frac16 \left( R(0)-R(0,1) \right)^3. 
\end{align}

\subsubsection{Symmetric representations}

The normalized grand canonical correlation functions of 
the Wilson line operators transforming in the symmetric representation 
are described by the matrix (\ref{XHg}). 
The traces of the normal ordered operators read
\begin{align}
\label{tr_2ptGH1}
\Tr(\mathcal{X}_H^{(n,-n)} \varrho)
&=\sum_{k=1}^{\infty}s_1^ks_2^k R(0),\\
\label{tr_2ptGH2}
\Tr(\mathcal{X}_H^{(n,-n)} \varrho)^2
&=\sum_{k=1}^{\infty} s_1^ks_2^k
\Bigl[
(k-1)R(0,0)+\sum_{l=1}^{k} 2(k-l+1)R(0,ln)
\Bigr],\\
\label{tr_2ptGH3}
\Tr(\mathcal{X}_H^{(n,-n)} \varrho)^3
&=\sum_{k=1}^{\infty}
s_1^ks_2^k
\Bigl[
\frac{(k-1)(k-2)}{2}R(0,0,0)
\nonumber\\
&+\sum_{l=1}^{k-1}\frac{3(k-l)(k-l+1)}{2}
\left(
R(0,0,ln)+R(0,ln,ln)
\right)
\nonumber\\
&+\sum_{l_1=1}^{k-1}\sum_{l_2=1}^{l_1-1}
3(k-l_1)(k-l_1+1)R(0,l_2n,l_1n)
\Bigr]. 
\end{align}

The normalized grand canonical 2-point function of the Wilson line operators in the rank-$2$ symmetric representation is given by 
\begin{align}
\label{g2pt_sym2}
&
\langle \mathcal{W}_{(2)}\mathcal{W}_{\overline{(2)}}\rangle^{\mathrm{GC}}
\nonumber\\
&=R(0)-\frac12 R(0,0)-2R(0,1)-R(0,2)
\nonumber\\
&+R(0,0,1)+R(0,1,1)+2R(0,1,2)-\frac12R(0,0,1,1)-R(0,1,1,2)
\nonumber\\
&+\frac12 (R(0)-R(0,1))^2. 
\end{align}

\subsection{Recursion formula}
We observe that 
the grand canonical partition function (\ref{grand_G}) obeys a differential equation
\begin{align}
\label{Xi_diff}
\frac{\partial}{\partial \mu}\Xi(\mu;u;\xi;q)&=R(0;\mu;u;\xi;q)\Xi(\mu;u;\xi;q). 
\end{align}
Recalling that $R(0;\mu;u;\xi;q)$ is the generating function for the spectral zeta function $Z_{l}(u;\xi;q)$, 
we obtain a recursion relation 
\footnote{
Similar recursion relations for the unflavored Schur indices have been discussed in \cite{Pan:2021mrw,Beem:2021zvt}. 
}
\begin{align}
\label{Xi_recursion}
\mathcal{Z}(N)&=\frac{1}{N}\sum_{l=1}^{N}(-1)^{l+1} Z_l \mathcal{Z}(N-l). 
\end{align}
For example, 
\begin{align}
\mathcal{Z}(1)&=Z_1, \\
\mathcal{Z}(2)&=\frac12 (Z_1\mathcal{Z}(1)-Z_2), \\
\mathcal{Z}(3)&=\frac13 (Z_1\mathcal{Z}(2)-Z_2\mathcal{Z}(1)+Z_3), \\
\mathcal{Z}(4)&=\frac14 (Z_1\mathcal{Z}(3)-Z_2\mathcal{Z}(2)+Z_3\mathcal{Z}(1)-Z_4). 
\end{align}
Also we have the differential equation (\ref{Xi_diff}) for the Schur line defect correlation functions. 
It follows that 
\begin{align}
\frac{\partial }{\partial\mu}\Xi_{E/H}^{(n_1,\cdots,n_k)}
&=\left[
-\sum_{l=1}^{\infty} Z_l^{E/H}(n_1,\cdots,n_k) (-\mu)^l
\right]
\Xi_{E/H}^{(n_1,\cdots,n_k)}. 
\end{align}
This leads to a recursion relation for the canonical partition function of the line defect correlation function
\begin{align}
\label{XiEH_recursion}
\mathcal{Z}_{E/H}(N)^{\{n_j\}}&=\frac{1}{N}\sum_{l=1}^{N}(-1)^{l+1} Z_l^{E/H} \mathcal{Z}_{E/H}(N-l)^{\{n_j\}}. 
\end{align}

\section{Large $N$ correlators}
\label{sec_largeN}
In this section we study the large $N$ limits of the Schur line defect correlators in $\mathcal{N}=4$ $U(N)$ SYM theory. 
They are interesting in the context of the AdS/CFT correspondence \cite{Maldacena:1997re} 
as they should capture the spectrum of the fundamental string and the excitations around the D-brane configuration in string theory. 
The Wilson loop operator in the fundamental representation for $\mathcal{N}=4$ $U(N)$ SYM theory 
is proposed to be dual to a fundamental string \cite{Maldacena:1998im,Rey:1998ik} 
(also see \cite{Drukker:1999zq,Erickson:2000af,Drukker:2000rr,Yamaguchi:2006te}). 
It was argued in \cite{Drukker:2005kx} that 
the Wilson loop operators in higher-dimensional representations would be dual to certain D-brane configurations, 
as the D-branes can be viewed as the effective description of multi-string configurations. 
For the antisymmetric (resp. symmetric) representations they are conjecturally dual to the configuration 
with D5-branes \cite{Yamaguchi:2006tq,Gomis:2006sb,Rodriguez-Gomez:2006fmx,Hartnoll:2006hr} 
(resp. D3-branes \cite{Drukker:2005kx,Gomis:2006sb,Gomis:2006im,Rodriguez-Gomez:2006fmx,Yamaguchi:2007ps}). 

\subsection{Closed-form formula}

\subsubsection{Charged Wilson lines}
For the flavored 2-point function of the Wilson line operators of charges $n$ and $-n$ 
we find that the large $N$ limit is simply given by
\begin{align}
\label{large_N_chargen}
\langle W_{n}W_{-n}\rangle^{U(\infty)}
&=
-n \frac{(n)_{q^{\frac12}t^2,1} (n)_{q^{\frac12}t^{-2},1}} 
{(n)_{q,1}} \mathcal{I}^{U(\infty)}
\nonumber\\
&=\frac{n (1-q^{n})}{(1-q^{\frac{n}{2}}t^{2n}) (1-q^{\frac{n}{2}} t^{-2n})} \mathcal{I}^{U(\infty)}. 
\end{align}
where $\mathcal{I}^{U(\infty)}$ is the large $N$ limit of the Schur index of $\mathcal{N}=4$ $U(N)$ SYM theory 
\cite{Kinney:2005ej}
\begin{align}
\mathcal{I}^{U(\infty)}&=
\prod_{n=1}^{\infty} \frac{1-q^n}{(1-q^{\frac{n}{2}} t^{2n}) (1-q^{\frac{n}{2}} t^{-2n})}. 
\end{align}
We do not have a direct derivation of this expression, but have checked it for various $n$ by using our exact closed-form expression. 

In particular, the flavored 2-point function of the Wilson line operators of unit charge, 
i.e. transforming as the fundamental representation for $\mathcal{N}=4$ $U(N)$ SYM theory in the large $N$ limit is 
\begin{align}
\label{large_N_charge1}
\langle W_{1}W_{-1}\rangle^{U(\infty)}
&=\frac{1-q}{(1-q^{\frac12}t^2) (1-q^{\frac12}t^{-2})} \mathcal{I}^{U(\infty)}. 
\end{align}
The expression (\ref{large_N_charge1}) can be also found in \cite{Gang:2012yr}. 
The half-BPS Wilson loop in the fundamental representation in $\mathcal{N}=4$ $U(N)$ SYM theory 
is holographically dual to a fundamental string wrapping $AdS_2$ in $AdS_5$. 
The large $N$ index (\ref{large_N_charge1}) counts the fluctuation modes of the fundamental string wrapping $AdS_2$ in $AdS_5$ \cite{Faraggi:2011bb}. 

More generally, 
we find that all the large $N$ odd-point functions vanish
\begin{align}
\label{largeN_oddpt}
\langle W_{n_1}\cdots W_{n_{2k+1}}\rangle^{U(\infty)}&=0
\end{align}
and that the most even-point functions also vanish except for the following form:
\begin{align}
\label{largeN_factorization}
&\frac{1}{\mathcal{I}^{U(\infty)}}\langle (W_{n_1}W_{-n_1})^{m_1}\cdots (W_{n_k}W_{-n_k})^{m_k}\rangle^{U(\infty)}\notag\\
&=m_1!\biggl(\frac{\langle W_{n_1}W_{-n_1}\rangle^{U(\infty)}}{\mathcal{I}^{U(\infty)}}\biggr)^{m_1}\cdots m_k!\biggl(\frac{\langle W_{n_k}W_{-n_k}\rangle^{U(\infty)}}{\mathcal{I}^{U(\infty)}}\biggr)^{m_k}, 
\end{align}
where $0<n_1<n_2<\cdots<n_k$. 

It is also intriguing to study the large charge limit as 
the holographic dual of the large representations have been also investigated e.g. in \cite{Yamaguchi:2006te,Lunin:2006xr,DHoker:2007mci,Okuda:2008px,Gomis:2008qa}. 
For example, when $n\rightarrow \infty$ while keeping $N$ finite, 
we find that 
\begin{align}
\langle W_{\infty}W_{-\infty}\rangle^{U(N)}&=N\mathcal{I}^{U(N)}. 
\end{align}

\subsubsection{Antisymmetric Wilson lines}
For the large $N$ correlation function of the Wilson line operators in the rank-$m$ antisymmetric representation in $\mathcal{N}=4$ $U(N)$ SYM theory, we start with Newton's identity (\ref{e_p}). 
Combining our observations (\ref{largeN_oddpt}) and (\ref{largeN_factorization}), the antisymmetric correlation function at large $N$ behaves as
\begin{align}
\langle W_{(1^m)}W_{\overline{(1^m)}}\rangle^{U(\infty)}
&=\sum_{\begin{smallmatrix}
\lambda\\
|\lambda|=m\\
\end{smallmatrix}
}
\sum_{\begin{smallmatrix}
\lambda'\\
|\lambda'|=m\\
\end{smallmatrix}
}
\biggl\langle \prod_{i=1}^{r}
\prod_{j=1}^{r'}
\frac{(-1)^{r+r'}}
{\lambda_i^{m_i} {\lambda'}_j^{m_j} (m_i!) (m_j'!)}W_{\lambda_i }^{m_i} W_{-\lambda_j'}^{m_j'}\biggr\rangle^{U(\infty)} \\
&=\sum_{\begin{smallmatrix}
\lambda\\
|\lambda|=m\\
\end{smallmatrix}
}
\prod_{i=1}^{r}
\frac{1}{\lambda_i^{2m_i} m_i!}
\biggl( \frac{\langle W_{\lambda_i } W_{-\lambda_i}\rangle^{U(\infty)}}{\mathcal{I}^{U(\infty)}}\biggr)^{m_i}\mathcal{I}^{U(\infty)}, 
\end{align}
where $\lambda$ is a partition of $m$ with $m=\sum_{i=1}^{r}\lambda_i m_i$, $\lambda_1>\lambda_2>\cdots>\lambda_r$ 
and $\lambda'$ is that with $m=\sum_{j=1}^{r'}\lambda'_j m_j'$, $\lambda_1'>\lambda_2'>\cdots>\lambda_r'$ . 
Using (\ref{large_N_chargen}), we finally obtain the closed-form expression 
\begin{align}
\label{large_N_asym_m0}
\langle W_{(1^m)}W_{\overline{(1^m)}}\rangle^{U(\infty)}
&=\sum_{\begin{smallmatrix}
\lambda\\
|\lambda|=m\\
\end{smallmatrix}
}
\prod_{i=1}^r \frac{1}{\lambda_i^{m_i} (m_i!)}
\Biggl(\frac{(1-q^{\lambda_i})}{(1-q^{\frac{\lambda_i}{2}}t^{2\lambda_i}) (1-q^{\frac{\lambda_i}{2}}t^{-2\lambda_i})}\Biggr)^{m_i}
\mathcal{I}^{U(\infty)}. 
\end{align}

For example, we have 
\begin{align}
&
\langle W_{(1^2)}W_{\overline{(1^2)}}\rangle^{U(\infty)}
\nonumber\\
&=
\frac12 \Biggl[
\underbrace{
\left(
\frac{1-q}{(1-q^{\frac12}t^2) (1-q^{\frac12}t^{-2})}
\right)^2
}_{\tiny \yng(1,1)}
+
\underbrace{
\frac{1-q^2}{(1-qt^4) (1-qt^{-4})}
}_{\tiny \yng(2)}
\Biggr]\mathcal{I}^{U(\infty)}, 
\end{align}
\begin{align}
\label{largeN_rank3}
&
\langle W_{(1^3)}W_{\overline{(1^3)}}\rangle^{U(\infty)}
\nonumber\\
&=
\frac16 \Biggl[
\underbrace{
\left(
\frac{1-q}{(1-q^{\frac12}t^2) (1-q^{\frac12}t^{-2})}
\right)^3
}_{\tiny \yng(1,1,1)}
+
\underbrace{
3
\frac{1-q^2}{(1-qt^4) (1-qt^{-4})}
\frac{1-q}{(1-q^{\frac12}t^2) (1-q^{\frac12}t^{-2})}
}_{\tiny \yng(2,1)}
\nonumber\\
&
\underbrace{
+2
\frac{1-q^3}{(1-q^{\frac32}t^6) (1-q^{\frac32}t^{-6})}
}_{\tiny \yng(3)}
\Biggr]\mathcal{I}^{U(\infty)}, 
\end{align}
\begin{align}
&
\langle W_{(1^4)}W_{\overline{(1^4)}}\rangle^{U(\infty)}
\nonumber\\
&=
\frac{1}{24} \Biggl[
\underbrace{
\left(
\frac{1-q}{(1-q^{\frac12}t^2) (1-q^{\frac12}t^{-2})}
\right)^4
}_{\tiny \yng(1,1,1,1)}
+
\underbrace{
6
\frac{1-q^2}{(1-qt^4) (1-qt^{-4})}
\left(
\frac{1-q}{(1-q^{\frac12}t^2) (1-q^{\frac12}t^{-2})}
\right)^2
}_{\tiny \yng(2,1,1)}
\nonumber\\
&+
\underbrace{
3
\left(
\frac{1-q^2}{(1-qt^4) (1-qt^{-4})}
\right)^2
}_{\tiny \yng(2,2)}
+
\underbrace{
8
\frac{1-q^3}{(1-q^{\frac32}t^6) (1-q^{\frac32}t^{-6})}
\frac{1-q}{(1-q^{\frac12}t^2) (1-q^{\frac12}t^{-2})}
}_{\tiny \yng(3,1)}
\nonumber\\
&
\underbrace{
+
6\frac{1-q^4}{(1-q^2t^8) (1-q^2t^{-8})}
}_{\tiny \yng(4)}
\Biggr]\mathcal{I}^{U(\infty)}. 
\end{align}

Also we find that it can be expressed as
\begin{align}
\label{large_N_asym_m}
&
\langle W_{(1^m)}W_{\overline{(1^m)}}\rangle^{U(\infty)}
\nonumber\\
&=
\Biggl[
\sum_{n=0}^{m}
\frac{1}{(q^{\frac12}t^2;q^{\frac12}t^2)_{n} (q^{\frac12}t^{-2};q^{\frac12}t^{-2})_{m-n}}
-\sum_{n=0}^{m-1}
\frac{1}{(q^{\frac12}t^2;q^{\frac12}t^2)_{n} (q^{\frac12}t^{-2};q^{\frac12}t^{-2})_{m-n-1}}
\Biggr]
\mathcal{I}^{U(\infty)}. 
\end{align}
The half-BPS Wilson loop in the rank-$m$ antisymmetric representation in $\mathcal{N}=4$ $U(N)$ SYM theory 
is holographically dual to a D5-brane with $AdS_2\times S^4$ geometry and $m$ fundamental strings, \textit{D5-brane giant} \cite{Yamaguchi:2006tq}. 
The number $m$ of fundamental strings cannot be greater than $N$, the amount of electric flux. 
The large $N$ indices should compute the spectra of the fluctuation modes of the D5-brane giant \cite{Faraggi:2011ge}. 

{When the representation of the Wilson line operators is very large, 
they will be appropriately described by a D5-brane with fluxes.
We also find that the large $m$ limit of the flavored 2-point function (\ref{large_N_asym_m}) agrees with 
\begin{align}
\label{large_N_m_asym}
\langle W_{(1^{m=\infty})}W_{\overline{(1^{m=\infty})}}\rangle^{U(\infty)}
&=\prod_{n=1}^{\infty}
\frac{1-q^n}
{(1-q^{\frac{n}{2}}t^{2n})^2 (1-q^{\frac{n}{2}}t^{-2n})^2}. 
\end{align}
In fact, the expression (\ref{large_N_m_asym}) can be also found in \cite{Gang:2012yr} and shown to agree with the holographic calculation in \cite{Faraggi:2011ge}. 

\subsubsection{Symmetric Wilson lines}
We also find 
that the large $N$ limit of the flavored 2-point function of the Wilson line operators 
in the rank-$m$ symmetric representation for $\mathcal{N}=4$ $U(N)$ SYM theory is 
that in the rank-$m$ antisymmetric representation:
\begin{align}
\label{large_N_sym_antisym}
\langle W_{(m)}W_{\overline{(m)}}\rangle^{U(\infty)}=\langle W_{(1^m)}W_{\overline{(1^m)}}\rangle^{U(\infty)}.
\end{align}
This follows from the vanishing theorem (\ref{largeN_oddpt}) and Newton's identities (\ref{e_p}) and (\ref{h_p}). 

The half-BPS Wilson loop in the rank-$m$ symmetric representation in $\mathcal{N}=4$ $U(N)$ SYM theory 
is holographically dual to a D3-brane with the geometry $AdS_2\times S^2$ and $m$ fundamental strings, \textit{D3-brane dual giant} \cite{Drukker:2005kx,Gomis:2006sb}. 
Unlike the D5-brane giant, there is no upper bound on the fundamental string charge $m$ for the D3-brane dual giant. 

As the large $N$ correlators of the symmetric Wilson line operators coincide with those of the antisymmetric Wilson line operators, 
the spectra of the fluctuation modes of the D3-brane dual giant will match with that for the D5-brane giant. 
This would demonstrate the large $N$ duality between a particle outside the droplet 
corresponding to the D5-brane giant 
and a hole inside the droplet corresponding to the D3-brane dual giant \cite{Okuyama:2006jc}. 
The large $N$ (anti)symmetric 2-point functions (\ref{large_N_sym_antisym}) admit a graphical notation for contracted tensors. 
For rank-$m$ 2-point function, we consider a tensor product of $m$ copies of $\langle W_{1} W_{-1}\rangle^{U(\infty)}$ 
and take a trace of it by closing the $m$ in-arrows and $m$ out-arrows. 
We identify the trace of $n$ products with 
the normalized large $N$ 2-point function of the charged Wilson line operators 
$\langle W_{n}^{\infty} W_{-n}^{\infty}\rangle$ $=$ $\frac{1}{n} \langle W_{n} W_{-n}\rangle^{U(\infty)}$. 
There exist $m!$ contractions. 
The large $N$ rank-$m$ (anti)symmetric 2-point function is obtained by 
summing over all possible permutations. 
We illustrate examples in Figure \ref{fig_large2} for $m=2$ 
and Figure \ref{fig_large3} for $m=3$. 
\begin{figure}
\begin{center}
\includegraphics[width=14.5cm]{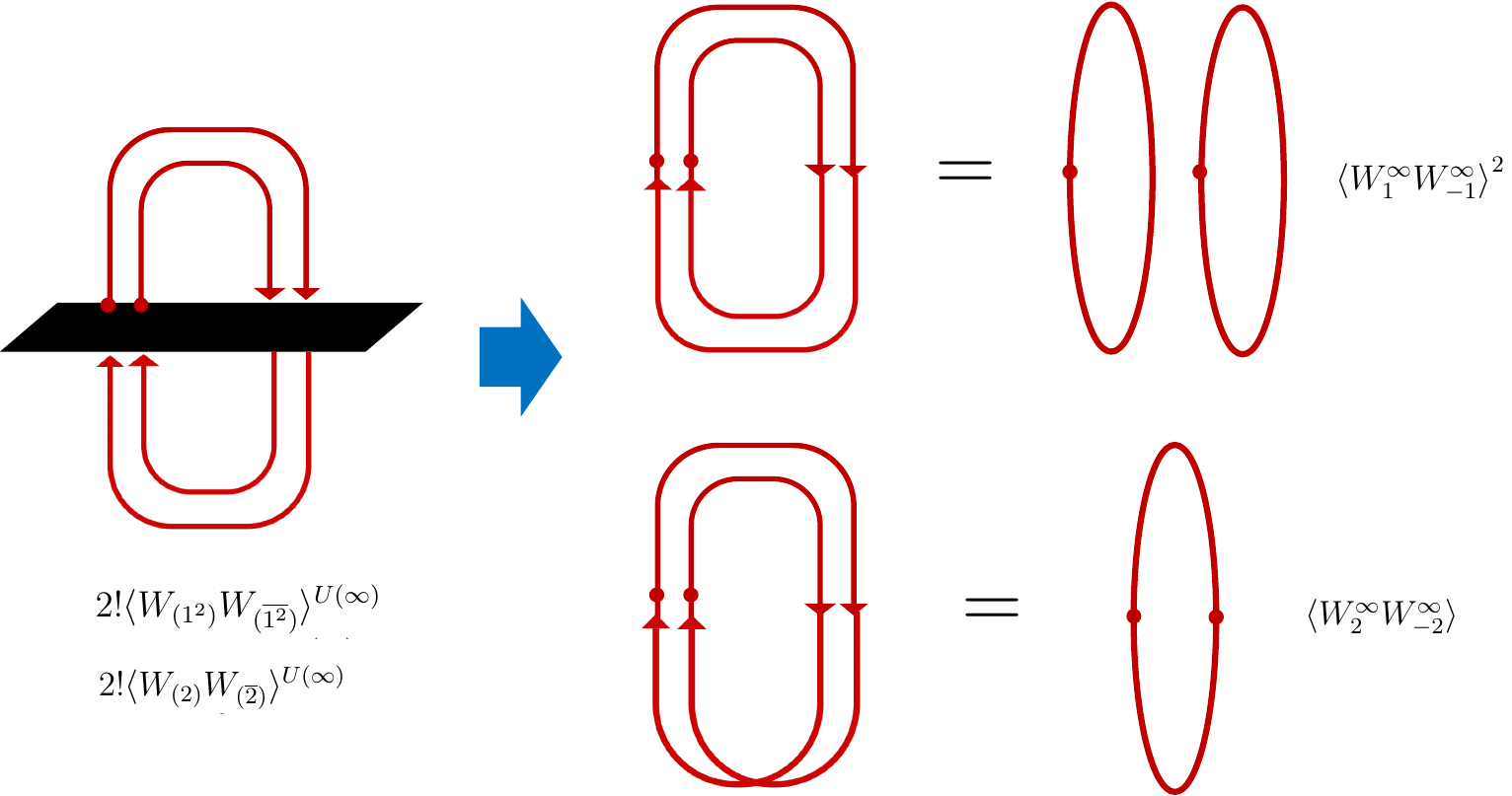}
\caption{
The graphical representation of the large $N$ rank-$2$ (anti)symmetric 2-point function. 
There is a unique way for each of contractions. 
}
\label{fig_large2}
\end{center}
\end{figure}
\begin{figure}
\begin{center}
\includegraphics[width=15.5cm]{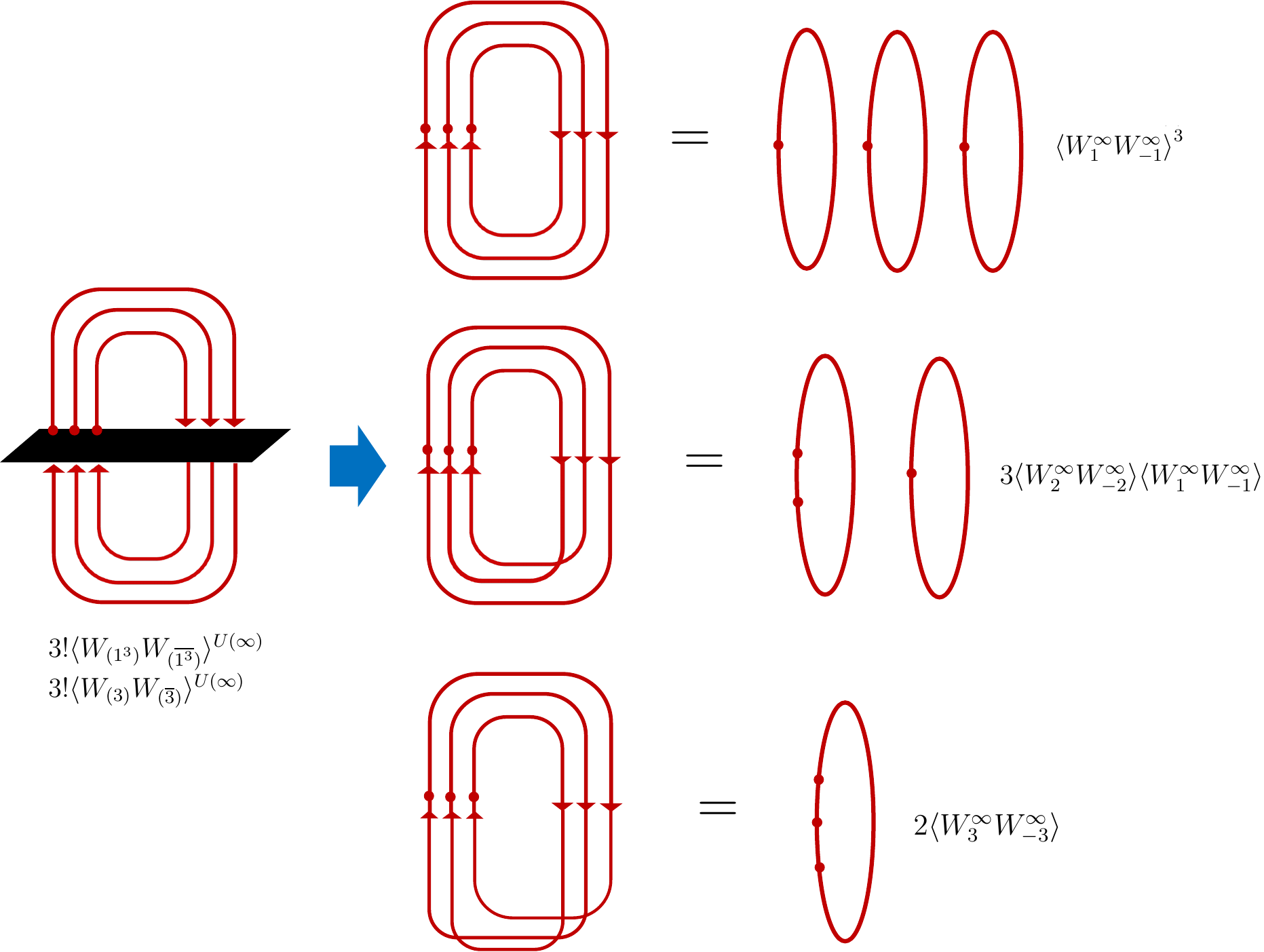}
\caption{
The graphical representation of the large $N$ rank-$3$ (anti)symmetric 2-point function. 
While for the top diagram there is a unique contraction, 
there are three for the middle and two for the bottom. 
These combinatorial factors determine the coefficients in (\ref{largeN_rank3}).  
}
\label{fig_large3}
\end{center}
\end{figure}
We leave it future work to examine the fluctuation modes on the D3-brane dual giant in detail 
and compare them with those from the gravity side as studied in \cite{Faraggi:2011bb}. 

Using the conjectures (\ref{large_N_asym_m0}) and (\ref{large_N_sym_antisym}), 
the generating function for the large $N$ limit of the 2-point functions of the Wilson line operators in the rank-$m$ 
(anti)symmetric representation is given by
\begin{align}
\sum_{m=0}^{\infty} (s_1s_2)^m 
\frac{\langle W_{(m)} W_{\overline{(m)}} \rangle^{U(\infty)}}
{\mathcal{I}^{U(\infty)}}
&=\sum_{m=0}^{\infty} (s_1s_2)^m 
\frac{\langle W_{(1^m)} W_{\overline{(1^m)}} \rangle^{U(\infty)}}
{\mathcal{I}^{U(\infty)}} \\
&=
\exp \Biggl[ \sum_{n=1}^\infty \frac{(s_1 s_2)^n}{n} \frac{1-q^n}{(1-q^{n/2} t^{2n})(1-q^{n/2} t^{-2n})}\Biggr]\\
&=
\frac{1-s_1s_2}{(s_1s_2;q^{\frac12}t^2)_{\infty} (s_1s_2;q^{\frac12}t^{-2})_{\infty}}. 
\end{align}
In the unflavored limit $t \to 1$, our result precisely reduces to the previous result in \cite{Drukker:2015spa}.

\subsection{Plane partition diamonds}
The large $N$ limit of the unflavored Schur index of $\mathcal{N}=4$ $U(N)$ and $SU(N)$ SYM theory 
are identified with a generating function for the overpartition \cite{MR2034322} and the 3-colored partitions \cite{MR1634067}. 
Here we discuss the combinatorial interpretation of the Schur line defect correlator. 

When the flavored fugacity $t$ is turned off, the 2-point function (\ref{large_N_m_asym}) can be written as
\begin{align}
\label{large_N_m_sym1}
&\langle W_{(m=\infty)}W_{\overline{(m=\infty)}}\rangle^{U(\infty)}(t=1;q)
=\prod_{n=1}^{\infty}
\frac{1-q^n}
{(1-q^{\frac{n}{2}})^4}
=\frac{(-q^{\frac12};q^{\frac12})_{\infty}}{(q^{\frac12};q^{\frac12})_{\infty}^3}. 
\end{align}
This admits an expansion 
\begin{align}
&\langle W_{(m=\infty)}W_{\overline{(m=\infty)}}\rangle^{U(\infty)}(t=1;q)=\sum_{n=0}d(n) q^{\frac{n}{2}}
\nonumber\\
&=1+4q^{1/2}+13q+36q^{3/2}+90q^{2}+208q^{5/2}+455q^3
+948q^{7/2}+1901q^4+\cdots. 
\end{align}
The coefficient $d(n)$ is identified with the number of the Schmidt type partitions 
referred to as the \textit{plane partition diamonds} of $n$ \cite{MR1868964,MR4370530}, 
that is the partitions of $n=a_1+a_4+a_7+\cdots$ 
whose parts $a_i$ lie on the graph 
which is made up of chains of rhombi 
in such a way that the set $(a_{3i-2}, a_{3i-1}, a_{3i}, a_{3i+1})$ corresponds to the four vertices of the $i$-th rhombus 
with the conditions 
\begin{align}
a_{3i-2}\ge a_{3i-1}\ge a_{3i+1},\qquad 
a_{3i-2}\ge a_{3i}\ge a_{3i+1}. 
\end{align}

For example, 
$d(1)$ counts the $4$ plane partition diamonds 
$\{a_1=1\}$, $\{a_1=1,a_2=1\}$, $\{a_1=1,a_3=1\}$ and $\{a_1=1,a_2=1,a_3=1\}$ 
and $d(2)$ counts the $13$ plane partition diamonds 
$\{a_1=1,a_2=1,a_3=1,a_4=1\}$, 
$\{a_1=1,a_2=1,a_3=1,a_4=1,a_5=1\}$, 
$\{a_1=1,a_2=1,a_3=1,a_4=1,a_6=1\}$, 
$\{a_1=1,a_2=1,a_3=1,a_4=1,a_5=1,a_6=1\}$, 
$\{a_1=2\}$, 
$\{a_1=2,a_2=2\}$, 
$\{a_1=2,a_3=2\}$, 
$\{a_1=2,a_2=2,a_3=2\}$, 
$\{a_1=2,a_2=1\}$, 
$\{a_1=2,a_3=1\}$, 
$\{a_1=2,a_2=1,a_3=1\}$
$\{a_1=2,a_2=2,a_3=1\}$ and 
$\{a_1=2,a_2=1,a_3=2\}$. 

Let us study the degeneracy of the excitation modes of the D3-branes wrapping the $AdS_2\times S^2$ 
(or equivalently the D5-branes wrapping the $AdS_2\times S^4$). 
The growth of the number $d(n)$ of operators with large scaling dimension can be studied from the infinite product (\ref{large_N_m_sym1}). 
Making use of the Meinardus Theorem \cite{MR62781}, 
we get the asymptotic growth
\begin{align}
\label{asym_largeNm}
d(n)\sim \frac{7}{96 n^{3/2}}
\exp\left[
\frac{7^{1/2}}{3^{1/2}} \pi n^{1/2}
\right]. 
\end{align}
The exact numbers $d(n)$ and the values $d_{\textrm{asymp}}(n)$ obtained from the formula (\ref{asym_largeNm}) are listed as follows: 
\begin{align}
\label{asym_Nm_table}
\begin{array}{c|c|c}
n&d(n)&d_{\textrm{asymp}}(n) \\ \hline 
10&6955&8982.37 \\
100&4.66051\times 10^{16}&5.05848\times 10^{16} \\
1000&1.80784\times 10^{60}&1.85552\times 10^{60} \\
5000&4.77308\times 10^{140}&4.82904\times 10^{140} \\
10000&1.86714\times 10^{201}&1.88260\times 10^{201} \\
\end{array}
\end{align}
It should be compared with 
the asymptotic growth of the number of the states 
in the absence of the line operators, which is equal to the number of the overpartitions is given by \cite{Hatsuda:2022xdv}
\begin{align}
\overline{p}(n)\sim \frac{1}{8n}
\left(
1-\frac{1}{\pi n^{1/2}}
\right) \exp\left[ \pi n^{1/2} \right]. 
\end{align}
It would be interesting to elucidate 
the combinatorial aspects of the enumeration of the operators in the large $N$ limit 
and their asymptotic behaviors from the holographically dual supergravity. 

\subsection*{Acknowledgements}
The authors would like to thank Kimyeong Lee, Hai Lin and Masatoshi Noumi for useful discussions and comments. 
The work of Y.H. is supported
in part by JSPS KAKENHI Grant No. 18K03657 and 22K03641. 
The work of T.O. is supported by the Startup Funding no. 4007012317 of the Southeast University. 

\appendix

\section{Definitions and notations}
\label{app_def}

\subsection{$q$-shifted factorial}
We have used the following notation of $q$-shifted factorial: 
\begin{align}
\label{qpoch_def}
(a;q)_{0}&:=1,\qquad
(a;q)_{n}:=\prod_{k=0}^{n-1}(1-aq^{k}),\qquad 
(q;q)_{n}:=\prod_{k=1}^{n}(1-q^{k}),\quad 
\quad  n\ge1,
\nonumber \\
(a;q)_{\infty}&:=\prod_{k=0}^{\infty}(1-aq^{k}),\qquad 
(q;q)_{\infty}:=\prod_{k=1}^{\infty} (1-q^k), 
\nonumber\\
(a^{\pm};q)_{\infty}&:=(a;q)_{\infty}(a^{-1};q)_{\infty},
\end{align}
where $a$ and $q$ are complex variables. 

\subsection{Twisted Weierstrass functions}
We define the twisted Weierstrass function by 
\footnote{The $P_k\left[\begin{smallmatrix}\theta\\ \phi\\ \end{smallmatrix}\right](z,\tau)$ defined here is the same as 
$P_k\left[\begin{smallmatrix}\theta\\ \phi\\ \end{smallmatrix}\right](2\pi iz,\tau)$ in \cite{Mason:2008zzb}. }
\begin{align}
\label{tP1}
P_1\left[
\begin{matrix}
\theta\\
\phi\\
\end{matrix}
\right](z,\tau)&=
-{\sum_{n\in \mathbb{Z}}}' \frac{x^{n+\lambda}}{1-\theta^{-1}q^{n+\lambda}},
\end{align}
and 
\begin{align}
\label{tPk}
P_k\left[
\begin{matrix}
\theta\\
\phi\\
\end{matrix}
\right](z,\tau)
&=
\frac{(-1)^k}{(k-1)!}
\frac{1}{(2\pi i)^{k-1}} 
\frac{\partial^{k-1}}{\partial z^{k-1}}
P_1\left[
\begin{matrix}
\theta\\
\phi\\
\end{matrix}
\right](z,\tau)
\nonumber\\
&=
\frac{(-1)^k}{(k-1)!} {\sum_{n\in \mathbb{Z}}}' 
\frac{(n+\lambda)^{k-1} x^{n+\lambda}}{1-\theta^{-1}q^{n+\lambda}},
\end{align}
where $\phi=e^{2\pi i\lambda}$.

\section{Multiple Kronecker theta series}
\label{app_qxx}
The multiple Kronecker theta series (\ref{s_zeta}) plays a role of elementary blocks of the Schur index and the Schur line defect correlators. 
They can be written in terms of the twisted Weierstrass functions. 
In this appendix, we show several examples. 

\subsection{$Q(l_0,l_1;n_0,n_1)$}
In general the multiple Kronecker theta series (\ref{s_zeta}) for $k=1$ is expandable from the equation (\ref{ql0l1}). 
They show up in the closed-form expression of the Schur line defect 2-point functions. 

The simplest example is $l_0=l_1=1$. 
We have
\begin{align}
\label{q11_expand}
Q(1,1;0,n;u;\xi;q)
&=\sum_{p\in \mathbb{Z}}
\frac{\xi^{-2p-n}}{(1-uq^p) (1-uq^{p+n})}
\nonumber\\
&=\frac{1}{1-q^n} \sum_{p\in \mathbb{Z}}
\left(
\frac{\xi^{-2p-n}}{(1-uq^p)}
-\frac{q^n \xi^{-2p-n}}{1-uq^{p+n}}
\right)
\nonumber\\
&=\Bigl[ (n)_{q,\xi}+(-n)_{q,\xi} \Bigr]Q(1;0;u;\xi^2;q)
\nonumber\\
&=
\Bigl[
(n)_{q,\xi}+(-n)_{q,\xi}
\Bigr]P_1\left[
\begin{matrix}
\xi^2\\
1\\
\end{matrix}
\right]
(\nu,\tau), 
\end{align}
where we have assumed that $n$ is non-zero integer. 

When $l_0=2$, $l_1=1$ and $n\neq 0$ we have 
\begin{align}
\label{q21_expanda}
&Q(2,1;0,n;u;\xi;q)
\nonumber\\
&=
-\frac{\xi^{-n}}{1-q^n}\sum_{p\in \mathbb{Z}}
\frac{\xi^{-3p}}{(1-uq^p)^2}
+\frac{q^n\xi^{-n}}{(1-q^n)^2}\sum_{p\in \mathbb{Z}}
\frac{\xi^{-3p}}{1-uq^p}
-\frac{q^{2n}\xi^{-n}}{(1-q^n)^2}
\sum_{p\in \mathbb{Z}}
\frac{\xi^{-3p}}{1-uq^{p+n}}
\nonumber\\
&=(n)_{q,\xi}Q(2;0;u;\xi^{\frac32};q)
+(1-q^{-n}\xi^{-3n}) (-n)_{q,\xi}^2 Q(1;0;u;\xi^3;q), \\
\label{q21_expandb}
&Q(1,2;0,n;u;\xi;q)
\nonumber\\
&=-\frac{\xi^{-2n}}{(1-q^n)^2}\sum_{p\in \mathbb{Z}}\frac{\xi^{-3p}}{1-uq^p}
+\frac{q^n \xi^{-2n}}{1-q^n}\sum_{p\in \mathbb{Z}}\frac{\xi^{-3p}}{(1-uq^{p+n})^2}
+\frac{q^n\xi^{-2n}}{(1-q^n)^2}
\sum_{p\in \mathbb{Z}}\frac{\xi^{-3p}}{1-uq^{p+n}}
\nonumber\\
&=(-n)_{q,\xi}Q(2;0;u;\xi^{\frac32};q)
+(1-q^n\xi^{3n})(n)_{q,\xi}^2 Q(1;0;u;\xi^3;q). 
\end{align}
It follows that 
\begin{align}
&
\label{q21_expand}
Q(2,1;0,n;u;\xi;q)+Q(1;2;0,n;u;\xi;q)
\nonumber\\
&=\frac{1}{u} \Bigl[
(n)_{q,\xi}+(-n)_{q,\xi}
\Bigr]P_2\left[
\begin{matrix}
q\xi^3\\
1\\
\end{matrix}
\right](\nu,\tau)-\Biggl[
\frac{(n)_{q,\xi}^2}{(n)_{q\xi,1}}
+\frac{(-n)_{q,\xi}^2}{(-n)_{q\xi,1}}
\Biggr]P_1\left[
\begin{matrix}
\xi^3\\
1\\
\end{matrix}
\right](\nu,\tau). 
\end{align}

For $l_0+l_1=3$ there are three types. 
When $n\neq 0$, we have 
\begin{align}
&Q(3,1;0,n;u;\xi;q)
\nonumber\\
&=(n)_{q,\xi}Q(3;0;u;\xi^{\frac43};q)
-q^{-n}\xi^{-3n}(-n)_{q,\xi}^2 Q(2;0;u;\xi^2;q)
\nonumber\\
&+(1-q^{-n}\xi^{-4n})(-n)_{q,\xi}^3 Q(1;0;u;\xi^4;q)
\nonumber\\
&=(n)_{q,\xi}Q(3,0;u;\xi^{\frac43};q)-q^n\xi^n (n)_{q,\xi}^2 Q(2;0;u;\xi^2;q)
-q^{2n}\xi^{2n} \frac{(n)_{q,\xi}^3}{(n)_{q\xi^4,1}} Q(1;0;u;\xi^4;q)
,\\
&Q(2,2;0,n;u;\xi;q)
\nonumber\\
&=\Bigl[
(n)_{q,\xi}^2+(-n)_{q,\xi}^2
\Bigr]Q(2;0;u;\xi^2;q)
-2q^{-n}\xi^{-n}(1-q^{-n}\xi^{-4n})(-n)_{q,\xi}^3 Q(1;0;u;\xi^4;q) \nonumber\\
&=\Bigl[
(n)_{q,\xi}^2+(-n)_{q,\xi}^2
\Bigr]Q(2;0;u;\xi^2;q)
-2\Bigl[q^n\xi^n (n)_{q,\xi}^3+q^{-n}\xi^{-n}(-n)_{q,\xi}^3\Bigr]Q(1;0;u;\xi^4;q)
,\\
&Q(1;3;0,n;u;\xi;q)
\nonumber\\
&=(-n)_{q,\xi}Q(3;0;u;\xi^{\frac43};q)
-q^{n}\xi^{3n}(n)_{q,\xi}^2 Q(2;0;u;\xi^2;q)
\nonumber\\
&+\Bigl[
(n)_{q,\xi}^2+(-n)_{q,\xi}^2
\Bigr]Q(2;0;u;\xi^2;q)
\left(
1-q^n \xi^{4n}
\right)(n)_{q,\xi}^3 Q(1;0;u;\xi^4;q)
\nonumber\\
&=(-n)_{q,\xi}Q(3,0;u;\xi^{\frac43};q)-q^{-n}\xi^{-n} (-n)_{q,\xi}^2 Q(2;0;u;\xi^2;q)
-q^{-2n}\xi^{-2n} \frac{(-n)_{q,\xi}^3}{(-n)_{q\xi^4,1}} Q(1;0;u;\xi^4;q). 
\end{align}
In terms of the twisted Weierstrass function they can be written as
\begin{align}
\label{q31_expand}
&Q(3,1;0,n;u;\xi;q)+Q(1,3;0,n;u;\xi;q)
\nonumber\\
&=
\frac{(n)_{q,\xi}+(-n)_{q,\xi}}{2u^2}\Bigl(
P_2\left[
\begin{matrix}
q^2\xi^3\\
1\\
\end{matrix}
\right](\nu,\tau)
+2P_3\left[
\begin{matrix}
q^2\xi^3\\
1\\
\end{matrix}
\right](\nu,\tau)
\Bigr)
\nonumber\\
&-\frac{q^n\xi^n(n)_{q,\xi}^2+q^{-n}\xi^{-n}(-n)_{q,\xi}^2}{u}
P_2\left[
\begin{matrix}
q\xi^4\\
1\\
\end{matrix}
\right](\nu,\tau)
-
\Bigl(
\frac{q^{2n}\xi^{2n} (n)_{q,\xi}^3}{(n)_{q\xi^4,1}}
+\frac{q^{-2n}\xi^{-2n} (-n)_{q,\xi}^3}{(-n)_{q\xi^4,1}}
\Bigr)
P_1\left[
\begin{matrix}
\xi^4\\
1\\
\end{matrix}
\right](\nu,\tau),
\end{align}
\begin{align}
\label{q22_expand}
&Q(2,2;0,n;u;\xi;q)
\nonumber\\
&=\frac{(n)_{q,\xi}^2+(-n)_{q,\xi}^2}{u}
P_2\left[
\begin{matrix}
q\xi^4\\
1\\
\end{matrix}
\right](\nu,\tau)
+2q^n \xi^n \frac{(n)_{q,\xi}^3}{(n)_{q\xi^4,1}}
P_1\left[
\begin{matrix}
\xi^4\\
1\\
\end{matrix}
\right](\nu,\tau). 
\end{align}

\subsection{$Q(l_0,l_1,l_2;0,n_0,n_1,n_2)$}
\label{app_qxxx1}
We present several examples of the multiple Kronecker theta series (\ref{s_zeta}) for $k=2$. 
They appear in the Schur line defect 3-point functions of the charged Wilson line operators. 
We assume that $n_1$ and $n_2$ are non-zero integers. 

For $l_0=l_1=l_2=1$ the function (\ref{s_zeta}) is given by
\begin{align}
\label{q111_expand1}
&
Q(1,1,1;0,n_1,n_2;u;\xi;q)
\nonumber\\
&=
\Bigl[
(n_1)_{q,\xi}(n_2)_{q,\xi}
+(-n_1)_{q,\xi}(-n_1+n_2)_{q,\xi}
+(-n_2)_{q,\xi}(-n_2+n_1)_{q,\xi}
\Bigr]Q(1;0;u;\xi^3;q)
\nonumber\\
&
=\Bigl[
(n_1)_{q,\xi}(n_2)_{q,\xi}
+(-n_1)_{q,\xi}(-n_1+n_2)_{q,\xi}
+(-n_2)_{q,\xi}(-n_2+n_1)_{q,\xi}
\Bigr]P_1\left[
\begin{matrix}
\xi^2\\
1\\
\end{matrix}
\right](\nu,\tau). 
\end{align}
When $l_0=2$, $l_1=l_2=1$ we have  the expansion
\begin{align}
&
Q(2,1,1;0,n_1,n_2;u;\xi;q)
\nonumber\\
&=(n_1)_{q,\xi}(n_2)_{q,\xi}Q(2;0;u;\xi^2;q)
\nonumber\\
&-q^{-(n_1+n_2)}\xi^{-3(n_1+n_2)}(-q^{-n_1}-q^{-n_2}+2)
(-n_1)_{q,\xi}^2(-n_2)_{q,\xi}^2Q(1;0;u;\xi^4;q)
\nonumber\\
&+(-n_1)_{q,\xi}^2 (-n_1+n_2)_{q,\xi} Q(1;0;u;\xi^4;q)
\nonumber\\
&+(-n_2)_{q,\xi}^2 (-n_2+n_1)_{q,\xi} Q(1;0;u;\xi^4;q). 
\end{align}
This leads to  
\begin{align}
&
Q(2,1,1;0,n_1,n_2;u;\xi;q)
\nonumber\\
&=(n_1)_{q,\xi}(n_2)_{q,\xi}\frac{1}{u} 
P_2\left[
\begin{matrix}
q\xi^4\\
1\\
\end{matrix}
\right](\nu,\tau)
+\Bigl[
-q^{n_1+n_2}\xi^{n_1+n_2}(2-q^{-n_1}-q^{-n_2})
(n_1)_{q,\xi}^2(n_2)_{q,\xi}^2
\nonumber\\
&+(-n_1)_{q,\xi}^2 (-n_1+n_2)_{q,\xi}
+(-n_2)_{q,\xi}^2 (-n_2+n_1)_{q,\xi}
\Bigr]
P_1\left[
\begin{matrix}
\xi^4\\
1\\
\end{matrix}
\right](\nu,\tau). 
\end{align}
For $l_0=3$, $l_1=l_2=1$
\begin{align}
&
Q(3,1,1;0,n_1,n_2;u;\xi;q)
\nonumber\\
&=(n_1)_{q,\xi}(n_2)_{q,\xi} Q(3;0;u;\xi^{\frac53};q)
\nonumber\\
&-q^{-(n_1+n_2)}\xi^{-3(n_1+n_2)}
(-q^{-n_1}-q^{-n_2}+2) 
(-n_1)_{q,\xi}^2 (-n_2)_{q,\xi}^2 Q(2;0;u;\xi^{\frac52};q)
\nonumber\\
&+q^{-(n_1+n_2)}\xi^{-4(n_1+n_2)}
(-3q^{-n_1}-3q^{-n_2}+q^{-2n_1}+q^{-2n_2}+q^{-n_1-n_2}+3)
(-n_1)^3 (-n_2)^3 Q(1;0;u;\xi^5)
\nonumber\\
&+(-n_1)_{q,\xi}^3(-n_1+n_2)_{q,\xi}Q(1;0;u;\xi^5)
\nonumber\\
&+(-n_2)_{q,\xi}^3(-n_2+n_1)_{q,\xi}Q(1;0;u;\xi^5). 
\end{align}
For $l_0=4$, $l_1=l_2=1$
\begin{align}
&
Q(4,1,1;0,n_1,n_2;u;\xi;q)
\nonumber\\
&=(n_1)_{q,\xi}(n_2)_{q,\xi} Q(4;0;u;\xi^{\frac32};q)
\nonumber\\
&-q^{-(n_1+n_2)}\xi^{-3(n_1+n_2)}
(-q^{-n_1}-q^{-n_2}+2) 
(-n_1)_{q,\xi}^2 (-n_2)_{q,\xi}^2 Q(3;0;u;\xi^2;q)
\nonumber\\
&+q^{-(n_1+n_2)}\xi^{-4(n_1+n_2)}
(-3q^{-n_1}-3q^{-n_2}+q^{-2n_1}+q^{-2n_2}+q^{-n_1-n_2}+3)
(-n_1)^3 (-n_2)^3 Q(2;0;u;\xi^3)
\nonumber\\
&-q^{-(n_1+n_2)}\xi^{-5(n_1+n_2)}
(-6q^{-n_1}-6q^{-n_2}+4q^{-2n_1}+4q^{-2n_2}+4q^{-n_1-n_2}
\nonumber\\
&-q^{-3n_1}-q^{-3n_2}-q^{-2n_1-n_2}-q^{-2n_1-n_2}+4
)(-n_1)_{q,\xi}^4 (-n_2)_{q,\xi}^4 Q(1;0;u;\xi^6;q)
\nonumber\\
&+(-n_1)_{q,\xi}^4(-n_1+n_2)_{q,\xi}Q(1;0;u;\xi^6)
\nonumber\\
&+(-n_2)_{q,\xi}^4(-n_2+n_1)_{q,\xi}Q(1;0;u;\xi^6). 
\end{align}
For $l_0=2$, $l_1=2$, $l_2=1$
\begin{align}
&
Q(2,2,1;0,n_1,n_2;u;\xi;q)
\nonumber\\
&=(n_1)_{q,\xi}^2 (n_2)_{q,\xi} Q(2;0;u;\xi^\frac{5}{2};q)
\nonumber\\
&-q^{-(2n_1+n_2)}\xi^{-3(n_1+n_2)-2n_1}(q^{-n_1}+2q^{-n_2}-3)(-n_1)_{q,\xi}^3(-n_2)_{q,\xi}^2 Q(1;0;u;\xi^3;q)
\nonumber\\
&+(-n_1)_{q,\xi}^2(-n_1+n_2)_{q,\xi} Q(2;0;u;\xi^{\frac52};q)
\nonumber\\
&+q^{n_2-n_1}\xi^{n_2-2n_1}(-3q^{-n_1}+2q^{-2n_1}+1)
(-n_1)_{q,\xi}^3(-n_1+n_2)_{q,\xi}^2 Q(1;0;u;\xi^5)
\nonumber\\
&+(-n_2)^2(-n_2+n_1)^2 Q(1;0;u;\xi^5;q). 
\end{align}
For $l_0=3$, $l_1=2$, $l_2=1$
\begin{align}
&
Q(3,2,1;0,n_1,n_2;u;\xi;q)
\nonumber\\
&=(n_1)_{q,\xi}^2 (n_2)_{q,\xi} Q(3;0;u;\xi^2;q)
\nonumber\\
&-q^{-(2n_1+n_2)}\xi^{-3(n_1+n_2)-2n_1}(q^{-n_1}+2q^{-n_2}-3)(-n_1)_{q,\xi}^3(-n_2)_{q,\xi}^2 Q(2;0;u;\xi^3;q)
\nonumber\\
&-q^{-(2n_1+n_2)}\xi^{-4(n_1+n_2)-2n_1}
(-4q^{-n_1}-8q^{-n_2}+q^{-2n_1}+3q^{-2n_2}+2q^{-n_1-n_2}+6)
\nonumber\\
&\times (-n_1)_{q,\xi}^4(-n_2)_{q,\xi}^3 Q(1;0;u;\xi^3;q)
\nonumber\\
&+(-n_1)_{q,\xi}^3(-n_1+n_2)_{q,\xi} Q(2;0;u;\xi^3;q)
\nonumber\\
&+q^{n_2-n_1}\xi^{2n_2-3(n_1+n_2)}(-4q^{-n_1}+3q^{-n_2}+	1)
(-n_1)_{q,\xi}^4(-n_1+n_2)_{q,\xi}^2 Q(1;0;u;\xi^6;q)
\nonumber\\
&+(-n_2)^3(-n_2+n_1)^2 Q(1;0;u;\xi^6;q). 
\end{align}
For $l_0=4$, $l_1=2$, $l_2=1$
\begin{align}
&
Q(4,2,1;0,n_1,n_2;u;\xi;q)
\nonumber\\
&=(n_1)_{q,\xi}^2(n_2)_{q,\xi} Q(4;0;u;\xi^{\frac{7}{4}};q)
\nonumber\\
&-q^{-(2n_1+n_2)}\xi^{-3(n_1+n_2)-2n_1}(q^{-n_1}+2q^{-n_2}-3)
(-n_1)_{q,\xi}^3 (-n_2)_{q,\xi}^2 Q(3;0;u;\xi^{\frac73};q)
\nonumber\\
&-q^{-(2n_1+n_2)}\xi^{-4(n_1+n_2)-2n_1}
(-4q^{-n_1}-8q^{-n_2}+q^{-2n_1}+3q^{-2n_2}+2q^{-n_1-n_2}+6)
\nonumber\\
&\times (-n_1)_{q,\xi}^4(-n_2)_{q,\xi}^3 Q(2;0;u;\xi^{\frac72};q)
\nonumber\\
&-q^{-(2n_1+n_2)}\xi^{-5(n_1+n_2)-2n_1}
(10q^{-n_1}+20q^{-n_2}-5q^{-2n_1}-15q^{-2n_2}-10q^{-n_1-n_2}
\nonumber\\
&+q^{-3n_1}+4q^{-3n_2}+3q^{-n_1-2n_2}+2q^{-2n_1-n_2}-10)
(-n_1)_{q,\xi}^5 (-n_2)^4 Q(1;0;u;\xi^7;q)
\nonumber\\
&+(-n_1)_{q,\xi}^4(-n_1+n_2)_{q,\xi}Q(2;0;u;\xi^{\frac72};q)
\nonumber\\
&+q^{n_2-n_1}\xi^{n_2-2n_1} (-5q^{-n_1}+4q^{-n_2}+1)(-n_1)_{q,\xi}^5 (-n_1+n_2)_{q,\xi}^2 Q(1;0;u;\xi^7;q)
\nonumber\\
&+(-n_2)_{q,\xi}^4 (-n_2+n_1)_{q,\xi}^2 Q(1;0;u;\xi^7;q). 
\end{align}
For $l_0=2$, $l_1=2$, $l_2=1$
\begin{align}
&
Q(2,2,2;0,n_1,n_2;u;\xi;q)
\nonumber\\
&=(n_1)_{q,\xi}^2 (n_2)_{q,\xi}^2 Q(2;0;u;\xi^3;q)
\nonumber\\
&+2q^{-(2n_1+2n_2)}\xi^{-5(n_1+n_2)} (q^{-n_1}+q^{-n_2}-2)
(-n_1)_{q,\xi}^3 (-n_2)_{q,\xi}^3Q(1;0;u;\xi^2;q)
\nonumber\\
&+(-n_1)_{q,\xi}^2 (-n_2)_{q,\xi}^2 Q(2;0;u;\xi^3;q)
\nonumber\\
&+2q^{n_2-n_1}\xi^{n_2-2n_1}(-2q^{-n_1}+q^{-n_2}+1)
(-n_1)_{q,\xi}^3 (-n_2)_{q,\xi}^3 Q(1;0;u;\xi^6;q)
\nonumber\\
&+(-n_1)_{q,\xi}^2 (-n_2)_{q,\xi}^2 Q(2;0;u;\xi^3;q)
\nonumber\\
&+2q^{n_1-n_2}\xi^{n_1-2n_2}(-2q^{-n_2}+q^{-n_1}+1)
(-n_1)_{q,\xi}^3 (-n_2)_{q,\xi}^3 Q(1;0;u;\xi^6;q). 
\end{align}

\subsection{$Q(1,1,\cdots,1;\{n_i\})$}
\label{app_qxN}
The multiple Kronecker theta series 
with $l_0=l_1=\cdots =l_{k}=1$ appears in the $U(k+1)$ $(k+1)$-point function of the charged Wilson line operators. 
It can be expanded in terms of the Kronecker theta function (\ref{q1_tP1}) by the relation (\ref{q1x1}). 

For example, 
\begin{align}
Q(1,1;0,n;u;\xi;q)
&=\left[ (n)_{q,\xi}+(-n)_{q,\xi} \right]
Q(1;0;u;\xi^2;q), \\
Q(1,1,1;0,n_1,n_2;u;\xi;q)
&=
\Bigl[
(n_1)_{q,\xi}(n_2)_{q,\xi}
+(-n_1)_{q,\xi}(-n_1+n_2)_{q,\xi}
\nonumber\\
&+(-n_2)_{q,\xi}(-n_2+n_1)_{q,\xi}
\Bigr]Q(1;0;u;\xi^3;q), \\
Q(1,1,1,1;0,n_1,n_2,n_3;u;\xi;q)
&=\Bigl[
(n_1)_{q,\xi}(n_2)_{q,\xi}(n_3)_{q,\xi}
\nonumber\\
&+(-n_1)_{q,\xi}(-n_1+n_2)_{q,\xi}(-n_1+n_3)_{q,\xi}
\nonumber\\
&+(-n_2)_{q,\xi}(-n_2+n_1)_{q,\xi}(-n_2+n_3)_{q,\xi}
\nonumber\\
&+(-n_3)_{q,\xi}(-n_3+n_1)_{q,\xi}(-n_3+n_2)_{q,\xi}
\Bigr]
\nonumber\\
&\times Q(1;0;u;\xi^4;q). 
\end{align}

\section{Spectral zeta functions}
\label{app_spectralZ}

\subsection{$Z_{l}^{E}$}
The 2-point functions of the Wilson line operators 
transforming in the antisymmetric representation for $\mathcal{N}=2^*$ $U(N)$ SYM theory are 
captured by the spectral zeta functions $Z_l^{E}(n)$, $l\le N$. For $l=6$ we have
\begin{align}
Z_6^E&=
(1+s_1^6s_2^6)Q(6;0)
+6(s_1s_2+s_1^5s_2^5)
\Bigl[
Q(6;0)+Q(5,1;0,n)+Q(4,2;0,n)
\nonumber\\
&+Q(3,3;0,n)+Q(2,4;0,n)+Q(1,5;0,n)
\Bigr]
+(s_1^2s_2^2+s_1^4s_2^4)
\Bigl[
15Q(6;0)+24Q(5,1;0,n)
\nonumber\\
&+33Q(4,2;0,n)
+36Q(3,3;0,n)
+22Q(2,4;0,n)
+24Q(1,5;0,n)
+6Q(3,2,1;0,n,2n)
\nonumber\\
&+12Q(2,3,1;0,n,2n)
+18Q(1,4,1;0,n,2n)
+6Q(2,2,2;0,n,2n)
+12Q(1,3,2;0,n,2n)
\nonumber\\
&+6Q(1,2,3;0,n,2n)
\Bigr]
+s_1^3s_2^3 
\Bigl[
20Q(6;0)+36Q(5,1;0,n)
+54Q(4,2;0,n)
\nonumber\\
&+62Q(3,3;0,n)
+54Q(2,4;0,n)+36Q(1,5;0,n)
+12Q(3,2,1;0,n,2n)
\nonumber\\
&+30Q(2,3,1;0,n,2n)
+36Q(1,4,1;0,n,2n)
+12Q(2,2,2;0,n,2n)
\nonumber\\
&+30Q(1,3,2;0,n,2n)
+12Q(1,2,3;0,n,2n)
+6Q(1,2,2,1;0,n,2n,3n)
\Bigr]. 
\end{align}

\subsection{$Z_{l}^{H}$}
For the 2-point functions of the Wilson line operators transforming in the rank-$k$ symmetric representation 
for $\mathcal{N}=2^*$ $U(N)$ SYM theory, 
we need the terms with $s_1^ks_2^k$ in the spectral zeta functions $Z_l^H(n)$, $l\le N$. 

For $l\ge 4$ and $k=0,1,2$ we have 
\begin{align}
Z_4^H&=
Q(4;0)
+4s_1s_2\Bigl[
Q(4;0)+Q(3,1;0,n)+Q(2,2;0,n)+Q(1,3;0,n)
\Bigr]
\nonumber\\
&+s_1^2s_2^2\Bigl[
10Q(4;0)+16Q(3,1;0,n)+18Q(2,2;0,n)+16Q(1,3;0,n)
\nonumber\\
&+4Q(3,1;0,2n)+4Q(2,2;0,2n)+4Q(3,1;0,2n)
\nonumber\\
&+12(1,2,1;0,n,2n)
+8(2,1,1;0,n,2n)
+8(1,1,2;0,n,2n)
\Bigr], 
\end{align}
\begin{align}
Z_5^H&=
Q(5;0)
+5s_1s_2\Bigl[
Q(5;0)+Q(4,1;0,n)+Q(3,2;0,n)
\nonumber\\
&+Q(2,3;0,n)+Q(1,4;0,n)
\Bigr]
+s_1^2s_2^2
\Bigl[
15Q(5;0)+25Q(4,1;0,n)+30Q(3,2;0,n)
\nonumber\\
&+30Q(2,3;0,n)+25Q(1,4;0,n)
+5Q(4,1;0,n)+5Q(3,2;0,n)
\nonumber\\
&+5Q(2,3;0,n)+5Q(1,4;0,n)
+15Q(2,2,1;0,n,2n)
+15Q(1,2,2;0,n,2n)
\nonumber\\
&+10Q(2,1,2;0,n,2n)
+20Q(1,3,1;0,n,2n)
+10Q(3,1,1;0,n,2n)
\nonumber\\
&+10Q(1,1,3;0,n,2n)
\Bigr], 
\end{align}
\begin{align}
Z_6^H&=Q(6;0)
+6s_1s_2 \Bigl[
Q(6;0)+Q(5,1;0,n)+Q(4,2;0,n)+Q(3,3;0,n)
\nonumber\\
&+Q(2,4;0,n)+Q(1,5;0,n)
\Bigr]
+s_1^2s_2^2 
\Bigl[
21Q(6;0)+36Q(5,1;0,n)+45Q(4,2;0,n)
\nonumber\\
&+48Q(3,3;0,n)+45Q(2,4;0,1)+36Q(1,5;0,n)+6Q(5,1;0,2n)
\nonumber\\
&+6Q(4,2;0,2n)+6Q(3,3;0,2n)+6Q(2,4;0,2n)+6Q(1,5;0,2n)
\nonumber\\
&+24Q(1,3,2;0,n,2n)+24Q(2,3,1;0,n,2n)
+18Q(3,2,1;0,n,2n)
\nonumber\\
&+18Q(1,2,3;0,n,2n)
+12Q(3,1,2;0,n,2n)
+12Q(2,1,3;0,n,2n)
\nonumber\\
&+30Q(1,4,1;0,n,2n)
+12Q(4,1,1;0,n,2n)
+12Q(1,1,4;0,n,2n)
\nonumber\\
&+18Q(2,2,2;0,n,2n)
\Bigr]. 
\end{align}


\bibliographystyle{utphys}
\bibliography{ref}

\end{document}